\documentclass[%
preprintnumbers,
 nofootinbib,
 amsmath,amssymb,
 aps,
 prd,
 twocolumn
]{revtex4-2}

\usepackage{graphicx}
\usepackage{xcolor}
\usepackage{dcolumn}
\usepackage{bm}
\usepackage[T1]{fontenc} 
\usepackage{graphicx}
\usepackage{epsfig}
\usepackage{xcolor}
\usepackage{rotating}
\usepackage{amssymb}
\usepackage{subfigure}
\usepackage{dsfont}
\usepackage{psfrag}
\usepackage{amsmath,euscript,array,mathrsfs}
\usepackage{bbold}
\usepackage{epsf}
\usepackage[utf8]{inputenc}
\usepackage{placeins}
\usepackage{dcolumn}

\newcommand{\beq}{\begin{equation}}
\newcommand{\eeq}{\end{equation}}
\newcommand{\beqs}{\begin{eqnarray}}
\newcommand{\eeqs}{\end{eqnarray}}

\begin{document}

\preprint{RIKEN-iTHEMS-Report-23}
\preprint{ET-0164A-23}

\title{ First-order phase transitions  in Yang-Mills theories\\
and the density of state method }

\author{Biagio Lucini,}
\email{b.lucini@swansea.ac.uk}
\affiliation{Department of Mathematics, Faculty  of Science and Engineering,
Swansea University, Fabian Way, SA1 8EN Swansea, Wales, UK}
\affiliation{Swansea Academy of Advanced Computing, Swansea University,
Fabian Way, SA1 8EN, Swansea, Wales, UK}
\author{David Mason,}
\email{2036508@Swansea.ac.uk}
\affiliation{Department of Physics, Faculty  of Science and Engineering,
Swansea University,
Singleton Park, SA2 8PP, Swansea, Wales, UK}
\affiliation{Department of Mathematics, Faculty  of Science and Engineering,
Swansea University, Fabian Way, SA1 8EN Swansea, Wales, UK}
\author{Maurizio Piai,}
\email{m.piai@swansea.ac.uk}
\affiliation{Department of Physics, Faculty  of Science and Engineering,
Swansea University,
Singleton Park, SA2 8PP, Swansea, Wales, UK}
\author{Enrico Rinaldi,}
\email{erinaldi.work@gmail.com}
\affiliation{Interdisciplinary Theoretical \& Mathematical Science Program, RIKEN (iTHEMS), 2-1 Hirosawa, Wako, Saitama, 351-0198, Japan}
\author{Davide Vadacchino}
\email{davide.vadacchino@plymouth.ac.uk}
\affiliation{School of Mathematics and Hamilton Mathematics Institute, Trinity
College, Dublin 2, Ireland}
\affiliation{Centre for Mathematical Sciences, University of Plymouth, Plymouth, PL4 8AA, United Kingdom}
 
\date{\today}%

\begin{abstract}

When studied at finite temperature,
Yang-Mills theories in $3+1$ dimensions
  display the presence of 
confinement/deconfinement phase transitions,
which are known to be of first order---the $SU(2)$ gauge theory being the exception.
Theoretical as well as phenomenological considerations
indicate that it is essential to establish a precise
characterisation of these
physical systems in proximity of such phase transitions.
We present and test a new method to study the critical 
region of parameter space in non-Abelian quantum field theories on 
the lattice, based upon the Logarithmic Linear Relaxation (LLR)
algorithm. We apply this method to the $SU(3)$ Yang Mills lattice gauge theory,
and perform extensive calculations with one fixed choice of lattice size.
We identify the critical temperature,
and measure interesting physical quantities 
near the transition. Among them, we determine the free energy of the model in the critical region, exposing for the first time its multi-valued nature with a numerical calculation from first principles, providing this novel evidence in support of a first order phase transition.
This study sets the stage for future high precision 
measurements, by demonstrating the potential of the method.
\end{abstract}

\maketitle

\tableofcontents

\section{Introduction}
\label{sec:introduction}

The  characterisation of phase transitions is a central topic of study
in theoretical physics, both 
for  reasons of principle and in view of applications.
In the proximity of second-order phase transitions, 
for critical values of the control parameters, the correlation length diverges,
hence such  systems can be classified in
universality classes,
distinguished by the value of  quantities
that are independent of the microscopic details.
But this is atypical, while many important physical systems undergo  first-order phase transitions,
which admit no notion of universality.
It is then essential to specify the details of the theory, and  identify computational strategies
optimised to the precise determination of model-dependent physical observables.
The latent heat is a particularly important example, as it determines the strength of the transition.

A case in point, within fundamental physics, is provided by the history of electroweak baryogenesis.
One of the three conditions identified by Sakharov~\cite{Sakharov:1967dj} to explain the matter-antimatter asymmetry
in the observable universe requires the dynamics to be out of equilibrium.
Hence,  the electroweak phase transition should be of first order and strong enough,
if it is to play a central role in these phenomena.
Testing this hypothesis required developing a programme of dedicated calculations.
The final outcome of this challenging endeavour 
is that electroweak baryogenesis cannot work within the standard model (SM)
of particle physics;
it  was demonstrated  non-perturbatively~\cite{Kajantie:1996mn,Karsch:1996yh,Gurtler:1997hr,
Rummukainen:1998as,Csikor:1998eu,Aoki:1999fi,
DOnofrio:2015gop}  that a line of first-order phase transitions
ends at a critical point, and that  the transition disappears into a cross-over,  except for unrealistically
 light Higgs boson masses, $m_h < {\cal O} (70)$ GeV.
This result still stands nowadays as 
prominent  evidence for new physics.
(See  Refs.~\cite{Laine:1998jb,Morrissey:2012db} for reviews, and also Ref.~\cite{Gould:2022ran}
for a recent non-perturbative update.)

New physics is needed 
also to explain the origin of dark matter, the existence of which
is supported by both observational astrophysics and cosmology. This  evidence
 motivates proposals postulating the existence of
hidden  sectors,  
comprised of  (dark) particles carrying no SM quantum numbers,  feebly coupled
to  SM particles (see, e.g., Refs.~\cite{Strassler:2006im,Cheung:2007ut,Hambye:2008bq,Feng:2009mn,
Cohen:2010kn,Foot:2014uba,
Bertone:2016nfn}).
Hidden sector dark matter scenarios find concrete realisations 
as composite dark matter (as, for example, in Refs.~\cite{DelNobile:2011je,
Hietanen:2013fya,Cline:2016nab,Cacciapaglia:2020kgq,Dondi:2019olm,
Ge:2019voa,Beylin:2019gtw,Yamanaka:2019aeq,Yamanaka:2019yek,Cai:2020njb})
and strongly interacting dark 
matter (see, e.g., Refs.~\cite{Hochberg:2014dra,Hochberg:2014kqa,Hochberg:2015vrg,Bernal:2017mqb,Berlin:2018tvf,
Bernal:2019uqr,Tsai:2020vpi,Kondo:2022lgg}).
Loosely inspired by quantum chromodynamics (QCD), 
their microscopic description  consists of new confining gauge theories,
 with or without matter field content.

If the new dark  sector undergoes a
first-order phase transition in the early universe, it would yield a relic stochastic background of 
gravitational waves~\cite{Witten:1984rs,Kamionkowski:1993fg,Allen:1996vm,Schwaller:2015tja,
 Croon:2018erz,Christensen:2018iqi}, potentially accessible  to 
a number of present and future gravitational-wave experiments~\cite{Seto:2001qf,
 Kawamura:2006up,Crowder:2005nr,Corbin:2005ny,Harry:2006fi,
 Hild:2010id,Yagi:2011wg,Sathyaprakash:2012jk,Thrane:2013oya,
 Caprini:2015zlo,
 LISA:2017pwj,
 LIGOScientific:2016wof,Isoyama:2018rjb,Baker:2019nia,
 Brdar:2018num,Reitze:2019iox,Caprini:2019egz,
 Maggiore:2019uih}.
Model-independent studies of the properties of such 
cosmological confinement phase transitions and their imprint on the stochastic gravitation 
background may adopt either of  two complementary theoretical strategies for investigation.  (See, e.g., 
Fig.~1 of Ref.~\cite{Huang:2020crf} but also 
Refs.~\cite{Halverson:2020xpg,Kang:2021epo,Reichert:2021cvs,Reichert:2022naa}.)
Either one  models the bubble nucleation rates by using the results of direct non-perturbative calculation of
latent heat, surface tension and other relevant dynamical quantities;
or one builds and constrains an effective description, such 
as the Polyakov loop model~\cite{Huang:2020crf,Kang:2021epo,Pisarski:2000eq,
Pisarski:2001pe,Pisarski:2002ji,Sannino:2002wb,
Ratti:2005jh,Fukushima:2013rx,Fukushima:2017csk,Lo:2013hla,Hansen:2019lnf}, or matrix 
models~\cite{Halverson:2020xpg,Meisinger:2001cq,Dumitru:2010mj,
Dumitru:2012fw, Kondo:2015noa,Pisarski:2016ixt,Nishimura:2017crr,Guo:2018scp,
KorthalsAltes:2020ryu,Hidaka:2020vna}, 
supplementing it by  thermodynamic information computed, again, non-perturbatively.

Either way, one arrives at a characterisation of the phase transition in terms of a set of 
parameters: critical temperature, $T_c$, percolation temperature, $T_{\ast}$,
strength of the transition, $\alpha$, inverse duration of  the transition, $\beta/H_{\ast}$, bubble wall velocity, $v_W$,
and number of degrees of freedom after the transition, $g_{\ast}$.
These are then used as input in the cosmological evolution,
 via existing software packages such as, for example, PTPlot~\cite{Caprini:2019egz},
 to obtain
 the power-spectrum of relic stochastic gravitational waves, $h^2 \Omega_{GW}$, 
 that can be compared with detector reach.

Hence, the first step towards 
calculating the power spectrum of gravitational waves requires  precise non-perturbative treatment of  the 
 dynamics, which can be provided by numerical simulations of lattice gauge theories.
The finite-temperature behaviour of many lattice gauge theories has been studied in the past;
for example, for $SU(N_c)$ see Refs.~\cite{Lucini:2002ku,Lucini:2003zr,Lucini:2005vg,Panero:2009tv,Datta:2010sq,Lucini:2012wq},
for $Sp(N_c)$ see Ref.~\cite{Holland:2003kg},
and for $G_2$ see Ref.~\cite{Pepe:2005sz,Pepe:2006er,Cossu:2007dk,Bruno:2014rxa}.
But these pioneering works were somewhat limited in scope, while 
dedicated high-precision measurements of specific observables,
in particular of the latent heat,
present technical challenges.
A handful of dedicated  lattice calculations has started to appear,
 focused on stealth dark matter 
with $SU(4)$ gauge dynamics~\cite{Appelquist:2015yfa,Appelquist:2015zfa,LatticeStrongDynamics:2020jwi},
or on  $Sp(4)$ gauge theories~\cite{Maas:2021gbf,Zierler:2021cfa,Kulkarni:2022bvh}.
A complementary approach to the study of the relevant out-of-equilibrium dynamics near criticality,
 bubble nucleation, and bubble wall velocity makes use of the non-perturbative tools provided by 
 gauge-gravity dualities~\cite{Maldacena:1997re,Gubser:1998bc,Witten:1998qj,Aharony:1999ti}, which can be generalised to strongly coupled systems exhibiting confinement and chiral symmetry 
 breaking~\cite{Witten:1998zw,
 Klebanov:2000hb,Maldacena:2000yy,Chamseddine:1997nm,Butti:2004pk,
 Brower:2000rp,Karch:2002sh,Kruczenski:2003be,Sakai:2004cn,Sakai:2005yt}---we  refer the reader to Refs.~\cite{Bigazzi:2020phm,Ares:2020lbt,Bea:2021zsu,Bigazzi:2021ucw,Henriksson:2021zei,Ares:2021ntv,Ares:2021nap,Morgante:2022zvc} and references therein for interesting examples  along these lines.

The history of the studies of $SU(3)$ gauge theories is quite interesting as a
general  illustration of how the 
field has been  evolving.
The recent Ref.~\cite{borsanyi:2022xml} critically summarises this history,
  discusses  the technical difficulties
intrinsic to current state-of-the-art lattice calculations, and 
 addresses some of  the challenges with the extrapolation to the continuum 
limit in proximity of the phase transition.
Among the salient points in such history, is the fact
that the pure gauge theory undergoes a first order phase transition, which 
has been accepted for a while~\cite{Svetitsky:1982gs,Yaffe:1982qf}.
Finite temperature lattice studies of the theory coupled to heavy quarks
have given encouraging results~\cite{Saito:2011fs,Ejiri:2019csa,Kiyohara:2021smr,
Fromm:2011qi,Cuteri:2020yke,
Borsanyi:2021yoz} but are still ongoing. More generally, considerable
activity is taking part in QCD, see Ref.~\cite{Aarts:2023vsf} for a recent summary. 
The thermodynamics of pure $SU(3)$ Yang-Mills theories  
has been studied intensively~\cite{Kajantie:1981wh,Celik:1983wz,Kogut:1983mn,
Svetitsky:1983bq,Gottlieb:1985ug,Brown:1988qe,Fukugita:1989yb,Bacilieri:1989ir,
Alves:1990pn,
Boyd:1995zg,Boyd:1996bx,
Borsanyi:2012ve,Shirogane:2016zbf}, and we know that the phase transition is not strong,
hence difficult to characterise.

Our interest in the characterisation of the confinement/deconfinement phase transition
originates in the ongoing research programme on $Sp(N_c)$ lattice gauge theories~\cite{Bennett:2017kga,Bennett:2019jzz,Bennett:2019cxd,Bennett:2020hqd,Bennett:2020qtj,
Bennett:2022yfa,Bennett:2022ftz,Bennett:2022gdz,Bennett:2023wjw,Bennett:2023rsl} 
and their composite bound states. Our long-term aim is to measure observable quantities, such
 as the latent heat at the transition,  which  have potential  implications for dark matter and for
 stochastic gravitational-wave
detection. We  approach this goal by exploiting a recent proposal,
which is based upon the density of states and provides
an alternative to Monte-Carlo importance sampling methods: the
Logarithmic Linear Relaxation (LLR) algorithm
~\cite{Langfeld:2012ah,Langfeld:2013xbf,Langfeld:2015fua,Cossu:2021bgn}.
We  describe the method  in the body of the paper.
It is worth mentioning that the literature on  finite-temperature studies 
of $Sp(N_c)$ gauge theories 
is rather limited~\cite{Holland:2003kg}. As is
the application of the LLR algorithm:  Abelian
 gauge theories have been studied in details~\cite{Langfeld:2015fua}, 
while  in the non-Abelian case the properties of $SU(3)$ have been investigated  at
 zero-temperature~\cite{Cossu:2021bgn}, and preliminary
 finite-temperature results exist for $SU(4)$~\cite{Springer:2021liy,
   Springer:2022qos} and $SU(N_c)$~\cite{Springer:2023wok}. 

We hence take a conservative approach.
In this paper, we apply the LLR algorithm to the best understood $SU(3)$ Yang-Mills theory.
We study the theory with one representative lattice, 
to set benchmarks for the future
 large-scale task of performing
 infinite volume and continuum limit extrapolations.
We confirm the presence of a metastability compatible with the
established first-order phase transition arising in the system,
determine the corresponding pseudocritical temperature using known
definitions, and measure the discontinuity that leads to the latent heat. 
Our results are consistent with other approaches,
and we can achieve the desired  precision level. 
In parallel, we started also to explore $Sp(N_c)$ Yang-Mills theories, in particular $Sp(4)$, about which we will report 
in a separate publication.

The paper is organised as follow.
We describe the LLR algorithm and its relation to the density of state
in Sect.~\ref{sec:method}. This section builds upon the method 
introduced in Ref.~\cite{Langfeld:2015fua},
and serves the purpose of setting the notation and making the exposition self-contained.
Sect.~\ref{sec:lattice} summarises the basic properties of the lattice theory of interest, and 
the definitions of the relevant observables.
The main body of the paper consists of Sects.~\ref{sec:observables} and~\ref{sec:results}.
This work sets the stage for our future investigations, discussed briefly in Sect.~\ref{sec:outlook}.
We relegate to Appendix~\ref{app:details}, \ref{app:moredetails}, and~\ref{app:convergence} technical details 
about the algorithm we use, and the tests we performed to validate it.
Some partial, preliminary results of the research project we report upon in  this paper have 
been presented in contributions to 
Conference Proceedings~\cite{Mason:2022trc,Mason:2022aka}, but here we present updated results,
including also a comprehensive and self-contained discussion
 of the procedure we follow, and an extended set of observables.


\vskip1.0cm

\begin{table*}
\setlength{\tabcolsep}{3pt}
  \caption{Parameters of the LLR algorithm used for the numerical computation of the
    density of states as formulated in this
   work for $(2N - 1)$ overlapping subintervals, each of amplitude
   $\Delta_E$, covering the relevant action interval, $[E_{\mathrm{min}},
   E_{\mathrm{max}}]$.}
 \label{tab:LLR_parameters}
\begin{center}
  \begin{tabular}{| c | l | l |  } 
 \hline
Symbol & Name/Role & Description/Purpose \\
 \hline 
$E_{\mathrm{min}}$  & minimal action & lower limit of the relevant
                                      action interval\\
$E_{\mathrm{max}}$  & maximal action & upper limit of the relevant
                                      action interval\\
$\Delta_E$ & amplitude of subintervals & controls the local
                                         first-order expansion of
                                         $\log \rho (E)$ \\
$\bar{m}$ & number of NR steps & enables to refine the
                                             initial guess for the
                                             $a_n$ \\
$\tilde{m}$ & number of RM updates & controls the tolerance
  on the convergence of the $a_n$\\
$n_{Th}$ & number of thermalisation steps per RM update & controls
                                                          decorrelation
                                                          between two RM updates\\
$n_M$ & number of measurements per RM update
              & controls the accuracy of the expectation values in Eq.~(\ref{eq:rm2})\\
$n_P$ & number of action-constrained updates per RM update & $n_{Th} + n_M$\\
$n_S$ & number of RM updates between swaps & ensures ergodicity of the algorithm\\
$n_R$ & number of determinations of the $a_n$ & enables to estimate
                                                statistical errors\\
    \hline
\end{tabular}
\end{center}
\end{table*}

\section{\label{sec:method}Density of states}
We start by defining the density of states, a quantity that plays
a central role in our calculations, and discussing its numerical
determination. The path integral of a Quantum Field Theory (QFT), with
degrees of freedom expressed by  
the field(s) $\phi$ and Euclidean action $S[\phi]$, can be written as 
\begin{equation}
\label{eq:pathintegral}
Z(\beta)\equiv\int [ D \phi ] e^{-\beta S [ \phi ]} \ ,
\end{equation}
where the coupling $\beta$ (not to be confused with $\beta/H_{\ast}$) has been exposed.
The density of states, $\rho(E)$,
is the measure of the hypersurface in field-configuration space 
spanned by the fields when the constraint $S = E$ is imposed:
\begin{equation}
\rho(E)\equiv\int [ D \phi ] \delta(S[\phi]-E) \ .
\end{equation}

Using the density of states, the path integral of the theory can be rewritten as
\begin{equation}
\label{eq:partition_function}
Z(\beta)=\int d E \rho(E) e^{-\beta E} \ .
\end{equation}

This expression of the path integral is particularly convenient
 for observables, $O(E)$, that only depend on the action, 
 since their expectation value can be reformulated as a one-dimensional integral:
\begin{equation}\label{eq:vev_obs}
 \langle O \rangle_\beta =\frac{\int dE \rho(E) O(E) e^{-\beta E}}{\int dE
   \rho(E) e^{-\beta E}} \ . 
\end{equation}
Hence, knowing the density of states provides a route
to the computation of these observables. In addition, as we show
later in this section, using the density of states, one can also access
observables that have a more general dependency on the fields,
not expressible in terms of the action alone. 
 
The density of states can be  computed efficiently by using the Linear
Logarithmic Relaxation (LLR)
method~\cite{Langfeld:2012ah,Langfeld:2015fua,Lucini:2016fid,Cossu:2021bgn}. The
algorithm exposed in this work is based on a variation of the LLR
algorithm with the replica exchange method introduced
in Ref.~\cite{Lucini:2016fid}, the key difference being that in this work
we are going to replace the two non-overlapping half-shifted replica
sequences with a single sequence of half-overlapping consecutive
subintervals. The algorithm depends on a set of tunable parameters,
which we introduce and describe in this section. For reference,
these parameters are summarised in Tab.~\ref{tab:LLR_parameters}.   

As a first step, we divide  an interval of interest, $E_{\mathrm{min}} \leq
E \leq  E_{\mathrm{max}}$, into $2 N - 1$ overlapping subintervals of fixed 
width, $\Delta_E = (E_{\mathrm{max}} - E_{\mathrm{min}})/N$, where each
of the subintervals but the first and the last have an overlap of amplitude
$\Delta_E/2$ with the preceeding and the following subinterval. 
As we shall see below, the overlap can 
be exploited to ensure the ergodicity of the algorithm. The
subintervals are numbered with an index, 
$n$, ranging from $n = 1$  (corresponding to central action value
$E_{\mathrm{min}} + \Delta_E/2$) to $2 N-1$ (central action
$E_{\mathrm{max}} - \Delta_{E}/2$). In each  subinterval $1\leq n
\leq 2 N - 1$, the central action is $E_n=E_{\mathrm{min}}+ n \Delta_E/2$. 
We approximate the density of states, $\rho(E)$, with the piecewise
linear function $\log \tilde{\rho}(E)\sim \log {\rho}(E)$,  defined as  
\beq
\label{eq:piecewise}
\log \tilde{\rho}(E) \equiv a_n \left(E - E_n \right) + c_n\,,
\eeq
where  $E_n - \Delta_E/4 \le E \le E_n + \Delta_E/4$, for each
$n$. This choice provides a prescription to deal unambiguously with
the overlapping regions in the subinterval, assigning each half of the
overlap to the subinterval with the nearest central action.  The
purpose of the LLR algorithm is to calculate numerically the $a_n$ and
$c_n$ coefficients, assuming continuity of the function $\log \tilde{\rho}(E)$ in the
interval $[E_{\mathrm{\rm min}},  E_{\mathrm{\rm max}}]$. 

As a second step, given any observable, $O(E)$, we  define $2N-1$ 
 restricted expectation values, $\langle \langle O \rangle \rangle_{n}$, as follows:
 \begin{widetext}
\beqs
\label{eq:truncated}
\langle \langle O \rangle \rangle_{n}(a)&\equiv &
 \frac{1}{{\cal N}_n(a)}
\int_{E_n-\frac{\Delta_E}{2}}^{E_n+\frac{\Delta_E}{2}} O(E) \rho(E)
e^{-a E} dE \ ,
\eeqs
\end{widetext}
where the normalisation factor is given by
\beqs
\label{eq:truncatedN}
 {\cal N}_n(a)& \equiv & \int_{E_n-\frac{\Delta_E}{2}}^{E_n+\frac{\Delta_E}{2}} \rho(E)
e^{-a E} dE \,. 
\eeqs
One now sees that if $a=a_n$ of Eq.~(\ref{eq:piecewise}), 
then the exponential factor inside the integrals in these definitions is 
$e^{-a_n E}=e^{-a_nE_n+c_n}/\tilde{\rho}$, and the constant factor, 
$e^{-a_nE_n+c_n}$, cancels between numerator and denominator,
so that the weight factor in the integrals in Eqs.~(\ref{eq:truncated}) and~(\ref{eq:truncatedN})
is just $\rho(E)/\tilde{\rho}(E)$.
The main idea behind the algorithm is that we consider $\tilde{\rho}(E)$  to be a good approximation of $\rho(E)$ 
if such weight factor, $\rho(E)/\tilde{\rho}(E)$,  for the restricted expectation 
value in the interval $[E_n - \Delta_E/2, E_n + \Delta_E/2$], is
approximately unit. More generally, we are interested in
expectation values, where we need this factor to be constant (we will deal with the
subinterval-dependent normalisation constant below).
Hence, we determine the value of $a_n$ iteratively, by imposing the condition:
\begin{widetext}
\beqs
\label{eq:dangle}
  \langle \langle \Delta E \rangle \rangle_{n}(a_n)&=& \frac{1 }{{\cal N}_n(a_n)} 
  \int_{\tiny E_n-\frac{\Delta_E}{2}}^{\tiny E_n+\frac{\Delta_E}{2}} \left(E - E_n \right) \rho(E) e^{-a_n E} dE
  \, =\, 0\,,
\eeqs
\end{widetext}
for each $n$.
The resulting stochastic equation makes use of the highly non-trivial information encoded in $\rho(E)$.
As long as $\Delta_E$ is sufficiently small, by
Taylor expanding $\log \rho(E)$ around $E_n$ in  Eq.~(\ref{eq:dangle}), one sees that 
\begin{equation}
    a_n = \left. \frac{\mathrm{d} \log \rho(E)}{\mathrm{d} E} \right|_{E = E_n} \ .
\end{equation}

For the  third step, we adopt  a combination of Newton-Raphson (NR) and 
Robbins-Monro (RM) algorithms~\cite{Robbins&Monro:1951}
to solve iteratively Eq.~(\ref{eq:dangle}). In a first sequence of iterations, 
we start from an initial trial value 
$a_n^{(0)}$ and recursively update it, using the relation 
\begin{widetext}
\beqs
\label{eq:nr}
a^{(m+1)}_n&=&a^{(m)}_n
- \frac{\langle \langle \Delta E \rangle
  \rangle_n(a^{(m)}_n)} {\langle \langle (\Delta E )^2 \rangle
  \rangle_n(a^{(m)}_n)} \\
\label{eq:nr2}
 &=&
  a^{(m)}_n
- \frac{ \left\langle  (E-E_n)
\left[\theta\left(E-E_n+\frac{\Delta_E}{2}\right)-\theta\left(E-E_n-\frac{\Delta_E}{2}\right)\right]
  \right\rangle_{a^{(m)}_n}
  }
{ \left\langle  (E-E_n)^2 
\left[\theta\left(E-E_n+\frac{\Delta_E}{2}\right)-\theta\left(E-E_n-\frac{\Delta_E}{2}\right)\right]
  \right\rangle_{a^{(m)}_n}
  }  
  \\
  \label{eq:nr3}
  &\simeq&
  a^{(m)}_n
- \frac{12}{\Delta_E^2} \left\langle  (E-E_n) 
\left[\theta\left(E-E_n+\frac{\Delta_E}{2}\right)-\theta\left(E-E_n-\frac{\Delta_E}{2}\right)\right]
  \right\rangle_{a^{(m)}_n} \,.
  \eeqs
    \end{widetext}
The above relation finds the root using one NR iteration. In the last step, the approximation consists
of assuming the validity of a second-order expansion for the density
of states in the action interval we are considering, which has been
used to express the denominator of the correction term in closed form.
The purpose of the initial NR iterations is to set up a convenient starting point for the more refined RM algorithm. 
This proves to be convenient, especially when insufficient prior knowledge is available on $a_n$. 
In these cases, even with rough initialisations, for suitable choices of $\bar{m}$,  $\bar{m}$ steps of the NR algorithm allows us to approach a value $a_n^{ (\bar{m})}\sim a_n$, in proximity of the true value of $a_n$. 
We then refine the process, by defining a new trial initial value
$a_n^{(0)}\equiv a_n^{ (\bar{m})}$ and recursively updating it using the 
modified relationship 
\begin{widetext}
\beqs
\label{eq:rm}
a^{(m+1)}_n&=&a^{(m)}_n
- \frac{\alpha}{m + 1} \frac{\langle \langle \Delta E \rangle
  \rangle_n(a^{(m)}_n)} {\langle \langle (\Delta E )^2 \rangle
  \rangle_n(a^{(m)}_n)} 
  \\
\label{eq:rm2}
  &\simeq&
  a^{(m)}_n
- \frac{\alpha}{m + 1} \left(\frac{12}{\Delta_E^2} \right)\left\langle  (E-E_n)
\left[\theta\left(E-E_n+\frac{\Delta_E}{2}\right)-\theta\left(E-E_n-\frac{\Delta_E}{2}\right)\right]
  \right\rangle_{a^{(m)}_n} \,.
\eeqs
\end{widetext}
This defines the RM step, which differs from the
NR one by the damping factor $\alpha/(m+1)$ of the
calculated correction term. While not a strict requirement of the
algorithm, for convenience we  fix the positive
constant  to $\alpha = 1$. Again, Eq.~(\ref{eq:rm2})
is obtained using a quadratic approximation of the density of states for a
closed-form computation of $\langle \langle (\Delta E )^2 \rangle
  \rangle_n({a^{(m)}_n})$, which
assumes that the action interval is sufficiently small for the
approximation to be sufficiently accurate. While the validity of this
approximation is not crucial in the NR steps, since they
are only used to refine the initialisation, it is more
important to verify its accuracy for the RM steps, since
the latter determine the values of the $a_n$ used in the calculation of
the observables. The check is performed by verifying that with the
obtained values of the $a_n$ the action is uniformly distributed in
the subinterval $n$ (i.e., its histogram is flat within a
predetermined tolerance), with the dynamics being compatible with a
random walk. Since $\Delta_E$ is a parameter of the calculation, it is
always possible to restrict the subinterval width so that the
quadratic approximation holds. 

The right-hand-sides of Eqs.~(\ref{eq:nr3}) and~(\ref{eq:rm2}) are
evaluated by computing ordinary ensemble averages through importance sampling methods, in which  the action  is restricted to a small interval, and the weight
redefined according to Eq.~(\ref{eq:truncated}). The restriction can
be done by rejecting update proposals that lead the action outside
the subinterval of interest or---as we will do in this work---imposing
constraints in the update proposals, so that each new trial value for
the field variables to be updated, automatically respects the
subinterval constraint (see Appendix~\ref{app:details} for further
details). The recursion converges to $a_n$ in the limit $m \to \infty$. We
truncate the recursion at step $\tilde{m}$ and repeat the process from
the start ensuring different random evolutions for $a^{(m)}_n$,
changing the initialisation of the random number sequences used in the
process of generating the restricted averages. This yields a gaussian-distributed
set of final values $a^{(\tilde{m})}_n$, with average $a_n$ and standard
deviation  proportional to $1/{\sqrt{\tilde{m}}}$, hence trading a
truncation  systematics with an error that can be treated
statistically.  

Restricting the averages to subintervals leads to ergodicity
problems. To ensure ergodicity, we use the fact that at any given
RM step the values of the actions in neighbour intervals
have a finite probability of being in the overlapping region. When
that happens, we can propose a 
Metropolis step that swaps the configurations in the two subintervals,
\begin{eqnarray}
\label{eq:swap}
  P_{\mathrm{swap}} = \min \left( 1, e^{(a_n^{(m)} -
  a_{n-1}^{(m)})(E_n^{(m)}- E_{n-1}^{(m)} )} \right) \ .
\end{eqnarray}
For these swap moves to be possible, simulations in the subintervals
need to run in parallel, with the synchronisation implemented by a
controller process. This can be easily implemented with standard
libraries such as the Message Passing Interface (MPI). 
However, even with this prescription, residual ergodicity problems
can derive from the fact that $E_{\rm min}$ and $E_{\rm max}$ would
otherwise be hard action 
cutoffs. The resulting lack of ergodicity is prevented by extending the action range
outside the  $[E_{\rm min}, E_{\rm max}]$ interval with two
truncated gaussians, one peaked at $E_{\rm min} + \Delta_E/2$ and
truncated at $E_{\rm min} + \Delta_E$, providing a prescription for
dealing with $E < E_{\rm min}$, and the 
other peaked at  $E_{\rm max} - \Delta_E/2$  and truncated at $E_{\rm
  max} - \Delta_E$ accounting for moves covering $E >
E_{\rm max}$. Ergodicity is recovered by choosing those truncated gaussians to
coincide with the Boltzmann factors associated with $\beta_{\rm upper}$ and
$\beta_{\rm lower}$---the values of $\beta$ at which the average
actions correspond to $E_{\rm min} + \Delta_E/2$ and to $E_{\rm max} - 
\Delta_E/2$ . Appendix~\ref{app:moredetails} describes how this is
achieved in practice. 

In our implementation, we propose a swap move between all neighbour intervals having energies in the overlapping regions after a fixed number, $n_S$, of RM updates. 
Each RM update consists of $n_P = n_{Th}+n_{M}$ action-constrained updates. 
The $n_{Th}$ updates decorrelate the configurations between RM updates, then $n_{M}$ configurations are used for the calculation of expectation values in equation Eq.~(\ref{eq:rm2}).
This sequence of steps is repeated until we have performed $\tilde{m}$
RM updates. As discussed previously, this process leads to a
determination of the $a_n$ for all values of $n$. Repeating it with
different random number sequences, we get gaussian distributed
values of $a_n$, which can be used in a bootstrap analysis to provide a
determination of the statistical uncertainty on observables. 

Having determined  the values of interest for $a_n$,  continuity of
$\tilde{\rho}(E)$ at the boundaries of the subintervals requires
\begin{align}\label{eq:c_n}
c_n =  c_1 + \frac{\Delta_E}{4} a_1 + \frac{\Delta_E}{2} \sum_{k=2}^{n-1} a_k +
\frac{\Delta_E}{4} a_n \ , 
\end{align}
for all values of $n > 1$, and with the summation taking effect only when the upper
index is bigger or equal to the lower one, i.e. for $n \ge 3$. This
conditions leaves the value of $c_1$ as a free parameter. $c_1$ can be
fixed by imposing a known global normalisation condition. For instance,  
for $\beta = \infty$, the density of states must be equal to the
number of degenerate vacua of the system. Nevertheless, in
applications where the knowledge of the value of the path integral per
se is not interesting, as is the case of observables expressed as
ensemble averages, the normalisation of the density of states can be
fixed arbitrarily. In these cases, for simplicity, we choose $c_1=0$. 

Having discussed the rationale for the various components of the
method, for convenience, we now provide a summary of the algorithm.
\begin{itemize}
\item Divide the interval $[E_{\mathrm{min}}, E_{\mathrm{max}}]$ in
  $2N - 1$ half-overlapping subintervals of amplitude $\Delta_E$,
  centered at energies $E_1 = E_{\mathrm{min}} + \Delta_E/2$, $\dots$,
  $E_n = E_{\mathrm{min}} + n \Delta_E/2$, $\dots$, $E_{2N - 1} =
  E_{\mathrm{max}} - \Delta_E/2$. Define two half-Gaussians for
  prescribing rules for accepting/rejecting moves outside the interval
  $[E_{\mathrm{min}}, E_{\mathrm{max}}]$, with the correct Boltzmann
  distribution.
\item Repeat $n_R$ times with different random sequences:
  \begin{enumerate}
  \item Initialise the values of $a_n$,
  \item Perform $\bar{m}$ steps of the NR algorihm,
    Eq.~(\ref{eq:nr2}),
  \item Repeat $\tilde{m}$ times:
    \begin{enumerate}
    \item Perform $n_P$ action-constrained updates (see
      Appendix~\ref{app:details}),
    \item Update $a_n$ according to the RM prescription,
      Eq.~(\ref{eq:rm2}),
      \item Repeat (a)-(b) $n_S$ times, then,
    \item Propose a configuration swap according to Eq.~(\ref{eq:swap}).

    \end{enumerate}
  \end{enumerate}
\end{itemize}
Note that the swap step implies that the determination of $a_n$
happens in parallel in the calculation, and requires a synchronisation
of the parallel processes. The parameters used in the algorithm are referenced in
Tab.~\ref{tab:LLR_parameters}. This algorithm provides $n_R$
statistically independent determinations of $a_n$, and,
consequently, of $c_n$, up to a prescription for fixing $c_1$, as
described earlier in this section. 

As shown in Ref.~\cite{Langfeld:2015fua}, 
the use of the LLR algorithm
introduces a $\Delta_E$-dependent systematic error.
Therefore, finite volume estimates of the 
quantities above can only be obtained after an extrapolation towards $\Delta_E=0$
has been performed.
We devote Appendix~\ref{app:convergence} to a discussion of this process,
for the lattice parameters adopted in this study. 

We conclude by observing that the LLR method enables us to compute also
canonical ensemble averages at coupling $\beta$ of 
observables $B[\phi]$ that have a dependency on the field $\phi$ not leading to
an explicit dependency on the action, using the formula~\cite{Langfeld:2015fua}:
\begin{equation}\label{eq:vev_gen}
 \langle B[\phi] \rangle_\beta = \frac{1}{Z(\beta)} \sum_{n=1}^{2N - 1}
 \frac{\Delta_E}{2} \tilde{\rho}(E_n) \tilde{B}[\phi] \ , 
\end{equation}
 where
\begin{equation}
\tilde{B}[\phi] = \langle \langle B[\phi]\exp{\left( -\beta S[\phi] + a_n
     (S[\phi] - E_n)\right)} \rangle \rangle _n(a_n) \ .
\end{equation}

\section{Lattice system}
\label{sec:lattice}

We compute ensemble averages with the distribution in
the partition function of Eq.~\eqref{eq:pathintegral}, by discretising the degrees of freedom on a lattice. 
We focus on the action, $S$, of a four-dimensional Yang-Mills theory with non-Abelian gauge group $SU(N_c)$ in Euclidean space, discretised  as
\begin{equation}\label{eq:latticeaction}
    S \, = \, \sum_{p} \left(1 - \frac{1}{N_c}{\rm Re}{\rm Tr}(U_p)\right) 
\,,
\end{equation}
which enters Eq.~(\ref{eq:pathintegral}) with bare lattice coupling constant $\beta = \frac{2N_c}{g_0^2}$,
related to the bare gauge coupling $g_0^2$.  The summation runs over all the elementary plaquette 
variables, $U_p$, on the four-dimensional grid. 
Sampling of the link variables, $U$, representing the discretised gauge potential,
 entering the construction of the plaquette, are discussed in Appendix~\ref{app:details}. The measure is the product of integrals over the links.

We use hypercubic lattices with $\tilde{V}/a^4=  N_T \times N_L^3$  points and 
isotropic lattice spacing $a$, in both temporal and spatial directions.
We adopt periodic boundary conditions for the link variables in all directions. 
The thermodynamic temperature of the $SU(N_c)$ Yang-Mills 
theory is $T = \frac{1}{aN_T}$. The lattice spacing, $a$, is dynamically controlled 
by the coupling,  $\beta$, through the 
non-perturbative beta function of the theory, hence knowing $\beta$ 
we can determine the temperature, $T$.

The order parameter for confinement 
is the Polyakov loop, $l_p$, defined as
\begin{equation}\label{eq:Ploop}
l_p \equiv \frac{1}{N_c N_L^3} \sum_{\vec{x}} \mathrm{Tr}~\prod_{k=0}^{N_T-1} U_0 (k\, \hat{0}, \vec{x})\, ,
\end{equation}
with $ \hat{0}$  the unit vector in the time direction.
For $T<T_c$, the system lies in its confined phase, $\langle l_p\rangle_\beta=0$. 
For $T>T_c$ it lies in its deconfined phase,
$\langle l_p\rangle_\beta \in \mathbb{Z}_{N_c}$, and the $\mathbb{Z}_{N_c}$ symmetry of the 
action, Eq.~(\ref{eq:latticeaction}), is spontaneously broken.

As the phase transition is of first order, 
it is characterised by a 
discontinuity in the first derivative of the
free energy with respect to the temperature $T$, for $T=T_c$,
that is recast in terms of
the internal energy (density), defined as
\begin{equation}\label{eq:internalE1}
\varepsilon(T) \equiv \frac{\kappa T^2}{V} \frac{\partial \ln Z(T)}{\partial T}~\,,
\end{equation}
where $\kappa$ is the Boltzmann constant, which we set to $\kappa=1$,
and $V=N_L^3 a^3$  is the spatial volume.
At the phase transition, two distinct equilibrium states exist, 
with energies equal to $\epsilon_+$ and $\epsilon_-$. 
The magnitude of the discontinuity across the transition, $L_h\equiv |\varepsilon_+-\varepsilon_-|$, is known as
\emph{latent heat}. 
At exactly $T=T_c$, the system exhibits macroscopic configurations characterised 
by the presence of separating surfaces, 
on either side of which the order parameter has different values,
and hysteresis can be observed in the evolution of the system.
The configurations on either side of the transition have equal free energy, hence  the rate
of tunnelling is the same in either direction, giving rise to phase co-existence.

Yet, even when $T\simeq T_c$ the physical system can still tunnel between 
the confined and deconfined phases, as the finiteness of the lattice system 
allows for metastable states to be physically realised in portions of the space,
though they have finite life time.
These phenomena present standard lattice algorithms with intrinsic difficulties,
as the finiteness of the system implies that ensemble averages have non-trivial
contaminations from metastable states, which ultimately smoothen the aforementioned 
non-analyticity characterising the transition.
For example one still finds that  the susceptibility of the Polyakov loop,
\begin{equation}\label{eq:PloopSus}
\chi_l(\beta) \equiv \langle |l_p|^2 \rangle_\beta - \langle |l_p| \rangle^2_\beta ,
\end{equation}
is expected to be  maximal at $T=T_c$, though finite,
and  establishing the existence of a first-order phase transition requires non-trivial 
studies of the finite-volume scaling of $\chi_l$.
As we will see, the LLR algorithm removes this difficulty, as it
allows to access individually the physically stable and unstable states in configuration space,
hence removing the practical problems due to tunnelling and their effects on ensemble averaging.

Our aim is to characterise the phase transition in the $SU(3)$ Yang-Mills theory,
and ultimately determine the quantities 
$T_c$ and $L_h$, 
extrapolated to infinite volume and to the continuum limit. 
In the body of this paper,  we take a first step in this 
direction, by studying a fixed lattice with
$N_T=4$ and $N_L=20$.

\section{Methodology}
\label{sec:observables}

In this section, we provide precise relations between
the values of the two main observables, 
critical temperature, $T_c$,
and latent heat, $L_h$, and the density of states, $\rho(E)$,
or, rather, its numerical estimate
$\tilde\rho(E)$.

When studying a lattice theory by Monte Carlo sampling, 
a typical signal of tunnelling between vacua is the presence of hops 
in the simulation-time evolution of the value
of the action per plaquette, $u_p$, defined as
\beqs
u_p&\equiv&\frac{a^4}{6\tilde{V}} \sum_p \frac{1}{N_c}{\rm Re}{\rm Tr}(U_p)\,.
\eeqs
Thanks to the relation  ${E}=6\tilde{V} (1-u_{p})/a^4$,
the (partial) distribution function, $P_\beta(u_p)$, can be expressed
as a function of $E=S$, and defined in terms
of the density of states $\rho(E)$ as follows:
\begin{equation}\label{eq:plaq_distribution}
P_\beta\left(u_p=1-\frac{a^4}{6\tilde{V}}E\right)=\rho(E) \frac{e^{-\beta E}}{Z(\beta)}~.
\end{equation}
where $Z(\beta)$ is the partition function, 
Eq.~(\ref{eq:partition_function}). 
In proximity of the critical region of parameter space, for a system that undergoes a first-order phase transition between two possibly local vacua, we expect that the distribution function display a characteristic double peak structure. 
The two values of the energy, ${E_{\pm}}=6\tilde{V} (1-u_{p\,\pm})/a^4$, at which 
 $P_\beta(u_p)$ is maximal, 
determine the  energy of the two phases. 
On a finite volume $\tilde{V}$, we define the critical temperature $T=T_c$
(and hence $\beta=\beta_c$,
as we are keeping the volume fixed while changing the coupling $\beta$) 
 as the temperature at which
the system
tunnels between configurations in different phases 
with the same rate in either direction. 
Hence, the peaks must have equal height, and the relation
\begin{equation}
P_{\beta_c}(u_{p\,+}) = P_{\beta_c}(u_{p\,-})\,
\end{equation}
can be used to determine the critical coupling  $\beta_c$.

As the temperature, $T$, of the gauge theory is a function
of $\beta$, we recast the derivative with 
respect to $T$ in Eq.~(\ref{eq:internalE1}) as a derivative 
with respect to $\beta$, following Ref.~\cite{Lucini:2005vg}. Direct calculation of the energy density requires the computation of Karsch coefficients\cite{Karsch:1982ve}, which is outside the scope of this work. However, assuming a vanishing pressure gap, as motivated by Ref.~\cite{Shirogane:2016zbf}, the latent heat can be related to the plaquette jump, $\Delta \langle u_p \rangle_{\beta_c} = |u_{p+} - u_{p-}|$, via
\begin{equation}\label{eq:latheat}
\frac{L_h}{T_c^4} = - \left( 6 N_t^4 a
\frac{\partial\beta}{\partial a} 
{\Delta \langle u_p \rangle_\beta }\right)_{\beta=\beta_c}{}\, ,
\end{equation}
where $a(\partial\beta/\partial a)$ can be calculated by setting the scale ---see the discussion leading to 
Eq.~(35) of Ref.~\cite{Lucini:2005vg}. 
In this work we focus on the calculation of the plaquette jump.

We define the effective potential, $W_{\beta}(E)$, as
\begin{eqnarray}
W_{\beta}(E) \equiv - \log \langle \delta \left( S - E \right) \rangle_\beta = - \log P_{\beta}(E) \ .
\end{eqnarray}
At criticality, $W_{\beta}(E)$ displays two degenerate minima at the values of $E$ 
of the equilibrium states of the two coexisting phases.\footnote{
If we replace $E$, which depends on  $u_p$,
with the extremal $\langle E\rangle_{\beta}=\frac{6\tilde{V}}{a^4}(1-\langle u_p\rangle_{\beta})$, 
for each $\beta$, 
then $W_{\beta}(\langle E\rangle_{\beta})$ is independent of $\beta$, and is a pure functional of 
the response function $\langle E\rangle_{\beta}$. In other words, this is the 
Legendre transform of the logarithm of the partition function $Z(\beta)$, 
in which $\beta$ is the  external source.}

 The quantum system defined in Eq.~(\ref{eq:pathintegral}) can be said to have
 internal energy $E$ and entropy $s=\log \rho(E)$, following the prescription of  the microcanonical ensemble.
 We then define a temperature $t$  as $1/t(E)\equiv \partial s/\partial E$,
 in analogy with thermodynamics, 
and  the free energy $F$  as the Legendre transform
\begin{equation}
\label{eq:free_F}
F(t) \equiv E - t s\,,
\end{equation}

All the quantities mentioned above can be estimated 
from the approximate density of states
$\rho(E) \simeq \tilde\rho(E) $.
In particular, thanks to Eq.~(\ref{eq:piecewise}), 
the entropy for $E\simeq E_n$ is
\begin{widetext}
\begin{equation}\label{eq:entropy}
   s (E\simeq E_n)=  c_1+\frac{\Delta_E}{4} (a_1+a_n)+\frac{\Delta_E}{2}\,\sum_{k=2}^{n-1} 
    a_k
    +
    a_n\left(E - E_n\right)\,,
\end{equation}
\end{widetext}
and the corresponding temperature is hence
\begin{equation}\label{eq:microT}
t = \frac{1}{a_n}\,.
\end{equation}

The computation of $s(E)$ (and of $F(t)$), 
is affected by an  ambiguity on the value
of $c_1$, as  mentioned at the end of Section~\ref{sec:method}. 
This ambiguity can in principle be fixed by
requiring that $s(E)$  be positive for all $E$ and 
 vanish as $t\to 0$.

Our estimate $\tilde{\rho}$ is obtained by computing the sequence of values  
$\left\{a_n\right\}_{n=1}^{2N-1}$ using the LLR algorithm, as outlined in Sect. \ref{sec:method}. The LLR parameters chosen for this work are shown in Tab.~\ref{tab:parameter_values}. Note that in the specific application studied here NR iterations were considered unnecessary, hence $\bar{m}=0$. Once $\tilde{\rho}$ is known, all relevant observables are known as well, by using the relations reported in this section.
\begin{table}
\caption{\label{tab:parameter_values}The LLR parameters used for this study}
\begin{center}
\begin{tabular}{|c|c|c|c|c|c|c|}
\hline
Parameter & $\bar{m}$ & $\tilde{m}$ & $n_{Th}$ & $n_M$ & $n_S$ & $n_R$\\
\hline
Value & 0 & 500 & 200 & 500 & 1 & 20 \\
\hline
\end{tabular}
\end{center}
\end{table}

There are two further numerical details worth discussing here, before we move onto
presenting our results. 
Firstly, the LLR algorithm requires trial initial $a_n$ vales.
We have computed the average action for evenly spaced $\beta$ values. Linearly fitting this and inverting can give an initial estimate for the relation between $a_n$ and $E_n$. The guess is then refined through a small number of RM iterations.    
The starting values $\left\{a^{(0)}_n\right\}$ have been thus set
for each energy interval over the relevant energy range of the system.
The preliminary runs are also useful to locate the energy ranges
$[E_\mathrm{\rm min},\,E_\mathrm{\rm max}]$ that are relevant to the study of the phase transition. 
On the basis of preliminary analyses 
we set $a^4 E_\mathrm{\rm min}/(6\tilde{V})\approx 0.44$ and $a^4 E_\mathrm{\rm max}/(6\tilde{V})\approx 0.46$.

Second, for each $n$, the
coefficient $a_n$ is obtained by truncating
the sequence $\left\{a_n^{(m)}\right\}$ 
of RM updates
at a value of $m$ for which we expect
the asymptotic $1/\sqrt{m}$ behaviour of the standard deviation to
have set in. Due to the centrality of this behaviour for the correct
working of the algorithm, the corresponding test is the first
numerical result we report in the next section.

\section{Results}
\label{sec:results}

\begin{figure}[t]
\centering
\includegraphics[width=0.45\textwidth]{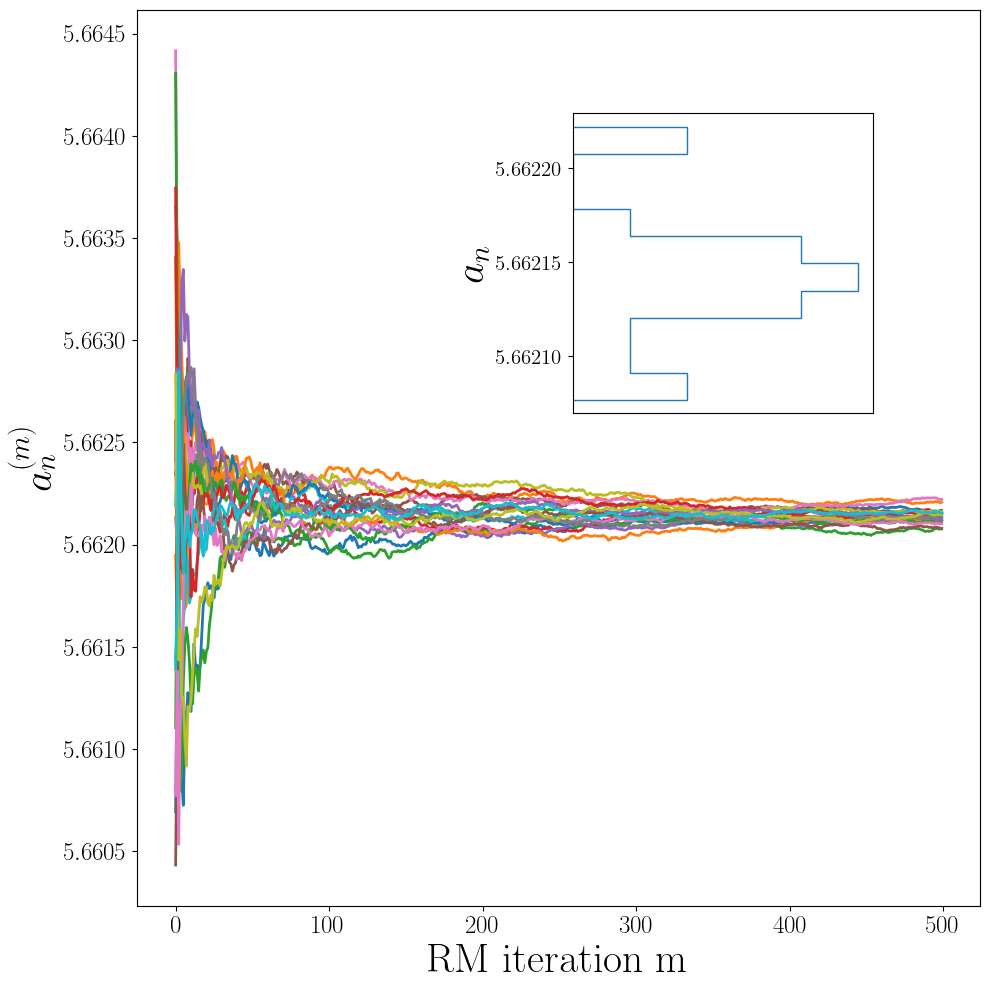}
\caption{\label{fig:RM_iterations}Twenty histories of $a_n^{(m)}$ as they are updated through 500 RM iterations, in different colours. 
This example is for a system with energy interval size $a^4 \Delta_E/(6\tilde{V})=0.0007$, and the interval centered at 
$u_p=0.540676$. 
 The inset shows the distribution of the final $a_n$ values.
 }
\end{figure}

In order to verify the convergence of the Robbins-Monro algorithm,
we  study the distribution of  the value of $a_n^{(m)}$,  for each energy interval,
as a function of the iteration number,
$m$. An example of this process is displayed in 
Fig.~\ref{fig:RM_iterations}, for the energy interval centred
at $a^4 E_n/(6\tilde{V})=0.459324$  $(u_p = 0.540676)$, with $a^4 \Delta_E/(6\tilde{V})=0.0007$. The figure shows  in 
different colours twenty independent  Robbins-Monro trajectories. 
The trajectories are characterised by large oscillations
at small $m$, followed by convergence to a common value at large $m$.
The distribution of the  asymptotic behaviour of the final values has 
standard deviation that scales as
$\sim 1/\sqrt{m}$. 
On the basis of extensive  test runs, we identified $\tilde{m} = 500$ iterations of the
RM algorithm as providing a good estimate of $a_n$. 
We verified that by this stage the twenty final estimates  are 
normally distributed around their average value.

\begin{table}
 \caption{The values of $\Delta_E$ used for this analysis, for each choice of the number 
 of intervals, $2N-1$, in the energy range $\left[E_\mathrm{\rm min},\,E_\mathrm{\rm max}\right]$. The four-dimensional
 lattice has space-time volume
 $\tilde{V}/a^4=4\times 20^3$. Numerical values have been rounded to 4 decimal places for convenience.}
 \label{tab:num_intervals_deltaE}
\begin{center}
\begin{tabular}{ |c|c|c|c| } 
 \hline
$2N-1$ & $\frac{a^4}{6\tilde{V}}\Delta_E$ & $\frac{a^4}{6\tilde{V}}E_{\rm min}$ & $\frac{a^4}{6\tilde{V}}E_{\rm max}$ \\ 
 \hline
$8$ & $0.0063$ & $0.4374$ & $0.4659$ \\ 
$15$ & $0.0030$ & $0.4380$ & $0.4619$ \\ 
$28$ & $0.0015$ & $0.4387$ & $0.4604$ \\ 
$55$ & $0.0007$ & $0.4391$ & $0.4601$ \\ 
$108$ & $0.0004$ & $0.4394$ & $0.4598$ \\ 
 \hline
\end{tabular}
\end{center}
\end{table}

\begin{figure}[t]
\centering
\includegraphics[width=0.45\textwidth]{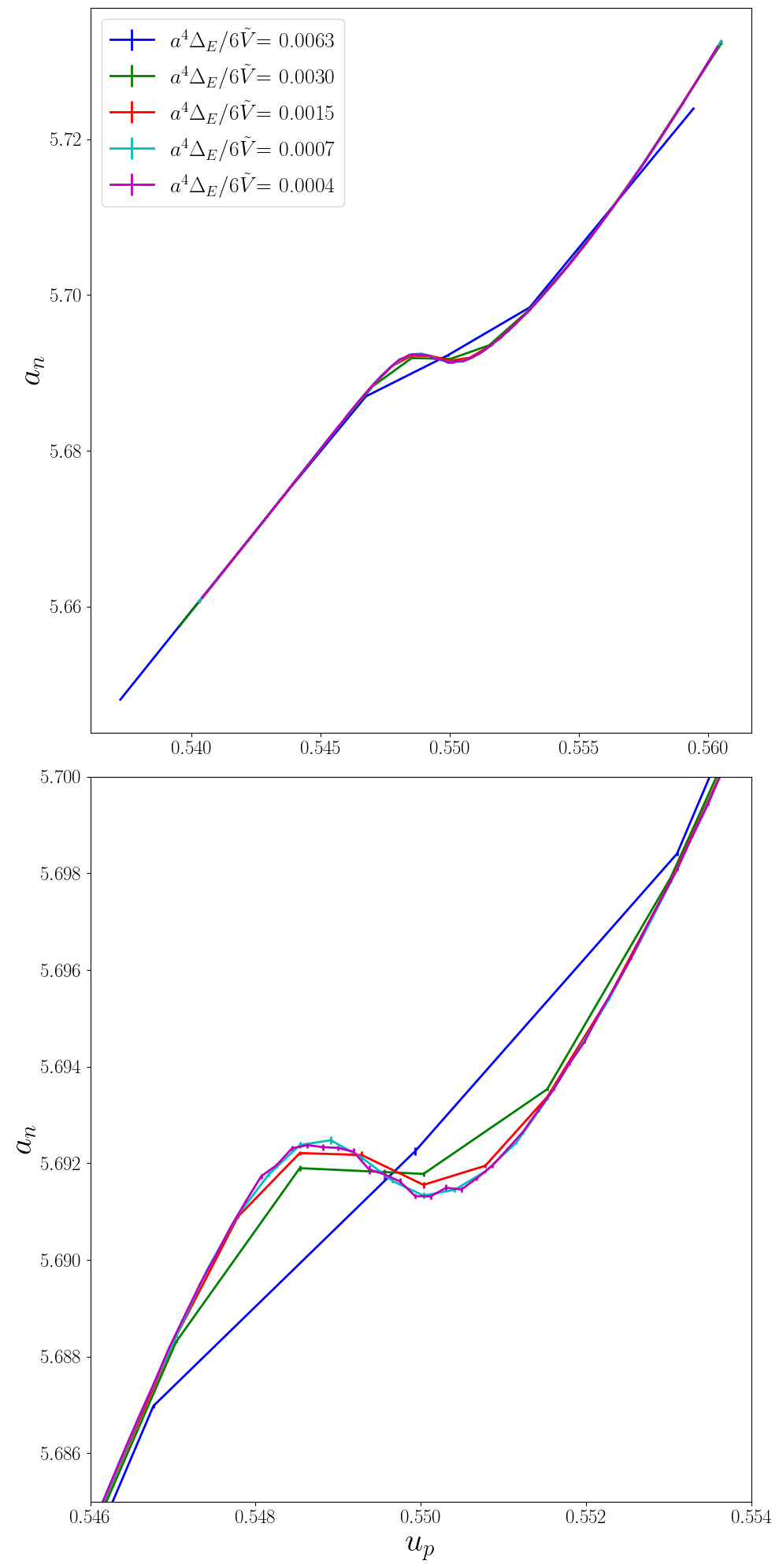}
\caption{\label{fig:ak_Ek} Measurements of $a_n$, 
 plotted against $u_p$ at the centre of the energy intervals. 
Errors are estimated by bootstrapping over 20 repeats.
Each value of $a_n$ is the final result of $500$ iterations of the Robins-Monro algorithm,
for a lattices with $\tilde{V}/a^4=4\times 20^3$. 
The different curves/colors refer to different choices of $\Delta_E$---see Table~\ref{tab:num_intervals_deltaE}---with 
the roughest results corresponding to the largest choices of $\Delta_E$.
The bottom panel is a detail of the top one, focusing on the region in which $a_n(u_p)$ is not invertible.}
\end{figure}

To extrapolate our results towards
the $\Delta_E\to 0$ limit, we vary the number of sub-intervals $2N-1$, and repeat the process of computing the estimates of $a_n^{(m)}$. The corresponding
values of $a^4 \Delta_E/(6\tilde{V})$, and of the intervals analysed, are reported in 
Tab.~\ref{tab:num_intervals_deltaE}.

In Fig.~\ref{fig:ak_Ek}, we show our measurements of $a_n$, with their uncertainty,
as a function of $u_p = 1 - a^4 E_n/(6\tilde{V})$. The different curves show the results for several different values 
of $\Delta_E$. 
For $\Delta_E$ sufficiently small ($a^4 \Delta_E/(6\tilde{V})\leq0.0030$), 
a characteristic limiting shape starts to
emerge in $a_n$ as a function of $u_p$, with the presence of  one 
local minimum, one local maximum, and an inflection point between them. 
The resulting  non invertibility of $a_n(u_p)$  is closely 
related to the qualitative features of $P_\beta(u_p)$, as discussed in Sect.~\ref{sec:observables}, and
to the presence of a first-order phase transition. 
Setting $\Delta_E$  to smaller values,  the curve $a_n(u_p)$ becomes
smoother, which reduces the
magnitude of the systematic error due to  $\Delta_E$ itself.

\begin{figure}[t]
\centering
\includegraphics[width=0.45\textwidth]{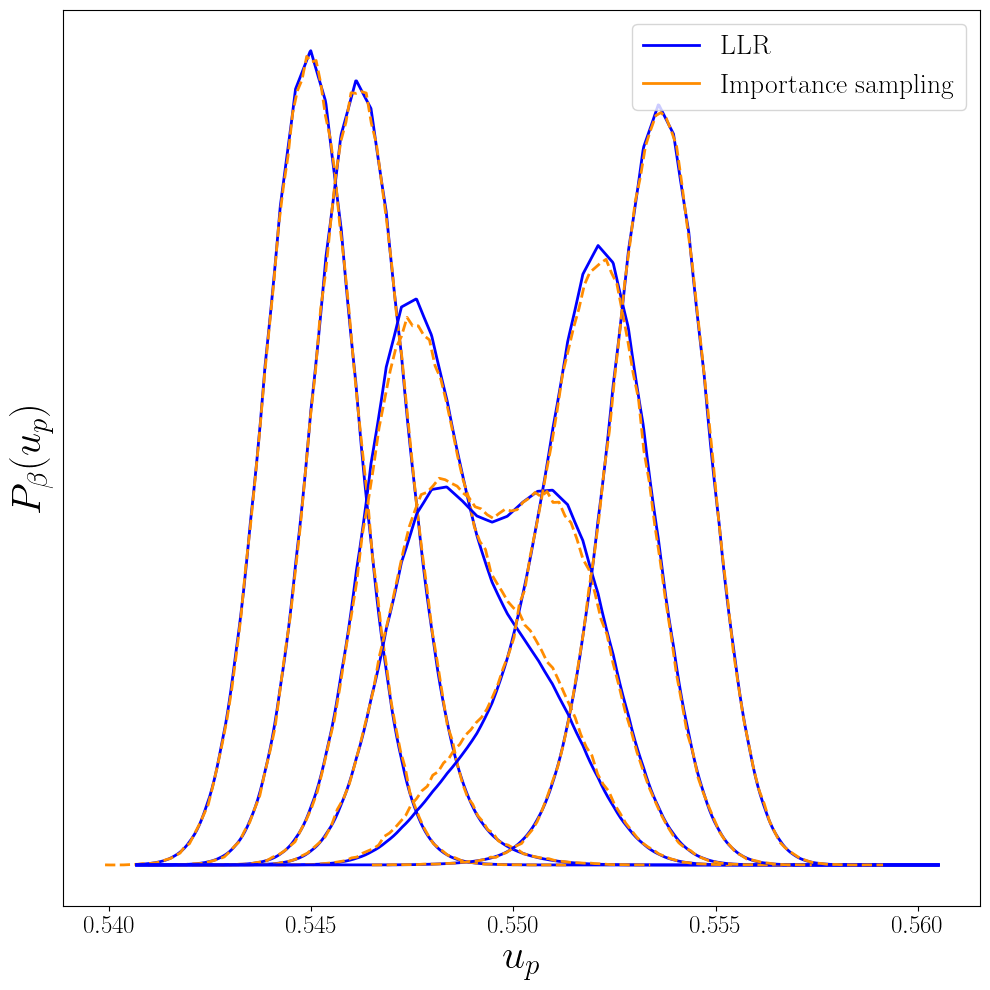}
\caption{\label{fig:plaq_distribution}  The probability distribution,  $P_\beta(u_p)$,
computed with  LLR method with $a^4\Delta_E/\tilde{V} = 0.0007$ and $\tilde{V}/a^4=4\times 20^3$ 
(solid blue), compared to the same probability distribution computed with the standard 
lattice method based upon importance sampling (dashed orange). 
The probability density for different values of the coupling $\beta$ are shown, from left to right: $\beta = 5.68000$, $\beta = 5.68500$, $\beta = 5.69000$, $\beta = 5.69187$, $\beta = 5.69500$ and $\beta = 5.70000$.}
\end{figure}

The probability distribution of the average plaquette
is obtained from Eq.~(\ref{eq:plaq_distribution}). 
Estimates of $P_\beta(u_p)$
 are displayed in
Fig.~\ref{fig:plaq_distribution}. The solid blue lines are our results, obtained using the 
LLR method. We compare them directly with the orange dashed lines 
obtained by using the standard importance sampling approach.
Agreement between the two is evident, yet
small discrepancies are visible in the neighbourhood of the
maxima and of the local minimum of $P_\beta(u_p)$. 
We show a number of examples displaying a single
peak, but  for $\beta=5.69187$
 two peaks of similar height are present. As explained
in Sect.~\ref{sec:observables}, this is the expected signal of
a first order phase transition. 

\begin{figure}[t]
\centering
\includegraphics[width=0.45\textwidth]{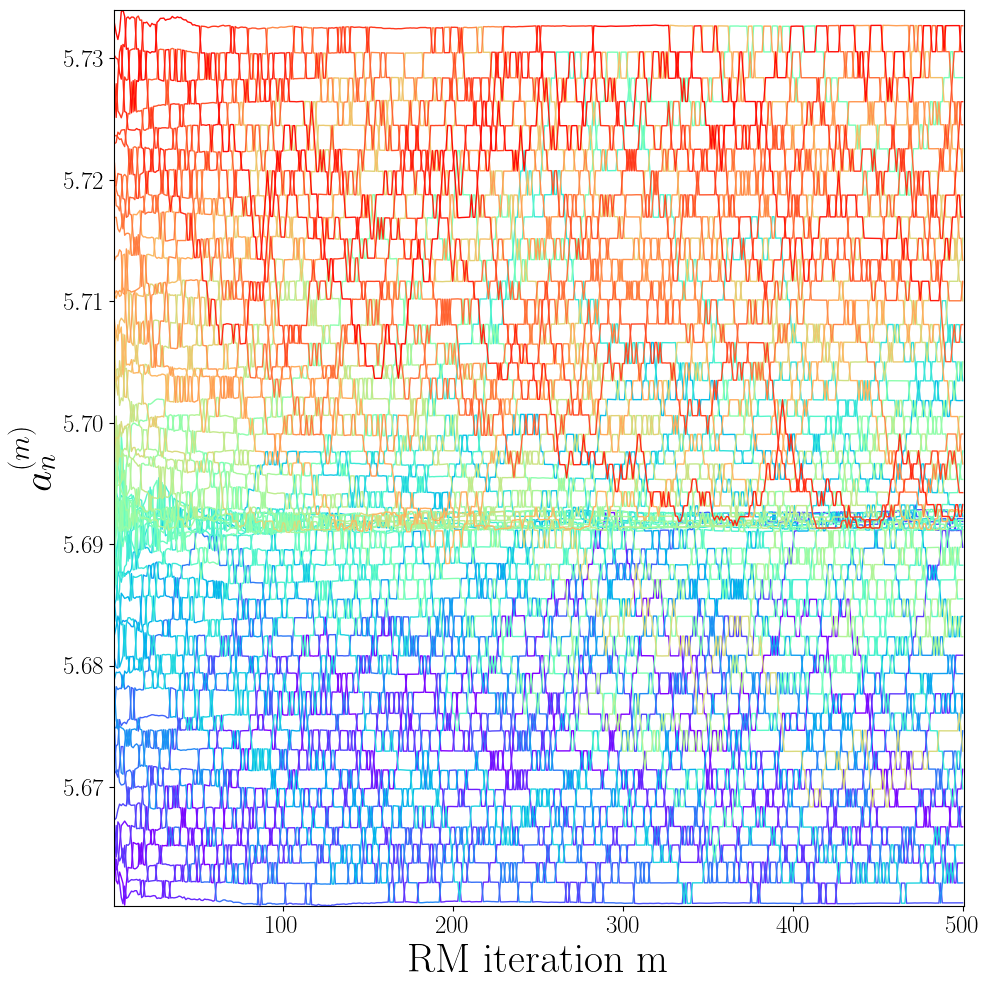}
\caption{\label{fig:Umbrella} The trajectories of the set $\{a_n\}_{n=1}^{2N-1}$ are shown for 500 Robbins-Monro updates for a single run with configuration swaps included, for $a^4 \Delta_E/(6\tilde{V})=0.0007$. The colours are determined by the energy interval that a given lattice system started in and follow them as they are exchanged between energy intervals.  
 }
\end{figure}

As discussed in Sect.~\ref{sec:method}, after each RM update,
configuration swaps are considered, to ensure ergodicity of the
algorithm. Figure \ref{fig:Umbrella} shows the evolution of the full
set of $\{a_n\}_{n=1}^{2N-1}$ against the RM iteration, m. Following
the track of the colours on the diagram shows how the configurations
are swapped. The clustering of values around $a_n\sim 5.692$ is due to
non invertibility of $a_n(u_p)$ in the critical region. The
diagram shows that, although in general terms there is an appreciable
rate of exchange of configurations, it appears to be less probable to
exchange configurations across the two different phases. 

\subsection{Critical T and latent heat}

The importance of the measurement of the critical value $\beta_c$ and of the position of the
 peaks of $P_{\beta_c}(u_p)$
($u_{p+}$ and $u_{p-}$) is
 explained in Sect.~\ref{sec:observables}.
In proximity of the transition at each value of $\beta$, a double gaussian 
function can be fitted to $P_\beta(u_p)$, using
the location of the local maxima and the width of the 
peaks as fitting parameters. The best-fit parameters
are functions of $\beta$. 
An estimate of $\beta_c$ can then be obtained by solving
the equation $|P_+-P_-|=0$, where $P_+=P_{\beta}(u_{p+})$ and $P_-=P_{\beta}(u_{p-})$, with the bisection method. 
The numerical values of $u_{p\,\pm}$ and $\beta_c$ can then used for the calculation of the 
latent heat through Eq.~(\ref{eq:latheat}).

\begin{figure}[t]
\centering
\includegraphics[width=0.45\textwidth]{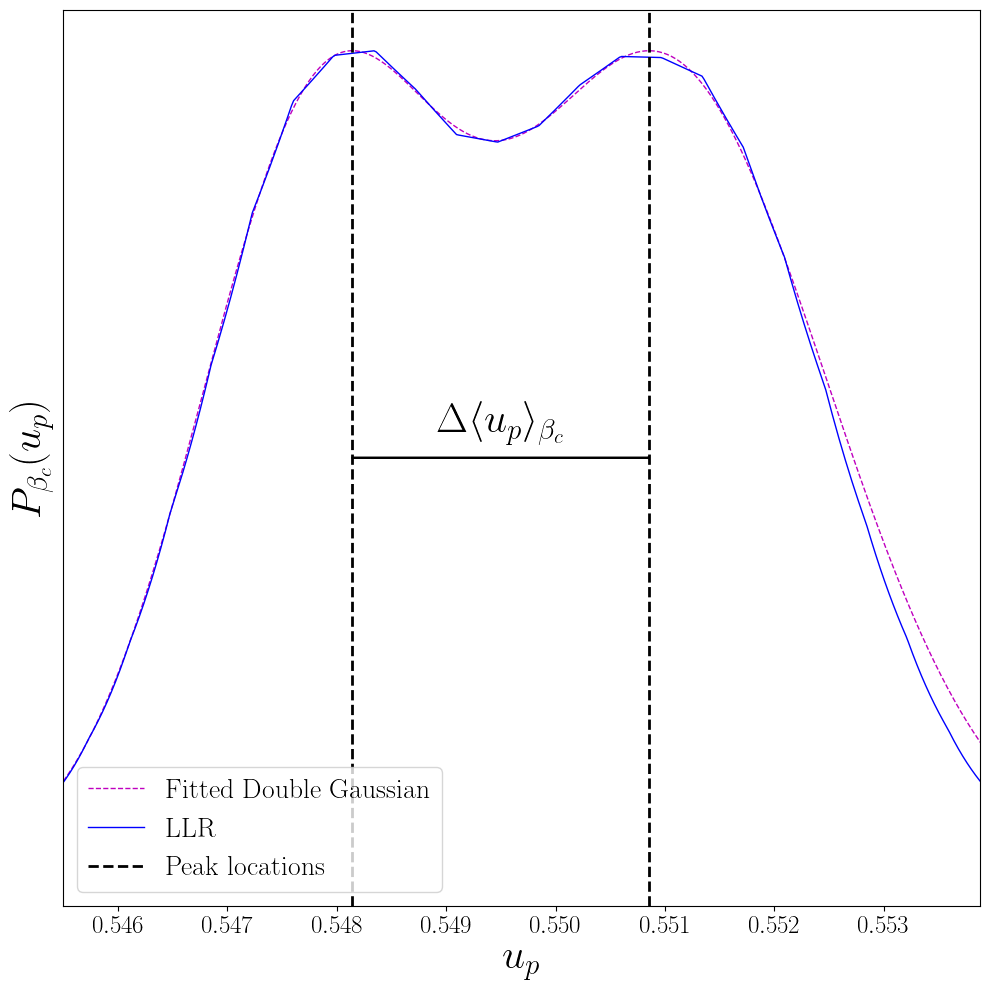}
\caption{\label{fig:DoublePeak_LatentHeat} The double peak structure of 
the plaquette probability distribution,  $P_\beta(u_p)$, for $\beta=\beta_c = 5.69196$
tuned to the critical point for this specific LLR run, when 
the two  peaks have equal height. A double gaussian fitted at the peaks 
of the plots is shown in magenta (dashed), and compared with the blue (solid) line 
representing the LLR numerical results. We also show explicitly the location of the
two peaks.}
\end{figure}

A representative example of the numerical results
obtained from the LLR method, 
displaying also a fitted double gaussian, is displayed in 
Fig.~\ref{fig:DoublePeak_LatentHeat}. The agreement
between the numerical and fitted curves is very good, with small deviations
only appearing at the boundaries of the interval of $u_p$
depicted in the plot, which are not of primary importance in the fitting procedure.

\begin{figure}[t]
\centering
\includegraphics[width=0.45\textwidth]{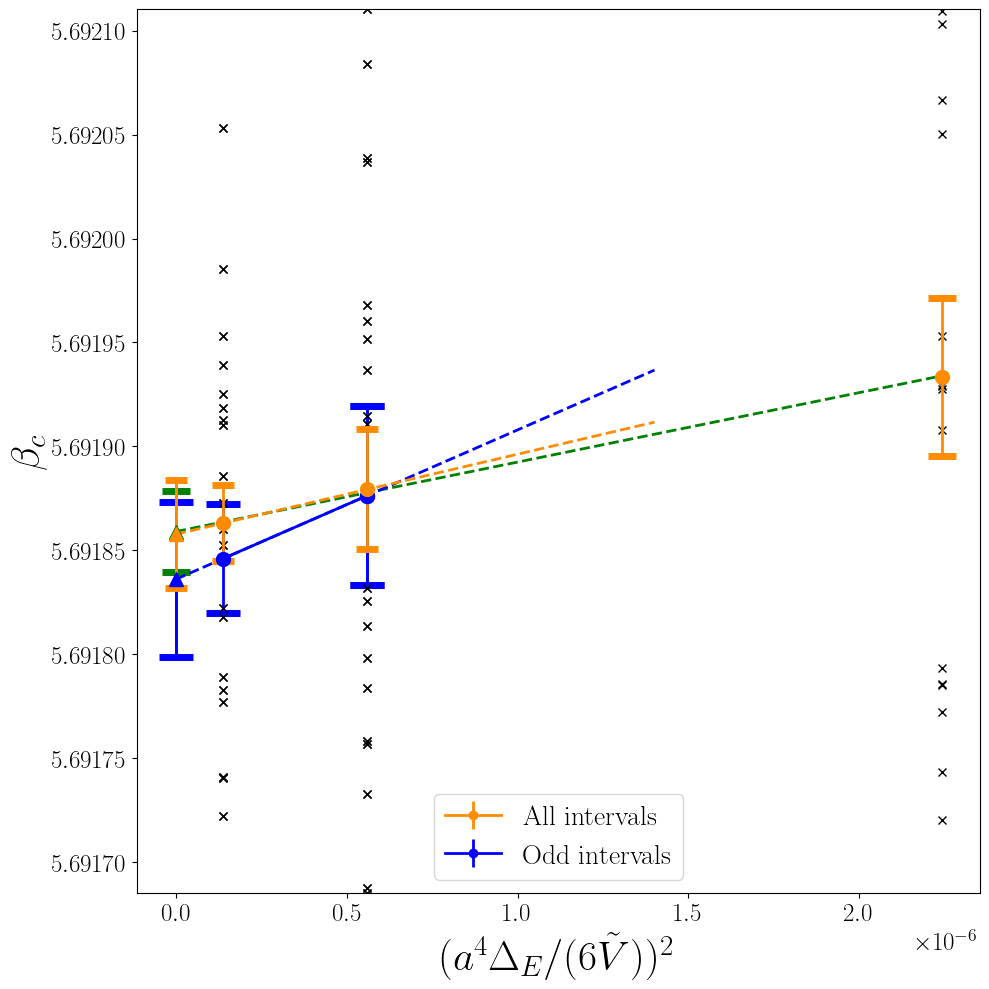}
\caption{\label{fig:betac_vs_deltaE} Estimates of the critical coupling
$\beta_c$ as a function of the square of the energy interval, $\Delta_E^2$. The black crosses show the values of the critical coupling determined for each of the 20 repeats when all of the energy intervals are used in the determination of the plaquette distribution. 
The orange circles are the mean values of the black crosses and the error is found by bootstrapping them. 
The orange dashed line and triangle is a $\Delta_E \to 0$ extrapolation of the two finest results when all intervals are included, while the green dashed line shows the extrapolation of all three points. 
The blue circles show results when only the odd numbered intervals are used  $\{a_{2n-1},E_{2n-1}\}_{n=1}^{N}$, an extrapolation to the $\Delta_E \to 0$ limit is shown by the blue line and the blue triangle. 
The coarsest point included in this graph only contains a small number of intervals in the critical region, making the double peak structure in the plaquette distribution difficult to resolve. All three final extrapolations are compatible with each other within errors.
}
\end{figure}

\begin{figure}[t]
\centering
\includegraphics[width=0.45\textwidth]{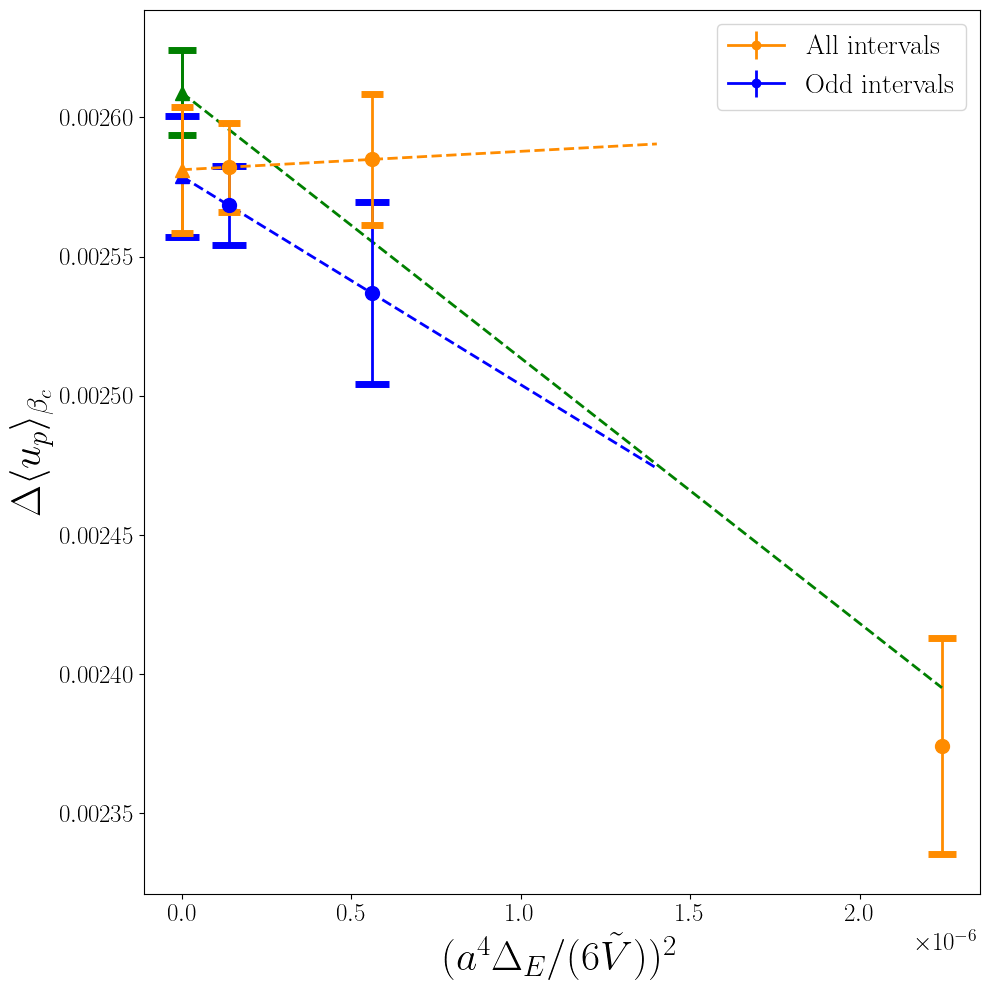}
\caption{\label{fig:dE_vs_deltaE} Estimates of the plaquette jump at
  the critical point, the difference between the plaquette values at
  the two peaks of the plaquette distribution when the double peak
  structure has peaks of equal height, as a function of the square of
  the energy interval, $\Delta_E^2$. The orange circles are the values
  found when all energy intervals are used to calculate the plaquette
  distribution, the error was found with a bootstrap procedure. The orange dashed line and triangle is an extrapolation of the two finest results when all intervals are included to the $\Delta_E \to 0$ limit, while the green dashed line shows the extrapolation of all three points. The blue circles show results when only the odd numbered intervals are used  $\{a_{2n-1},E_{2n-1}\}_{n=1}^{N}$, an extrapolation to the $\Delta_E \to 0$ limit is shown by the blue line and the blue triangle. The coarsest point included in this graph only contains a small number of intervals in the critical region, making the double peak structure difficult to resolve. All three final extrapolations are compatible with each other within errors.}
\end{figure}

Our estimates of $\beta_c$ and $\langle \Delta u_p \rangle_{\beta_c}$,
 for each choice of 
$\Delta_E$, are displayed in Figs.~\ref{fig:betac_vs_deltaE}
and~\ref{fig:dE_vs_deltaE}. The numerical values are reported in
Tab.~\ref{tab:latbet}. We perform a linear fit of the behaviour
of both $\beta_c$ and $\langle \Delta u_p \rangle_{\beta_c}$ as a function of $\Delta_E^2$. These fits are 
displayed in Figs.~\ref{fig:betac_vs_deltaE} 
and~\ref{fig:dE_vs_deltaE}. 
We found the $\chi ^2$ of the
linear fit to be much less than 1. 
The origin of its smallness lies in the large error in the determination of
$\beta_c$ at fixed $\Delta_E$. Several contributions to the errors of 
$\beta_c$ have been carefully analyzed and 
accounted for, except for those
originating from the correlation between
different subintervals (leading to correlations across each set of $a_n$)
and the error on the fit of the double gaussian itself. Calculations in
different subintervals would indeed be completely 
independent, were it not for the configuration swapping, which is necessary 
to achieve ergodicity in the sampling of configuration space. 
Since canonical observables, such as $P_\beta(u_p)$, are determined from several values 
of $a_n$, they are affected by these autocorrelations. In order to
approximately quantify the magnitude of these effects on the final
estimate of $\beta_c$, we have computed this quantity from only half the $a_n$
estimates, i.e. only computed using the non-overlapping odd numbered energy intervals $\{a_{2n-1},E_{2n-1}\}_{n=1}^{N}$. 
The coarsest example in the plot ($a^4 \Delta_E/\tilde{V} = 0.0015$) has also been treated separately as it only contains a small number of intervals in the critical region. Extrapolations both including and excluding this point have been carried out, as well as an extrapolation using the points with only half the energy intervals. All extrapolations agree with one another within errors.

\begin{table}
\caption{The values of the critical coupling, $\beta_c$, and the difference between the peaks of the probability distribution at this coupling, $\Delta\langle u_p\rangle_{\beta_c}$, for different energy interval sizes $\Delta_E$. The table shows results obtained in two ways. Either all intervals are included or only the odd intervals, $\{a_{2n-1},E_{2n-1}\}_{n=1}^{N}$. The symbol `$\to 0$' is used to denote the result of an extrapolation. The table contains three extrapolations. An extrapolation using the results when only odd intervals are considered, an extrapolation when all points with all the intervals are used and an extrapolation of the two finest interval sizes when all intervals are used. See also Ref.\cite{Lucini:2005vg}
 \label{tab:latbet}
}
\begin{center}
\begin{tabular}{|c|c|c|c|}
\hline
  & $\frac{a^4\Delta_E}{6\tilde{V}}$ & $\beta_c$ & $\Delta\langle u_p\rangle_{\beta_c}$ \\
\hline
Odd intervals            & $0.0007$ & $5.69188(4)$ & $0.00254(3)$ \\
Odd intervals            & $0.0004$ & $5.69185(3)$ & $0.00257(2)$ \\
All intervals            & $0.0015$ & $5.69193(4)$ & $0.00237(4)$ \\
All intervals            & $0.0007$ & $5.69188(3)$ & $0.00258(2)$ \\
All intervals            & $0.0004$ & $5.69186(2)$ & $0.00258(2)$ \\
Odd intervals            & $\to 0$  & $5.69184(4)$ & $0.00258(2)$ \\
All intervals all points & $\to 0$  & $5.69186(3)$ & $0.00258(2)$ \\
All intervals 2 points   & $\to 0$  & $5.69186(2)$ & $0.00261(2)$ \\
\hline
\end{tabular}
\end{center}
\end{table}

\subsection{Thermodynamic potentials}
As discussed in Sect.~\ref{sec:observables}, the LLR algorithm, through the estimation of $\rho(E)$, allows us to estimate
the thermodynamic potentials of the bulk system. We focus
our attention  on the free energy, $F$, defined in Eq.~(\ref{eq:free_F}), 
the entropy, $s$, defined in Eq.~(\ref{eq:entropy}), and the (microcanonical) temperature,
$t$, in Eq.~(\ref{eq:microT}). 
As we showed explicitly in Fig. \ref{fig:ak_Ek}, $a_n(u_p)$ and therefore $t(E)$ is not globally invertible.
Yet, we can study how $F$ evolves as a function of $t$, by piece-wise inverting $t(E)\leftrightarrow E(t)$.

\begin{figure}[t]
\centering
\includegraphics[width=0.45\textwidth]{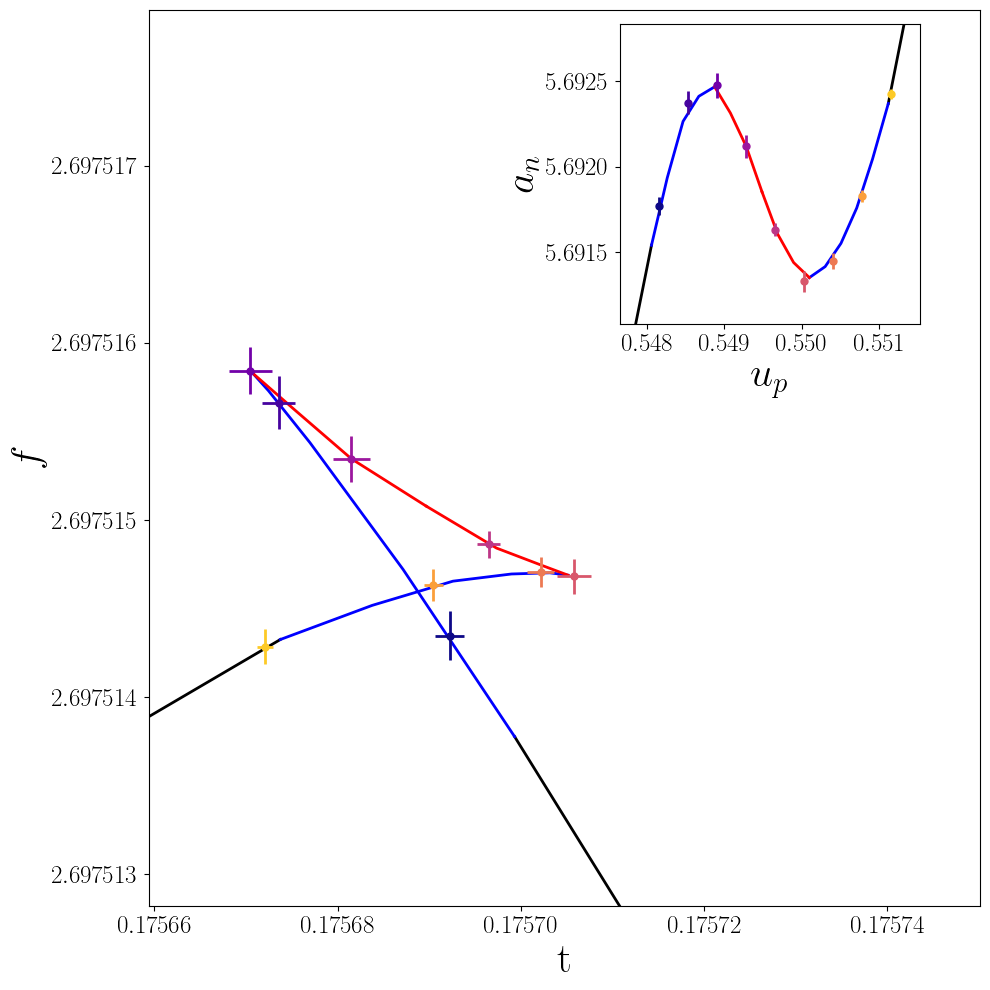}
\caption{\label{fig:FreeEnergy}The (subtracted) free energy, $f$,  as
a function of the (discretised, microcanonical) temperature $t=1/a_n$. $f$ is defined
 in the main body of the text.
The dots (with errors) represent the values of   $f$
corresponding to the centre of each energy interval used in the LLR algorithm, while
 the solid line is reconstructed by piece-wise linearly interpolation of $a_n(u_p)$. 
 The color coding of the points and solid lines 
 are chosen to match those in the inset, displaying our numerical results for $a_n(u_p)$.
 In black we show the regions of $t$ for which $f$ is single-valued, in blue we show the (meta-)stable 
 solutions within the region where $f$ is multi-valued, and in red the
 unstable (tachyonic) branch of solutions. The dots follow a colormap
 in the value average plaquette, with darker colors corresponding to
 smaller $u_p$.}
\end{figure}

In order to best expose the behaviour of $F(t)$, we consider 
a \emph{subtracted} free energy, defined as $f = a^4(F(t)+\Sigma t)/(\tilde{V})$.
The constant $\Sigma$, as we anticipated after Eq.~(\ref{eq:microT}), reflects the existence of an 
arbitrary additive 
constant in $s$, which we are now removing, with an approximate numerical procedure.
The subtracted free energy is displayed
in Fig.~\ref{fig:FreeEnergy}, as a function of (discretised) $t=1/a_n$.
For the purpose of producing this figure,  $\Sigma $ has been calculated as
 the average of the  entropy over the interval of (microcanonical) temperatures displayed in the plot---which would correspond 
to the average gradient of the curve. This rough estimate is not equivalent to imposing the third law of thermodynamics
($\lim_{t\rightarrow 0} s=0$), but suffices for our current purposes, and allows us to avoid the 
expensive process of repeating the LLR procedure for choices of $E_n$ that lie far away from the critical region.

In Fig.~\ref{fig:FreeEnergy}, the uncertainty in the numerical extraction of $a_n$
affects both axes of the plot. 
 The values of $f$
corresponding to the piece-wise linear interpolation are represented
in the main plot as colored lines.
The plot clearly shows the multi-valued nature of the free energy, the location of the
temperature corresponding to criticality in the thermodynamic limit, 
and details about stable, metastable  and tachyonic 
branches of configurations of the system.
The discontinuity in the first derivative is located at
$t_c \simeq 0.175690$, which is the temperature at which  the system undergoes a
first-order phase transition.

\begin{figure}[t]
\centering
\includegraphics[width=0.45\textwidth]{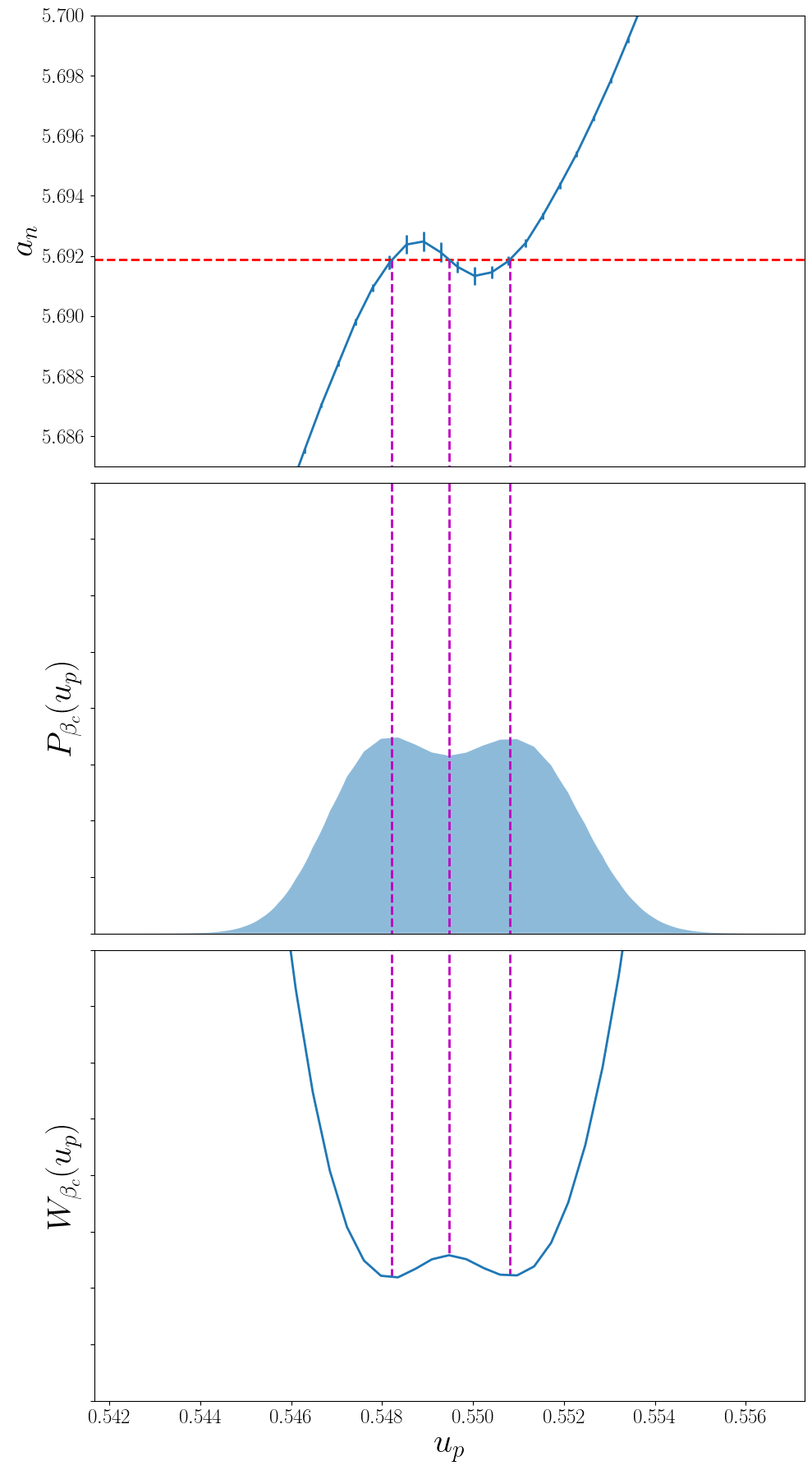}
\caption{\label{fig:Criticalpoint_aUV} 
Top panel: $a_n$ ($=1/t_n$) against 
$u_p$ at the centre of the energy intervals, in close proximity of the critical region.
Middle panel: the reconstructed plaquette distribution, $P_\beta(u_p)$, at critical
coupling $\beta_c = 5.69187$. 
Bottom panel: quantum effective potential 
for the plaquette at the critical coupling. The red line in the top plot shows 
the critical value of the coupling and its relation to the microcanonical 
temperature, $a_n$. The magenta (vertical, dashed) lines show the locations at 
which the red line intersects the curve  $a_n(u_p)$.}
\end{figure}

The relation between different branches of $F(t)$ and physical 
stability is illustrated in Fig.~\ref{fig:Criticalpoint_aUV}. The inverse temperature
$a_n$ is displayed as a function of $u_p=1-a^4E_n/6\tilde{V}$ in the top panel.
The plaquette probability distribution at the critical point,
$P_{\beta_c}(u_p)$, is depicted in the middle panel.
The corresponding effective potential is plotted in the bottom panel. 
While the two configurations corresponding to maxima of 
$P_{\beta_c}(u_p)$ are both absolute minima of the effective potential, 
a third configuration, corresponding to a local minimum of the probability, is a local maximum of the effective potential.

\section{Outlook}
\label{sec:outlook}

With this paper, we set the basis of a systematic research programme that exploits
the properties of the LLR method to yield future high precision measurements  
characterising lattice gauge theories in proximity of the
confinement/deconfinement phase transition.\footnote{See
  Refs.~\cite{Springer:2021liy,Springer:2022qos,Springer:2023wok} for
  early results of studies that use a similar approach.}
The method is powerful  and promises to yield 
information that is difficult to access otherwise,
as it modifies Monte Carlo sampling by restricting it to arbitrarily small energy windows.
It hence 
 provides numerical access to the details of the physics in regions of parameter space
exhibiting all the typical feature of first-order phase transitions:
 phase coexistence, metastability and/or instability of multiple branches of solutions, non-invertibility 
 and/or multi-valuedness of some state function.
 
 We showed how the information from these energy-bound Monte Carlo feeds into recursive
 relations (e.g., an implementation of the 
 Robbins-Monro algorithm) that can determine the density of states for any interesting range of energies.
 And we provided explicit relations between the density of states and 
  observables such as the critical temperature and  the latent heat.
Furthermore, we found that the results for the density of states can be recast in terms of an effective free
  energy and an effective potential that exhibit with spectacular level of resolution the details of the
  physics near the transition.
  
We  restricted this study to the $SU(3)$ lattice
 Yang-Mills theory, and performed it with 
one choice of lattice parameters, fixing $N_L=20$ and $N_T=4$.
The trademark   of the LLR algorithm is that we found clear evidence of the first-order nature of the transition,
without the need of a finite-volume study, and an extrapolation of the scaling to large volumes.
The physically interesting observables need to be extrapolated to the continuum and infinite-volume limits, 
with dedicated, extensive numerical work, which would allow for a direct comparison with results that use different numerical techniques. In the future we plan to repeat the process with larger values of both $N_T$ and $N_L$, which will provide us with control over lattice systematics.

We plan to apply this process to other theories, in particular
those based on the sequence of symplectic groups $Sp(2N)$,
which might play an important role in models of dark matter, and hence 
in the physics of the early universe, by yielding a potentially detectable  stochastic background of  gravitation waves.
 In particular, the precise measurement of the effective potential, $W_{\beta}(u_P)$, the results of which are 
exemplified in Fig.~\ref{fig:Criticalpoint_aUV}, can be used to obtain a precise determination not just of the parameter, $\alpha$, 
controlling the strength of the phase transition, 
but also of the inverse duration of the transition, $\beta/H_{\ast}$.
The latter is challenging to estimate from first principle, yet it
is necessary in the calculation of the power-spectrum of stochastic gravitational waves, $h^2\Omega_{GW}$.

There are still some limitations to what we are able to do at this stage of development of this technique, 
and we would like
to address them in the future. The first such challenge has to do with scalability and
 parallelisation of the algorithm
and software: the attentive reader will certainly be aware of the fact the energy constraint we are imposing is 
globally defined on the whole lattice configuration, a constraint 
that cannot be immediately parallelised, because it requires communication between different parallel subprocesses.
This obstruction can be circumvented by partitioning the system  in domains, and allowing for the information about the total energy to be shared across processes  living in separate domains. But optimisation of this process is a non-trivial open problem.
Related to scalability is also the fact that when we tested the algorithm on larger volumes, we found a weakening of the 
transition, which makes it more difficult to detect. Whether this is an intrinsic feature of the algorithm, or
a consequence of the choice of theory---$SU(3)$ is believed to undergo a weak first-order 
phase transition---is an open problem.

\begin{figure}[t]
\centering
\includegraphics[width=0.45\textwidth]{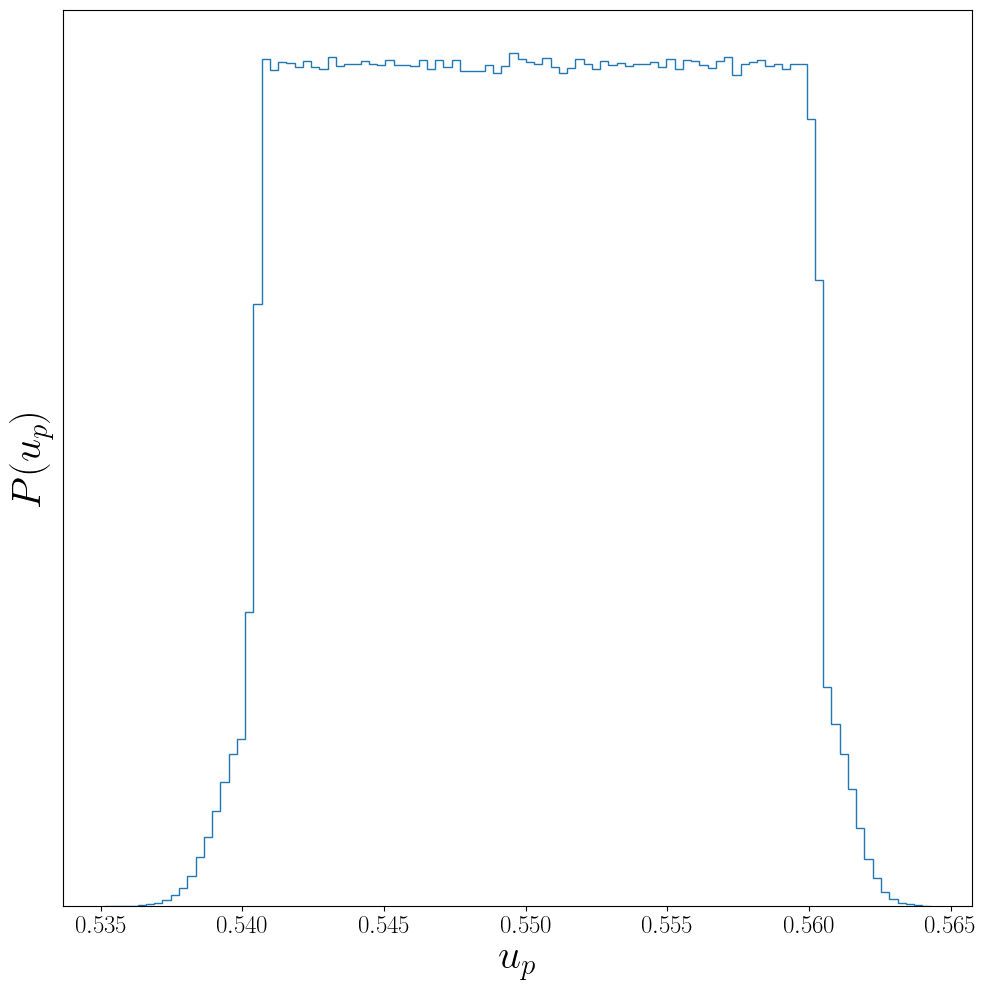}
\caption{\label{fig:fixed_a_plaquette_distribution} For one LLR run
  with $a^4\Delta_E/6\tilde{V} = 0.0007$, the distribution of measured
  average plaquette values for all intervals is plotted for
  configurations restricted to a given energy interval, updated with a
  fixed value of $a_n$. Within the plaquette range, $a^4
  E_{\mathrm{min}}/6\tilde{V} \leq 1 - u_p \leq a^4 E_{\mathrm{max}}/6\tilde{V}$, the distribution is approximately flat, while the boundary intervals have gaussian tails allowing configurations to escape the energy boundaries, resolving the residual ergodicity problems.}
\end{figure}

\begin{figure}[t]
\centering
\includegraphics[width=0.45\textwidth]{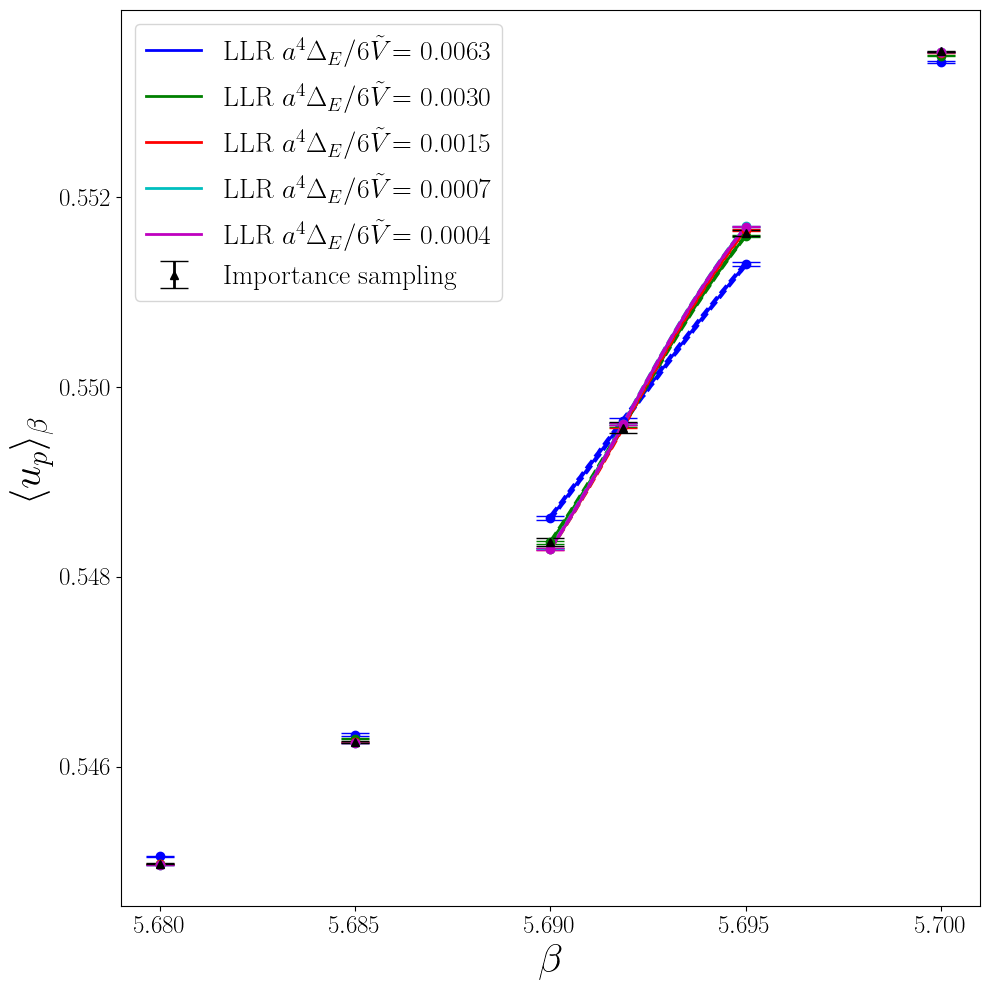}
\caption{\label{fig:PlaqVsBeta} The vacuum expectation value of the average plaquette for different couplings calculated using the LLR method for different $a^4 \Delta_E / (6\tilde{V})$ sizes, and compared to the measurement from importance sampling methods (black triangles). In both cases the lattice size is $\tilde{V}/a^4 = 4\times 20^3$. The dots are at a coupling with a direct comparison to the importance sampling. The solid lines are a finer scan around the critical region, containing 1000 points evenly spaced between 5.690 and 5.695. The errors on this line are calculated by bootstrapping over the repeats and are represented by the dashed curves.}
\end{figure}

Finally, a more conceptual set of questions arises in view of applications: 
we showed that we can compute an  effective potential, without the need to build an
 intermediate effective field theory treatment based on simplifying assumptions
for the functional dependence on the order parameter. It would be useful to understand how this feature
can be exploited for phenomenological purposes. For example, is the detailed knowledge of the effective potential going to 
improve current understanding of the amplitude of gravitational waves arising in the early universe?

All these and other interesting questions are left for what we foresee to become an interesting and original research programme, which we are planning to develop in the near and long-term future.


\begin{figure}[t]
\centering
\includegraphics[width=0.45\textwidth]{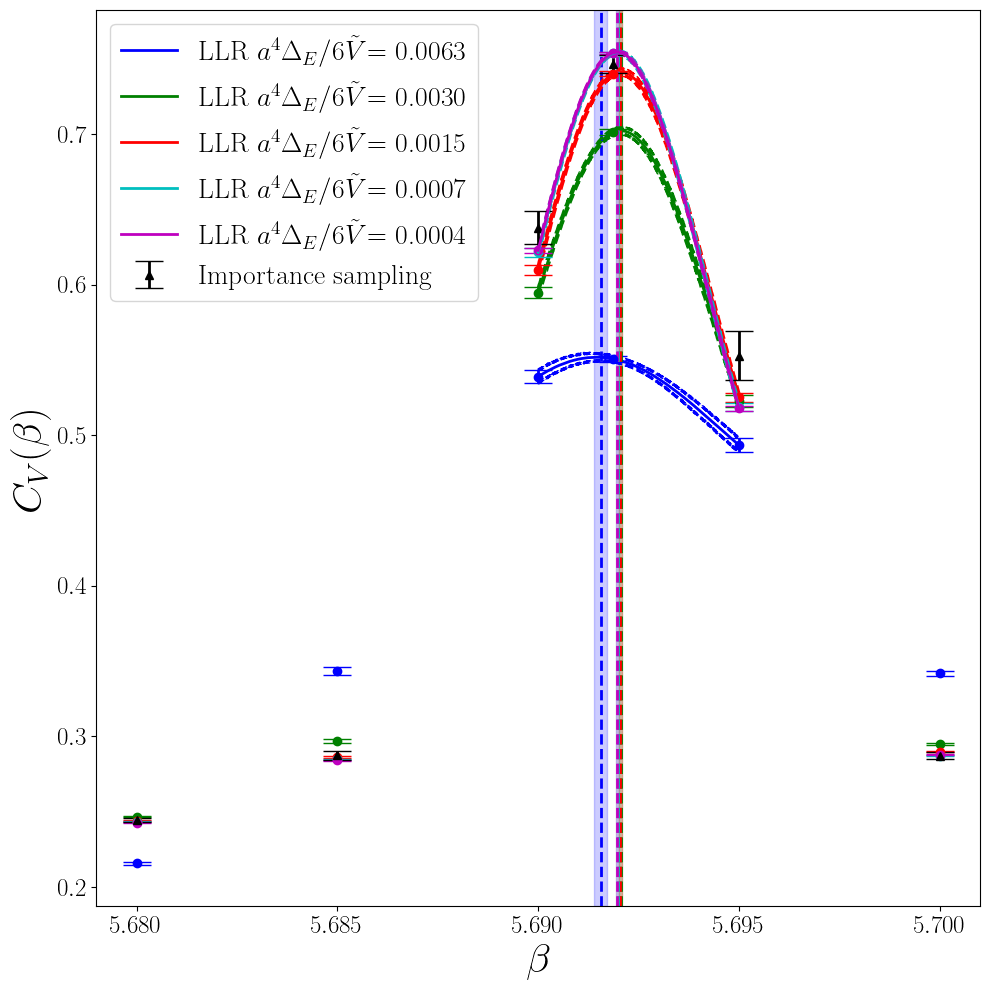}
\caption{\label{fig:SpecificHeatVsBeta} The specific heat for different couplings calculated using the LLR method for different values of $a^4 \Delta_E / (6\tilde{V})$ sizes, and compared to the measurement from importance sampling methods (black triangles). In both cases the lattice size is $\tilde{V}/a^4 = 4\times 20^3$. The dots are at a coupling with a direct comparison to the importance sampling. The solid lines are a finer scan around the critical region, containing 1000 points evenly spaced between 5.690 and 5.695. The errors on this line are calculated by bootstrapping over the repeats and are represented by the dashed curves. The extrema of the curve for all $a^4 \Delta_E / (6\tilde{V})$ sizes are shown by the vertical dashed lines, with the corresponding error represented by the shaded region. The extrema of all but the coarsest interval sizes overlap.}
\end{figure}

\begin{figure}[t]
\centering
\includegraphics[width=0.45\textwidth]{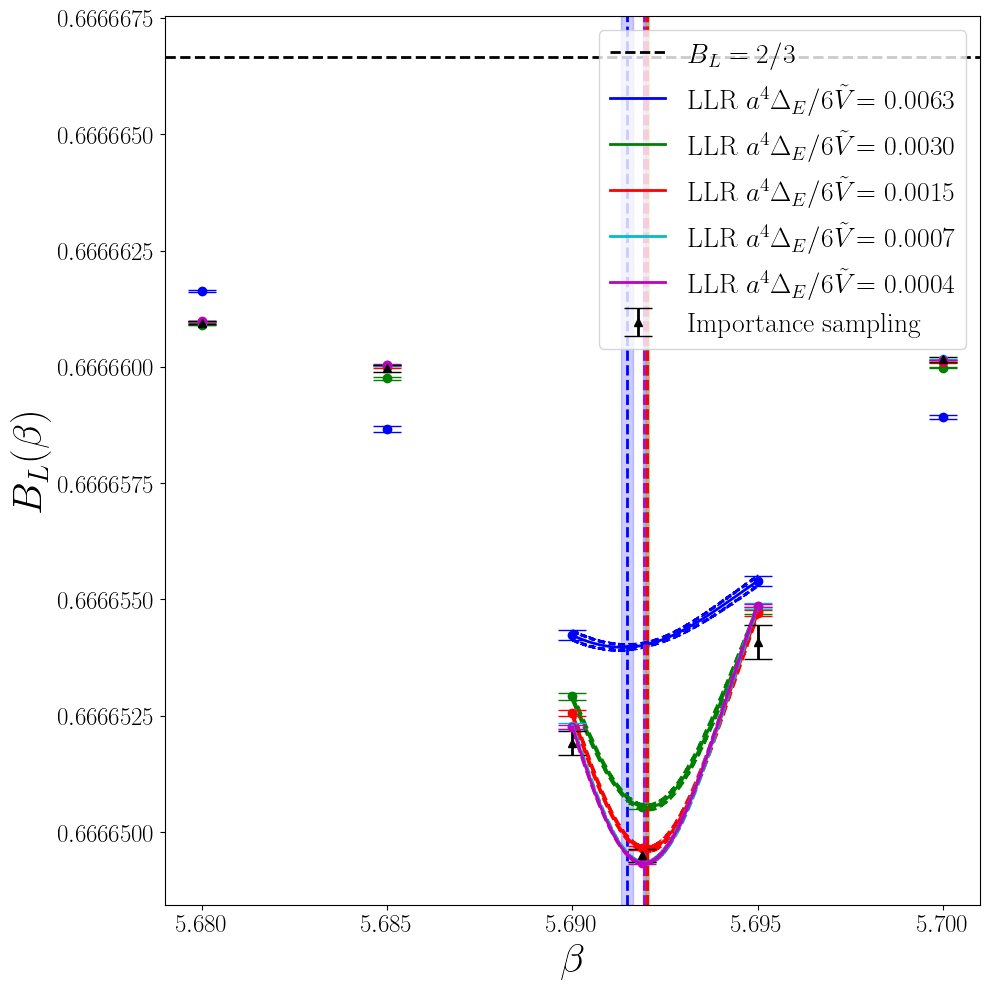}
\caption{\label{fig:BinderCumulantVsBeta}  The Binder cumulant for different couplings calculated using the LLR method for different values of $a^4 \Delta_E / (6\tilde{V})$ sizes, and compared to the measurement from importance sampling methods (black triangles). In both cases the lattice size is $\tilde{V}/a^4 = 4\times 20^3$. The dots are at a coupling with a direct comparison to the importance sampling. The solid lines are a finer scan around the critical region, containing 1000 points evenly spaced between 5.690 and 5.695. The errors on this line are calculated by bootstrapping over the repeats and are represented by the dashed curves. The extrema of the curve for all $a^4 \Delta_E / (6\tilde{V})$ sizes are shown by the vertical dashed lines, with the corresponding error represented by the shaded region. The extrema of all but the coarsest interval sizes overlap.}
\end{figure}

\begin{figure}[t]
\centering
\includegraphics[width=0.45\textwidth]{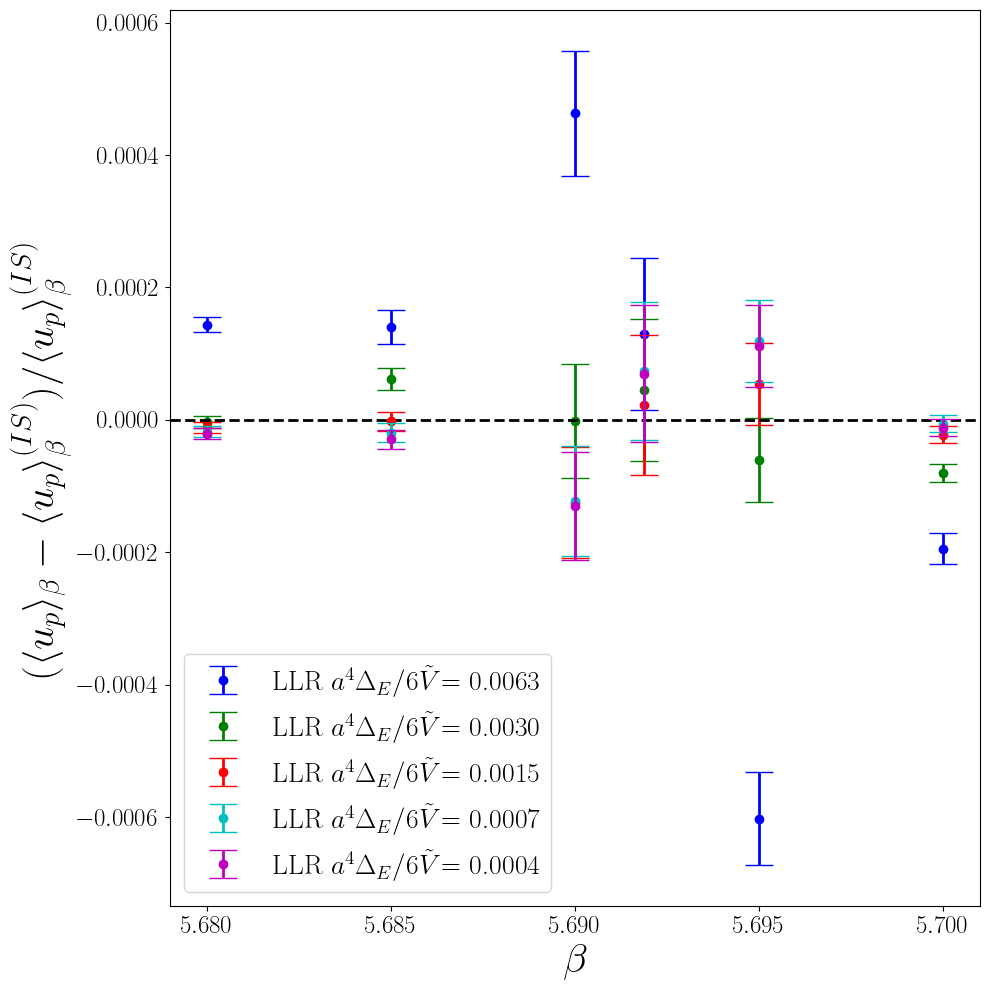}
\caption{\label{fig:PlaqVsBeta_comparison} The relative change for the vacuum expectation value of the average plaquette calculated between the LLR method and importance sampling (IS) method results is shown at different couplings. In both cases the lattice size is $\tilde{V}/a^4 = 4\times 20^3$. The coloured dots show the results for different values of $a^4 \Delta_E / (6\tilde{V})$. The errors were found by using bootstrap methods to calculate the error on LLR and importance sampling results separately, then propagating them.}
\end{figure}

\begin{figure}[t]
\centering
\includegraphics[width=0.45\textwidth]{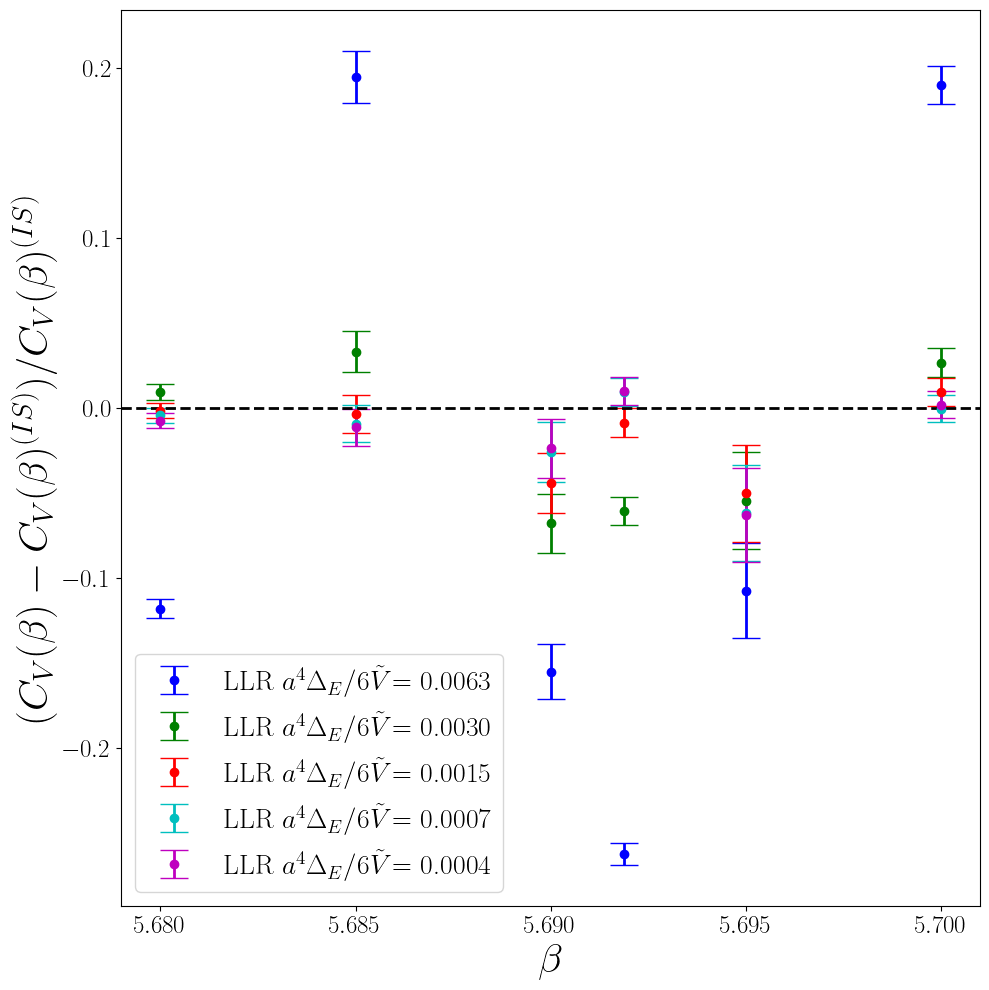}
\caption{\label{fig:SpecificHeatVsBeta_comparison}  The relative change for the specific heat calculated between the LLR method and importance sampling (IS) method results is shown at different couplings. In both cases the lattice size is $\tilde{V}/a^4 = 4\times 20^3$. The coloured dots show the results for different values of $a^4 \Delta_E / (6\tilde{V})$. The errors were found by using bootstrap methods on the LLR and importance sampling results separately, then propagating them.}
\end{figure}

\begin{figure}[t]
\centering
\includegraphics[width=0.45\textwidth]{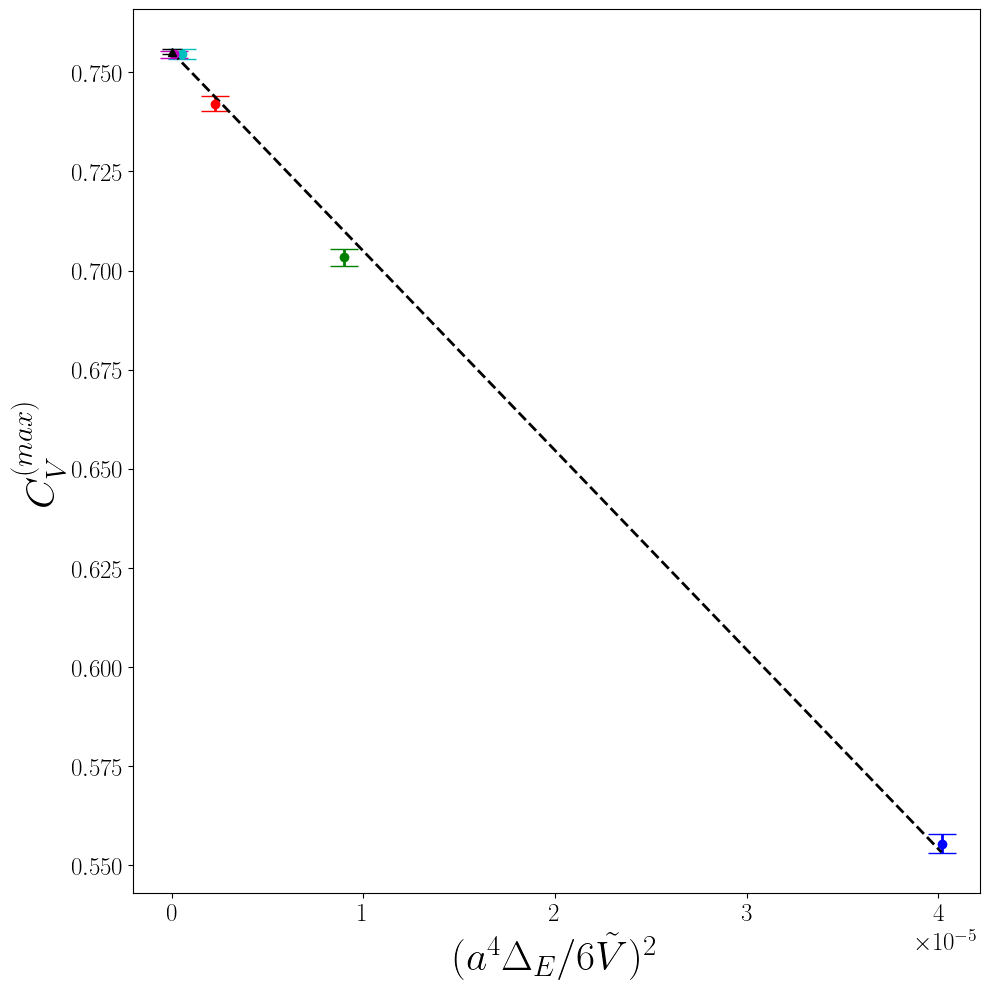}
\caption{\label{fig:dElim0_SpecificHeat} The maximum value of the specific heat calculated using the LLR method for couplings around the critical region, containing 1000 points evenly spaced between 5.690 and 5.695, is plotted against the $(a^4 \Delta_E / (6\tilde{V}))^2$ value it was calculated at. The errors on each point are found by bootstrapping the repeats. The black dot shows the limit of $(a^4 \Delta_E / (6\tilde{V}))^2 \to 0$ and its errors, found by a linear fit. The lattice size is $\tilde{V}/a^4 = 4\times 20^3$.}
\end{figure}

\begin{figure}[t]
\centering
\includegraphics[width=0.45\textwidth]{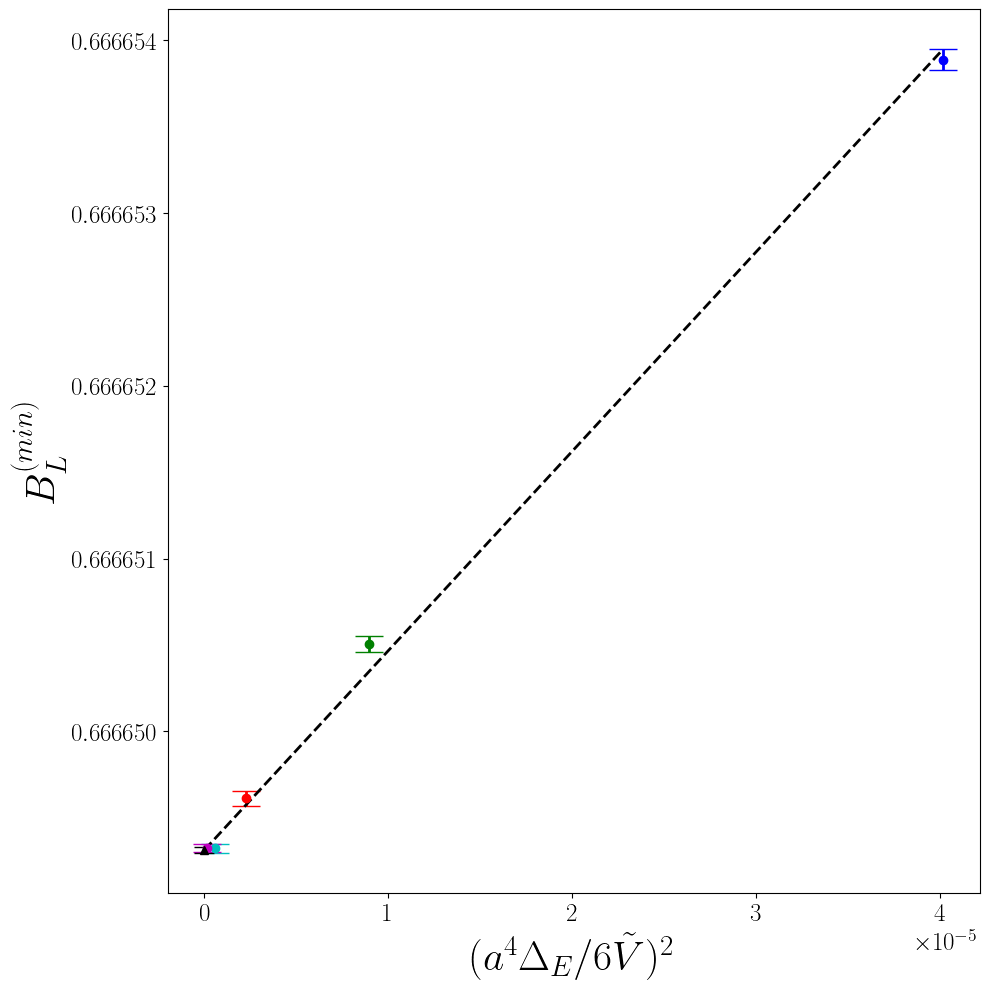}
\caption{\label{fig:dElim0_BinderCumulant} The minimum value of the Binder cumulant calculated using the LLR method for couplings around the critical region, containing 1000 points evenly spaced between 5.690 and 5.695, is plotted against the $(a^4 \Delta_E / (6\tilde{V}))^2$ value it was calculated at. The errors on each point are found by bootstrapping the repeats. The black dot shows the limit of $(a^4 \Delta_E / (6\tilde{V}))^2 \to 0$ and its errors, found by a linear fit. The lattice size is $\tilde{V}/a^4 = 4\times 20^3$.}
\end{figure}

\begin{figure}[t]
\centering
\includegraphics[width=0.45\textwidth]{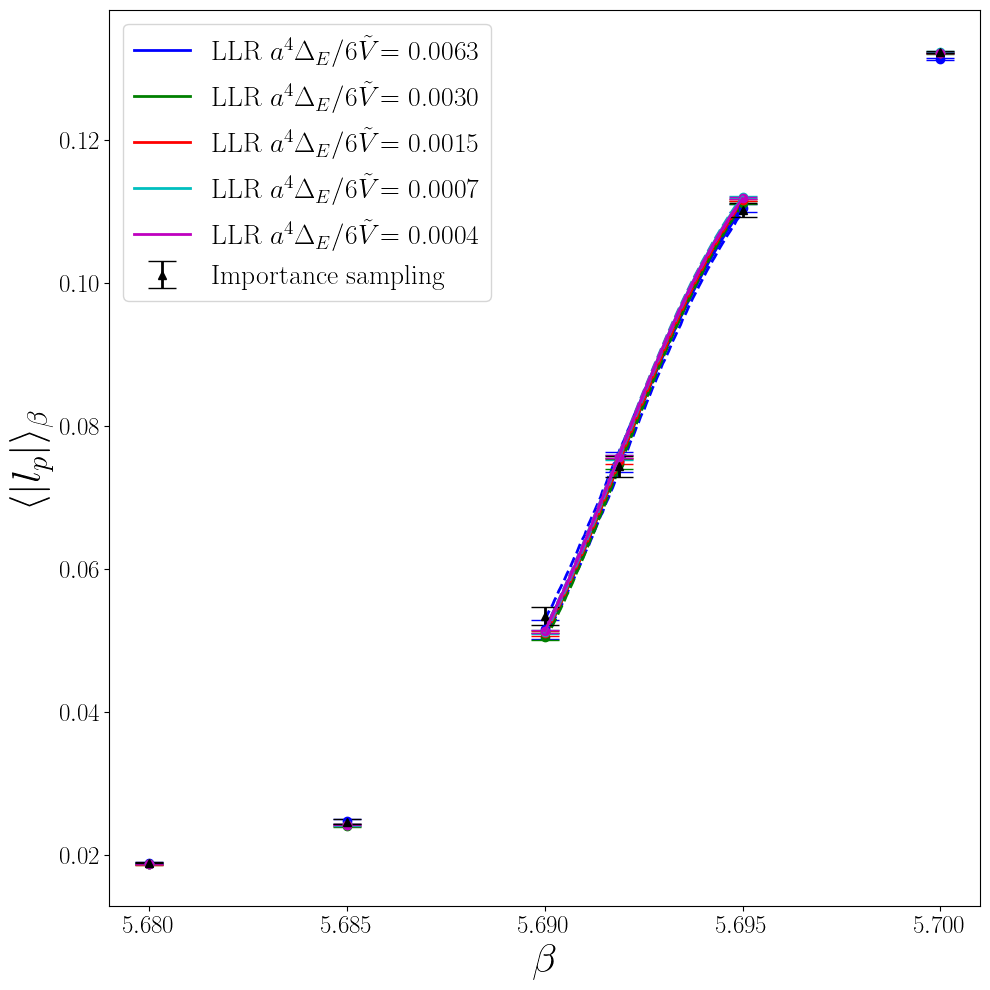}
\caption{\label{fig:PolyakovLoopVsBeta} The vacuum expectation value of the absolute value of the Polyakov loop for different couplings calculated using the LLR method for different values of $a^4 \Delta_E / (6\tilde{V})$, and compared to the measurement from importance sampling methods (black triangles). In both cases the lattice size is $\tilde{V}/a^4 = 4\times 20^3$. The dots are at a coupling with a direct comparison to the importance sampling. The solid lines are a finer scan around the critical region, containing 100 points evenly spaced between 5.690 and 5.695. The errors on this line are calculated by bootstrapping over the repeats and are represented by the dashed curves. For the values determined with the LLR method, the calculation requires measurements of the Polyakov loop on a set of configurations found using the restricted energy updates, with $a_n$ fixed to its final value. For these calculations 40000 fixed $a_n$ measurements were performed on all the intervals.}
\end{figure}

\begin{figure}[t!]
\centering
\includegraphics[width=0.45\textwidth]{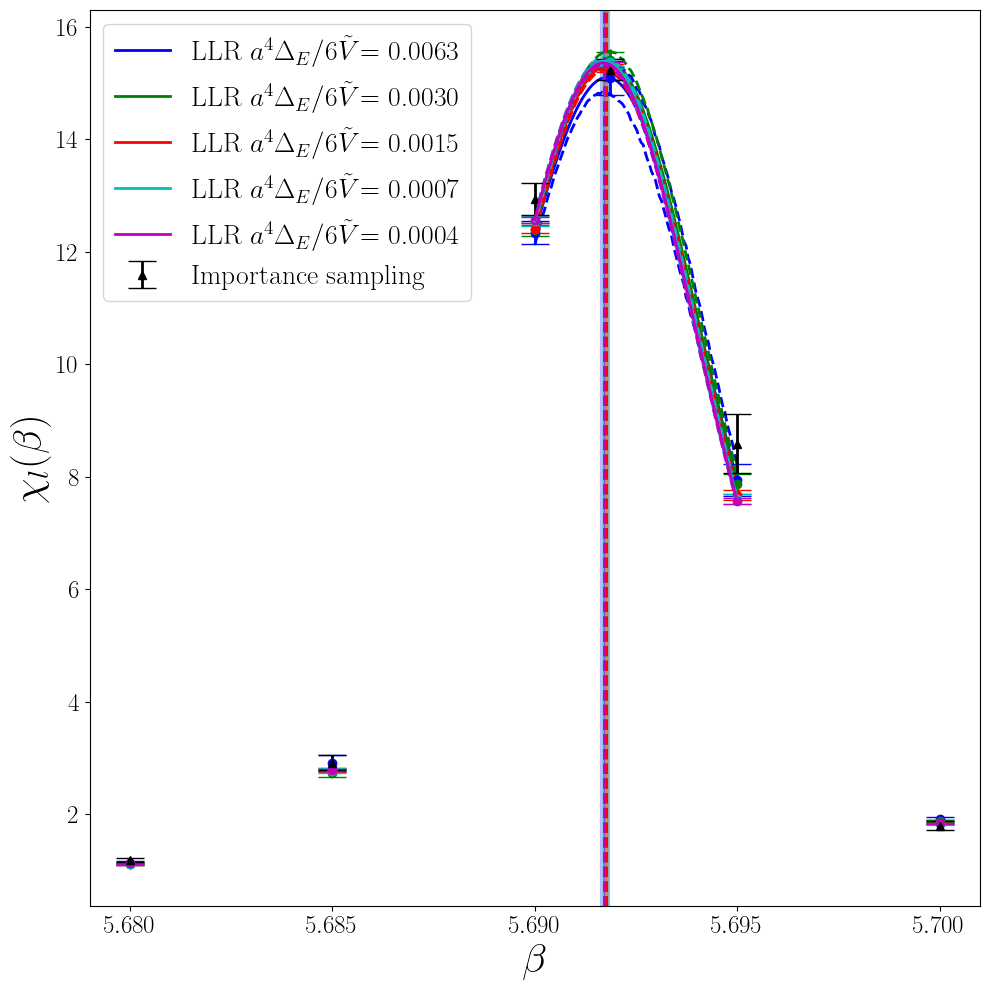}
\caption{\label{fig:PolyakovSusceptibilityVsBeta} The Polyakov loop susceptibility for different couplings calculated using the LLR method for different values of $a^4 \Delta_E / (6\tilde{V})$, and compared to the measurement from importance sampling methods (black triangles). In both cases the lattice size is $\tilde{V}/a^4 = 4\times 20^3$. The dots are at a coupling with a direct comparison to the importance sampling. The solid lines are a finer scan around the critical region, containing 100 points evenly spaced between 5.690 and 5.695. The errors on this line are calculated by bootstrapping over the repeats and are represented by the dashed curves. For the values determined with the LLR method, the calculation requires measurements of the Polyakov loop on a set of configurations found using the restricted energy updates, with $a_n$ fixed to its final value. For these calculations 40000 fixed $a_n$ measurements were performed on all the intervals. The extrema of the curve for all $a^4 \Delta_E / (6\tilde{V})$ sizes are shown by the vertical dashed lines, with the corresponding error represented by the shaded region. The extrema of all interval sizes overlap.}
\end{figure}

\begin{figure}[t!]
\centering
\includegraphics[width=0.45\textwidth]{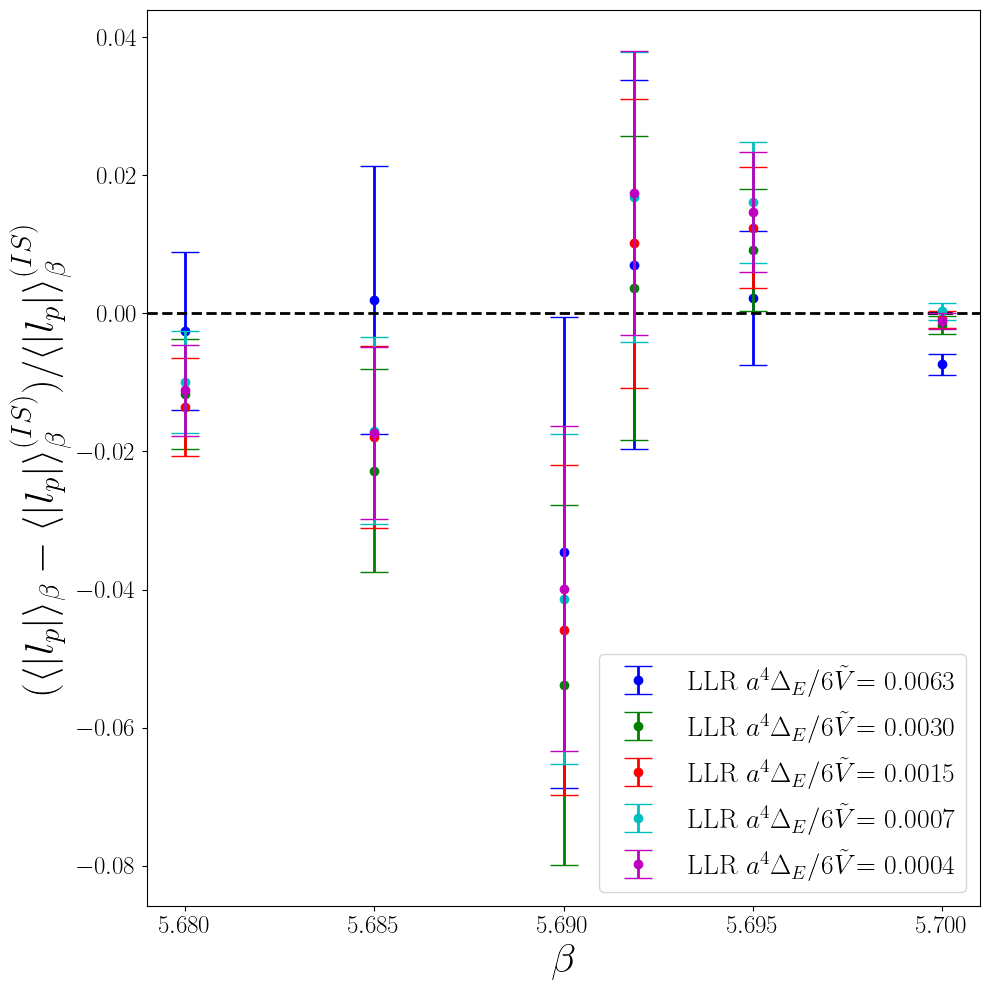}
\caption{\label{fig:PolyakovLoopVsBeta_comparison}  The relative change in the vacuum expectation value of the absolute value of the Polyakov loop, calculated between the LLR method and the importance sampling (IS) method results,  is shown at different couplings. In both cases the lattice size is $\tilde{V}/a^4 = 4\times 20^3$. The coloured dots show the results for different values of $a^4 \Delta_E / (6\tilde{V})$. The error is found by using bootstrap methods on  LLR and importance sampling results separately, then propagating them. For the values determined with the LLR method, the calculation requires measurements of the Polyakov loop on a set of configurations found using the restricted energy updates, with $a_n$ fixed to its final value. For these calculations 40000 fixed $a_n$ measurements were performed on all the intervals.}
\end{figure}

\begin{figure}[t!]
\centering
\includegraphics[width=0.45\textwidth]{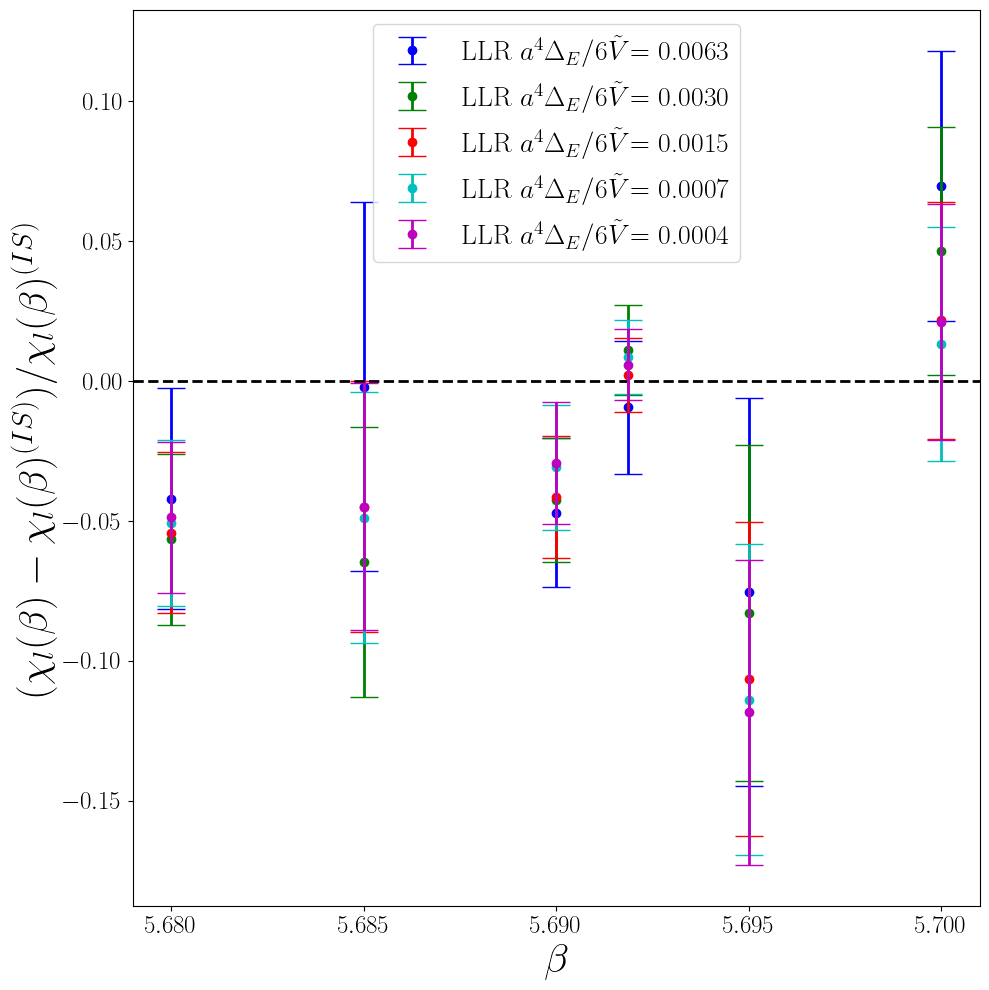}
\caption{\label{fig:PolyakovSusceptibilityVsBeta_comparsion} The relative change in the Polyakov loop susceptibility,  calculated between the  LLR method and importance sampling (IS) method results, is shown at different couplings. In both cases the lattice size is $\tilde{V}/a^4 = 4\times 20^3$. The coloured dots show the results for different values of $a^4 \Delta_E / (6\tilde{V})$. The error is found by using bootstrap methods on  LLR and importance sampling results separately, then propagating them. For the values determined with the LLR method, the calculation requires measurements of the Polyakov loop on a set of configurations found using the restricted energy updates, with $a_n$ fixed to its final value. For these calculations 40000 fixed $a_n$ measurements were performed on all the intervals.}
\end{figure}


\begin{acknowledgments}
We would like to thank David Schaich, Felix Springer, Jong-Wan Lee and Antonio Rago for discussions.
The work of DM is supported by a studentship awarded by the Data Intensive Centre for Doctoral Training, which is funded by the STFC grant ST/P006779/1.
The work of DV is partly supported by the Simons Foundation under the program “Targeted Grants to Institutes” awarded to the Hamilton Mathematics Institute. 
The work of BL and MP has been supported in part by the STFC Consolidated Grants No. ST/P00055X/1 and No. ST/T000813/1.
BL and MP received funding from the European Research Council (ERC) under the European Union’s Horizon 2020 research and innovation program under Grant Agreement No.~813942. 
The work of BL is further supported in part by the Royal Society Wolfson Research Merit Award WM170010 and by the Leverhulme Trust Research Fellowship No. RF-2020-4619.
Numerical simulations have been performed on the Swansea SUNBIRD cluster (part of the Supercomputing Wales project) and AccelerateAI A100 GPU system, and on the DiRAC Data Intensive service at Leicester. The Swansea SUNBIRD system and AccelerateAI are part funded by the European Regional Development Fund (ERDF) via Welsh Government. The DiRAC Data Intensive service at Leicester is operated by the University of Leicester IT Services, which forms part of the STFC DiRAC HPC Facility (www.dirac.ac.uk). The DiRAC Data Intensive service equipment at Leicester was funded by BEIS capital funding via STFC capital grants ST/K000373/1 and ST/R002363/1 and STFC DiRAC Operations grant ST/R001014/1. DiRAC is part of the National e-Infrastructure.

\vspace{1.0cm}

{\bf Open Access Statement}---For the purpose of open access, the authors have applied a Creative Commons 
Attribution (CC BY) licence  to any Author Accepted Manuscript version arising.

\vspace{1.0cm}

{\bf Research Data Access Statement}---The data generated for this manuscript can be downloaded from  Ref.~\cite{LMPRV}. The simulation code can be found from Ref.~\cite{LMPRV2}. 


\end{acknowledgments}


\appendix

\section{Constraint-preserving update proposals}
\label{app:details}

In this appendix, we present our strategy for  sampling random configurations 
from the probability density
\beq
\label{eq:bolzdistrib}
 d P_c(U) \propto dP(U)
 \theta(E-E_{n-1})
 \theta(E_{n+1}-E)\,,
\eeq
where $dP(U)$ is the unconstrained probability density associated to the 
link variable $U$, and the $\theta$ functions 
that implement the energy constraints, $E_{n-1} = E_n - \Delta_E /2 \leq E \leq E_n + \Delta_E /2 = E_{n+1}$.

The problem of sampling $dP(U)$
has been elegantly solved in Ref.~\cite{Creutz:1980zw} for the gauge group $SU(2)$,
and then generalised to $SU(N_c)$ 
gauge groups in Ref.~\cite{Cabibbo:1982zn}. In the case $N_c=2$,
\beq
\label{eq:bolzdistrib_1}
 d P(U) \propto dU \exp\left\{ -\frac{\beta}{2} {\rm Re}\left[{\rm Tr}\left(U \frac{}{}U_\sqcup \right)\right]\right\}\,,
\eeq
where $U_\sqcup$ is the \emph{staple} around $U$, and
$dU$ is the Haar measure of the gauge group.
Any $SU(2)$ matrix $U_p/k=U U_\sqcup/k$ can be parameterised as $U_p/k =u_0\mathbb{I}_2+i\,\vec{u}\cdot\vec\tau$,
where $\vec{\tau}$ are the Pauli matrices, $u_\mu$ are
real numbers satisfying the normalisation $\sum_\mu u_\mu^2=1$ and $k\equiv {\rm det}(U_\sqcup)$. The matrix
$U_p/k$ is obtained by first sampling $\vec{u}$ uniformly on a sphere of radius $\sqrt{1-u_0^2}$, and then $u_0$ from the probability distribution 
\beq
\label{eq:a0ditrib}
 d\tilde{P}(u_0) \sim d u_0 \sqrt{1-u_0^2} \exp\left\{\beta\, k\frac{}{} u_0\right\}~.
\eeq
We determine $u_0$ as
\beq
\label{eq:a0distrib_1}
u_0 = \frac{1}{\beta k}
    \log{\left( e^{-\beta k} + \xi ( e^{\beta k}-e^{-\beta k}) \right)},
\eeq
where $0\leq \xi\leq 1$ is a uniform random variable, and
then perform an
accept-reject step to correct for the presence of the factor $\sqrt{1-u_0^2}$
in Eq.~(\ref{eq:a0ditrib}).

We further generalise these ideas 
to take into account, in the Monte Carlo evolution, the presence of the constraints
$E_{n-1} \leq E \leq E_{n+1}$.
Consider the variation in the total energy $E$ due to the update of a specific link 
variable. Let
$E_i$ ($E_f$) be this energy contribution before (after) 
the update.  The energy constraints after the update are
\beqs
E_{n-1} &\leq&  E - E_i + E_f
 \,\leq\, E_{n+1}\,.
\eeqs
Since $E_{f}=2(d-1)-k u_0$, where $d$ is the number of space-time dimensions, the above constraint can be expressed as 
$u_{\rm min}< u_0 < u_{\rm max}$ where
\begin{align}\label{eq:xminmax}
 u_{\mathrm{min}}&=\max{ \left(\frac{2(d-1)+ (E-E_{n+1}) -E_i}{k},-1\right)}~,\\ 
 u_{\mathrm{max}}&=\min{\left(\frac{2(d-1) + (E-E_{n-1}) -E_i}{k},1\right)}~.
\end{align}

These constraints  can be enforced on the random sampling
of $u_0$ by setting
\beq
    u_0 =\frac{1}{\beta k}
    \log\left( e^{\beta k u_{\rm min}} + \xi ( e^{\beta k u_{\rm max}}-e^{\beta k u_{\rm min}}) \right) ,
\eeq
where, as in Eq.~(\ref{eq:a0distrib_1}), $0\leq \xi \leq 1$ is sampled uniformly, and an
accept-reject step is performed to correct for the presence 
of the factor $\sqrt{1-u_0^2}$.

The constrained heat-bath algorithm outlined above can be generalised to $SU(N_c)$ gauge groups 
following the Cabibbo-Marinari process suggested in Ref.~\cite{Cabibbo:1982zn}. The contribution
of each $SU(2)$ subgroup of a $SU(N_c)$ link variable to the total energy of the system
is additive. Thus, the constraint can be solved independently for each $SU(2)$
subgroup of each link variable. It is easy to show that
the constrained probability density of $U$ is invariant under $U\rightarrow \alpha_k U$, 
where $\alpha_k$ is an element of one of the $SU(2)$ subgroups of $SU(N_c)$. 

\section{Further technical details on the algorithm and parallelism}
\label{app:moredetails}
To improve the scalability of the LLR algorithm when moving to larger lattice sizes, domain decomposition was implemented, in which the full lattice is split into subdomains, which can be processed separately. The restricted heat-bath updates, discussed in Appendix~\ref{app:details}, require prior knowledge of the total action of the system and will change it's value. Therefore, the restricted heat-bath update cannot occur in multiple subdomains simultaneously. 

To circumvent this issue, in this work, domain decomposition is instead built out of a combination of restricted local heat-bath updates and the inherently micro-canonical over-relaxation updates. This ensures the value of the total action is only changed in one subdomain at a time. If we have a lattice with $N_D$ subdomains, during each sweep, one domain is updated with a local heat-bath update, while the other $N_D-1$ subdomains use the over-relaxation. After each sweep, the subdomain using the local heat-bath update is changed. One full lattice update is completed once each subdomain has been updated once using local heath-bath. Therefore, for each full update, each subdomain undergoes one local heat-bath update and $N_D-1$ over-relaxation updates. For this work we use $N_D = 4$.

As discussed in Sect.~\ref{sec:method}, there is a residual ergodicity problem, due to hard energy cutoffs at the boundaries $E_\mathrm{min}$ and $E_\mathrm{max}$. To avoid these problems, in the boundary intervals the boundary cutoffs are removed, allowing configurations in the first and final intervals to freely move into energies $E < E_\mathrm{min}$ and $E> E_\mathrm{max}$, respectively. This is done by simply replacing $E_{n+1}$ ($E_{n-1}$) in Eq.~(\ref{eq:xminmax}) with $6\tilde{V}/a^4$ (0) in the final (first) interval. 

Due to the removal of the hard energy cutoffs, the boundary intervals are no longer symmetric about the centre of the interval. In this case, Eq.~(\ref{eq:rm}) cannot be used to update $a_1$ and $a_{2N-1}$. Instead, we assume the boundaries are away from the critical region and the interval width is small, so the function $E_n(a_n)$ is approximately linear. In this case we approximate, $a_1 = 2a_2 - a_3$ and $a_{2N-1} = 2a_{2N-2} - a_{2N-3}$. 

The sampled energy distribution of the boundary intervals are expected to be gaussians centred at $E_{2N-1}$ and $E_1$, with hard cut-offs at $E_{2N-2}$ and $E_2$, respectively. The sampled plaquette distribution for all intervals, therefore, should be approximately flat within $E_\mathrm{min} \leq E \leq E_\mathrm{max}$ with gaussian tails on the boundaries. These expectations are confirmed by Fig.~\ref{fig:fixed_a_plaquette_distribution}.

\section{$\Delta_E \to 0$ limit}
\label{app:convergence}

In the calculation of the observables there is a systematic error which is proportional to the size of the energy interval squared, $\Delta_E^2$, see Ref.~\cite{Langfeld:2015fua}. To accurately represent an expectation value and it's error, we require that this systematic error be smaller than the statistical error, arising from repeating the determination of $\{a_n\}_{n=1}^{2N-1}$. To ensure $\Delta_E$ is sufficiently small, in this section we analyze the $\Delta_E \to 0$ limit for the average plaquette, the specific heat,
\begin{equation}
C_V(\beta) \equiv \frac{6\tilde{V}}{a^4}\left(\langle u_p^2 \rangle_\beta - \langle u_p \rangle_\beta^2\right),
\end{equation}
the Binder cumulant, 
\begin{equation}
B_L(\beta) \equiv 1 - \frac{\langle u_p^4 \rangle_\beta}{3 \langle u_p^2 \rangle_\beta^2},
\end{equation}
the ensemble average of the absolute value of the Polyakov loop, $\langle | l_p | \rangle_\beta$, and the Polyakov loop susceptibility, $\chi_l(\beta)$. We also take this limit for the maximum of the specific heat $C_V^{(max)}$ and the minimum of the Binder cumulant $B_L^{(min)}$.

The observables calculated from the LLR method are compared against expectation values measured on a lattice of the same size ($\tilde{V}/a^4=4\times20^3$) but obtained using standard importance sampling methods. The lattice was updated using 1 local heat bath update followed by 4 over-relaxation updates. At each coupling value, 500000 measurements were taken, and the errors were computed using bootstrap methods.     

The observables with explicit dependence on the energy, $\langle u_p \rangle_\beta$, $C_V(\beta)$ and $B_L(\beta)$, are calculated using Eq.~(\ref{eq:vev_obs}). The integral is computed over the entire possible energy range of the system. Since, the contribution from energy outside the range relevant for the problem is exponentially suppressed, the limits of the integral can be replaced with $E_{\rm min}$ and  $E_{\rm max}$. We then take the piecewise log linear approximation for the density of states, $\rho(E)\to\tilde{\rho}(E)$, giving 
\begin{equation}
 \langle O \rangle_\beta =\frac{1}{Z_\beta}\sum_{n=1}^{2N-1} \int_{E_n - \Delta_E/4}^{E_n + \Delta_E/4} dE \tilde{\rho}(E) O(E) e^{-\beta E}. 
\end{equation}
Using Eqs.~(\ref{eq:piecewise})~and~(\ref{eq:c_n}), and taking all terms with no explicit $E$ dependence outside the integral gives
\begin{equation}
 \langle O \rangle_\beta =\sum_{n=1}^{2N-1}\frac{\tilde{\rho}(E_n) e^{-\beta E_n}}{Z_\beta} \int_{- \Delta_E/4}^{\Delta_E/4} dE O(E+E_n) e^{E(a_n-\beta)}.
\end{equation}
By analytically solving the integral, inputting the desired coupling $\beta$ and the obtained $\{a_n \}$ values, we can therefore gain a numerical value for the expectation values. 

The Polyakov loop and susceptibility depend on the configuration of the lattice ($\phi$) rather than explicitly on the action, therefore they are calculated using Eq.~(\ref{eq:vev_gen}). After the $a_n$ values are found, a set of energy-restricted updates are carried out with $a_n$ remaining fixed at its final value. On these configurations the action, $S[\phi]$, and observables of interest, $B[\phi]$, are calculated, giving access to the expectation value $\tilde{B}[\phi]$. 

Figures \ref{fig:PlaqVsBeta}, \ref{fig:SpecificHeatVsBeta}, and
\ref{fig:BinderCumulantVsBeta} show the results for $\langle u_p
\rangle_\beta$, $C_V(\beta)$, and $B_L(\beta)$, respectively. In all
cases the LLR results appear to converge to the curve obtained for the smallest interval. The results for the two smallest interval sizes are clearly consistent with each other. 

The results for the smallest interval size follow the general trend of the values found using importance sampling. By plotting the relative change between expectation values of these observables and the importance sampling counter-parts, Figs.~\ref{fig:PlaqVsBeta_comparison} and \ref{fig:SpecificHeatVsBeta_comparison}, we see they are generally consistent within two standard deviations. 

As discussed in Sect.~\ref{sec:results}, for the smaller interval
sizes the structure of  $a_n(E_n)$ is not invertible, giving rise to a
probability distribution, $P_\beta(u_p)$, with a characteristic double
peak structure. However, for $a^4 \Delta_E / (6\tilde{V}) = 0.0063$,
the interval size is not sufficient to resolve this structure. As can
be seen from the plots Fig.~\ref{fig:PlaqVsBeta},
\ref{fig:SpecificHeatVsBeta}, and \ref{fig:BinderCumulantVsBeta},
the behaviour of the expectations of this ensemble is different. The
peaks in the specific heat and the dip of the Binder cumulant are much
shallower and the change in the plaquette much slower, making it
consistent with a weaker transition or even a second order transition.

The location of the extrema of the specific heat and Binder cumulant, in the limit of $\Delta_E \to 0 $, are shown in Figs.~\ref{fig:dElim0_SpecificHeat} and \ref{fig:dElim0_BinderCumulant}, respectively. 
A linear fit has been taken in $(a^4 \Delta_E/6\tilde{V})^2$, and the results have been extrapolated to $\Delta_E = 0$.
In both cases, the phase transition appears to become stronger as the critical region becomes better resolved with decreasing $\Delta_E$.
In both plots the two smallest interval sizes appear to be consistent with each other and the extrapolation.

\begin{table}
\caption{The values of the coupling, $\beta$, that correspond to the maximum of the specific heat, $\beta(C_V^{(max)})$, the minimum of the binder cumulant, $\beta(B_L^{(min)})$, and the maximum of the Polyakov loop susceptibility, $\beta(\chi_l^{(max)})$, found by measuring the observables at a $\beta$ values evenly spaced between $5.690$ and $5.695$. 
Measurements were carried out at 1000 $\beta$ values for $C_V$ and $B_L$, while for $\chi_l$ 100 values were scanned. 
These pseudocritical couplings are compared with the value presented in Sect.\ref{sec:results}, for the critical coupling, $\beta_c$.
These results are for a SU(3) lattice of size $\tilde{V}/a^4 = 4\times 20^3$, using the LLR method with interval sizes $a^4 \Delta_E / (6\tilde{V}) =
0.0007$ and $0.0004$.
 \label{tab:appbeta}
}
\begin{center}
\begin{tabular}{|c|c|c|c|c|}
\hline
  $\frac{a^4\Delta_E}{6\tilde{V}}$ & $\beta(C_V^{(max)})$ & $\beta(B_L^{(min)})$ & $\beta(\chi_l^{(max)})$ &
  $\beta_c$\\
\hline
0.0007 & 5.69198(3) & 5.69194(3) & 5.69170(3) & 5.69188(3) \\
0.0004 & 5.69197(2) & 5.69193(2) & 5.69170(2) & 5.69186(2) \\
\hline
\end{tabular}
\end{center}
\end{table}

The results for $\langle | l_p | \rangle_\beta$ and $\chi_l(\beta)$ are shown in Figs.~\ref{fig:PolyakovLoopVsBeta} and \ref{fig:PolyakovSusceptibilityVsBeta}. The relative change between  LLR and importance sampling results are shown in Figs.~\ref{fig:PolyakovLoopVsBeta_comparison} and~\ref{fig:PolyakovSusceptibilityVsBeta_comparsion}. Once more, the results converge to those of the smallest interval size and show good agreement with importance sampling ones. For these observables, the discrepancy between the largest interval and the others is small.   

We report in Tab.~\ref{tab:appbeta} the values of the pseudocritical couplings identified by the extrema of the observables discussed in this appendix and by the equal height of the peaks in the energy distribution for the two finest values of $\Delta_E$. Our results show consistency across the definitions we have studied and good agreement between the values at the two $\Delta_E$ for fixed observable. 
\par
In summary, all the tests reported in this Appendix show that when the
energy interval is small enough---$a^4 \Delta_E / (6\tilde{V}) =
0.0007$ and $0.0004$---there is no discernible difference with the
results of the extrapolation to zero interval, and general agreement
is found with importance sampling.



\bibliography{main}

\providecommand{\noopsort}[1]{}\providecommand{\singleletter}[1]{#1}%
\begin{thebibliography}{171}%
\makeatletter
\providecommand \@ifxundefined [1]{%
 \@ifx{#1\undefined}
}%
\providecommand \@ifnum [1]{%
 \ifnum #1\expandafter \@firstoftwo
 \else \expandafter \@secondoftwo
 \fi
}%
\providecommand \@ifx [1]{%
 \ifx #1\expandafter \@firstoftwo
 \else \expandafter \@secondoftwo
 \fi
}%
\providecommand \natexlab [1]{#1}%
\providecommand \enquote  [1]{``#1''}%
\providecommand \bibnamefont  [1]{#1}%
\providecommand \bibfnamefont [1]{#1}%
\providecommand \citenamefont [1]{#1}%
\providecommand \href@noop [0]{\@secondoftwo}%
\providecommand \href [0]{\begingroup \@sanitize@url \@href}%
\providecommand \@href[1]{\@@startlink{#1}\@@href}%
\providecommand \@@href[1]{\endgroup#1\@@endlink}%
\providecommand \@sanitize@url [0]{\catcode `\\12\catcode `\$12\catcode
  `\&12\catcode `\#12\catcode `\^12\catcode `\_12\catcode `\%12\relax}%
\providecommand \@@startlink[1]{}%
\providecommand \@@endlink[0]{}%
\providecommand \url  [0]{\begingroup\@sanitize@url \@url }%
\providecommand \@url [1]{\endgroup\@href {#1}{\urlprefix }}%
\providecommand \urlprefix  [0]{URL }%
\providecommand \Eprint [0]{\href }%
\providecommand \doibase [0]{https://doi.org/}%
\providecommand \selectlanguage [0]{\@gobble}%
\providecommand \bibinfo  [0]{\@secondoftwo}%
\providecommand \bibfield  [0]{\@secondoftwo}%
\providecommand \translation [1]{[#1]}%
\providecommand \BibitemOpen [0]{}%
\providecommand \bibitemStop [0]{}%
\providecommand \bibitemNoStop [0]{.\EOS\space}%
\providecommand \EOS [0]{\spacefactor3000\relax}%
\providecommand \BibitemShut  [1]{\csname bibitem#1\endcsname}%
\let\auto@bib@innerbib\@empty
\bibitem [{\citenamefont {Sakharov}(1967)}]{Sakharov:1967dj}%
  \BibitemOpen
  \bibfield  {author} {\bibinfo {author} {\bibfnamefont {A.~D.}\ \bibnamefont
  {Sakharov}},\ }\bibfield  {title} {\bibinfo {title} {{Violation of CP
  Invariance, C asymmetry, and baryon asymmetry of the universe}},\ }\href
  {https://doi.org/10.1070/PU1991v034n05ABEH002497} {\bibfield  {journal}
  {\bibinfo  {journal} {Pisma Zh. Eksp. Teor. Fiz.}\ }\textbf {\bibinfo
  {volume} {5}},\ \bibinfo {pages} {32} (\bibinfo {year} {1967})}\BibitemShut
  {NoStop}%
\bibitem [{\citenamefont {Kajantie}\ \emph {et~al.}(1996)\citenamefont
  {Kajantie}, \citenamefont {Laine}, \citenamefont {Rummukainen},\ and\
  \citenamefont {Shaposhnikov}}]{Kajantie:1996mn}%
  \BibitemOpen
  \bibfield  {author} {\bibinfo {author} {\bibfnamefont {K.}~\bibnamefont
  {Kajantie}}, \bibinfo {author} {\bibfnamefont {M.}~\bibnamefont {Laine}},
  \bibinfo {author} {\bibfnamefont {K.}~\bibnamefont {Rummukainen}},\ and\
  \bibinfo {author} {\bibfnamefont {M.~E.}\ \bibnamefont {Shaposhnikov}},\
  }\bibfield  {title} {\bibinfo {title} {{Is there a~ hot electroweak phase
  transition at $m_H \gtrsim m_W$?}},\ }\href
  {https://doi.org/10.1103/PhysRevLett.77.2887} {\bibfield  {journal} {\bibinfo
   {journal} {Phys. Rev. Lett.}\ }\textbf {\bibinfo {volume} {77}},\ \bibinfo
  {pages} {2887} (\bibinfo {year} {1996})},\ \Eprint
  {https://arxiv.org/abs/hep-ph/9605288} {arXiv:hep-ph/9605288} \BibitemShut
  {NoStop}%
\bibitem [{\citenamefont {Karsch}\ \emph {et~al.}(1997)\citenamefont {Karsch},
  \citenamefont {Neuhaus}, \citenamefont {Patkos},\ and\ \citenamefont
  {Rank}}]{Karsch:1996yh}%
  \BibitemOpen
  \bibfield  {author} {\bibinfo {author} {\bibfnamefont {F.}~\bibnamefont
  {Karsch}}, \bibinfo {author} {\bibfnamefont {T.}~\bibnamefont {Neuhaus}},
  \bibinfo {author} {\bibfnamefont {A.}~\bibnamefont {Patkos}},\ and\ \bibinfo
  {author} {\bibfnamefont {J.}~\bibnamefont {Rank}},\ }\bibfield  {title}
  {\bibinfo {title} {{Critical Higgs mass and temperature dependence of gauge
  boson masses in the SU(2) gauge Higgs model}},\ }\href
  {https://doi.org/10.1016/S0920-5632(96)00736-0} {\bibfield  {journal}
  {\bibinfo  {journal} {Nucl. Phys. B Proc. Suppl.}\ }\textbf {\bibinfo
  {volume} {53}},\ \bibinfo {pages} {623} (\bibinfo {year} {1997})},\ \Eprint
  {https://arxiv.org/abs/hep-lat/9608087} {arXiv:hep-lat/9608087} \BibitemShut
  {NoStop}%
\bibitem [{\citenamefont {Gurtler}\ \emph {et~al.}(1997)\citenamefont
  {Gurtler}, \citenamefont {Ilgenfritz},\ and\ \citenamefont
  {Schiller}}]{Gurtler:1997hr}%
  \BibitemOpen
  \bibfield  {author} {\bibinfo {author} {\bibfnamefont {M.}~\bibnamefont
  {Gurtler}}, \bibinfo {author} {\bibfnamefont {E.-M.}\ \bibnamefont
  {Ilgenfritz}},\ and\ \bibinfo {author} {\bibfnamefont {A.}~\bibnamefont
  {Schiller}},\ }\bibfield  {title} {\bibinfo {title} {{Where the electroweak
  phase transition ends}},\ }\href {https://doi.org/10.1103/PhysRevD.56.3888}
  {\bibfield  {journal} {\bibinfo  {journal} {Phys. Rev. D}\ }\textbf {\bibinfo
  {volume} {56}},\ \bibinfo {pages} {3888} (\bibinfo {year} {1997})},\ \Eprint
  {https://arxiv.org/abs/hep-lat/9704013} {arXiv:hep-lat/9704013} \BibitemShut
  {NoStop}%
\bibitem [{\citenamefont {Rummukainen}\ \emph {et~al.}(1998)\citenamefont
  {Rummukainen}, \citenamefont {Tsypin}, \citenamefont {Kajantie},
  \citenamefont {Laine},\ and\ \citenamefont
  {Shaposhnikov}}]{Rummukainen:1998as}%
  \BibitemOpen
  \bibfield  {author} {\bibinfo {author} {\bibfnamefont {K.}~\bibnamefont
  {Rummukainen}}, \bibinfo {author} {\bibfnamefont {M.}~\bibnamefont {Tsypin}},
  \bibinfo {author} {\bibfnamefont {K.}~\bibnamefont {Kajantie}}, \bibinfo
  {author} {\bibfnamefont {M.}~\bibnamefont {Laine}},\ and\ \bibinfo {author}
  {\bibfnamefont {M.~E.}\ \bibnamefont {Shaposhnikov}},\ }\bibfield  {title}
  {\bibinfo {title} {{The Universality class of the electroweak theory}},\
  }\href {https://doi.org/10.1016/S0550-3213(98)00494-5} {\bibfield  {journal}
  {\bibinfo  {journal} {Nucl. Phys. B}\ }\textbf {\bibinfo {volume} {532}},\
  \bibinfo {pages} {283} (\bibinfo {year} {1998})},\ \Eprint
  {https://arxiv.org/abs/hep-lat/9805013} {arXiv:hep-lat/9805013} \BibitemShut
  {NoStop}%
\bibitem [{\citenamefont {Csikor}\ \emph {et~al.}(1999)\citenamefont {Csikor},
  \citenamefont {Fodor},\ and\ \citenamefont {Heitger}}]{Csikor:1998eu}%
  \BibitemOpen
  \bibfield  {author} {\bibinfo {author} {\bibfnamefont {F.}~\bibnamefont
  {Csikor}}, \bibinfo {author} {\bibfnamefont {Z.}~\bibnamefont {Fodor}},\ and\
  \bibinfo {author} {\bibfnamefont {J.}~\bibnamefont {Heitger}},\ }\bibfield
  {title} {\bibinfo {title} {{Endpoint of the hot electroweak phase
  transition}},\ }\href {https://doi.org/10.1103/PhysRevLett.82.21} {\bibfield
  {journal} {\bibinfo  {journal} {Phys. Rev. Lett.}\ }\textbf {\bibinfo
  {volume} {82}},\ \bibinfo {pages} {21} (\bibinfo {year} {1999})},\ \Eprint
  {https://arxiv.org/abs/hep-ph/9809291} {arXiv:hep-ph/9809291} \BibitemShut
  {NoStop}%
\bibitem [{\citenamefont {Aoki}\ \emph {et~al.}(1999)\citenamefont {Aoki},
  \citenamefont {Csikor}, \citenamefont {Fodor},\ and\ \citenamefont
  {Ukawa}}]{Aoki:1999fi}%
  \BibitemOpen
  \bibfield  {author} {\bibinfo {author} {\bibfnamefont {Y.}~\bibnamefont
  {Aoki}}, \bibinfo {author} {\bibfnamefont {F.}~\bibnamefont {Csikor}},
  \bibinfo {author} {\bibfnamefont {Z.}~\bibnamefont {Fodor}},\ and\ \bibinfo
  {author} {\bibfnamefont {A.}~\bibnamefont {Ukawa}},\ }\bibfield  {title}
  {\bibinfo {title} {{The Endpoint of the first order phase transition of the
  SU(2) gauge Higgs model on a four-dimensional isotropic lattice}},\ }\href
  {https://doi.org/10.1103/PhysRevD.60.013001} {\bibfield  {journal} {\bibinfo
  {journal} {Phys. Rev. D}\ }\textbf {\bibinfo {volume} {60}},\ \bibinfo
  {pages} {013001} (\bibinfo {year} {1999})},\ \Eprint
  {https://arxiv.org/abs/hep-lat/9901021} {arXiv:hep-lat/9901021} \BibitemShut
  {NoStop}%
\bibitem [{\citenamefont {D'Onofrio}\ and\ \citenamefont
  {Rummukainen}(2016)}]{DOnofrio:2015gop}%
  \BibitemOpen
  \bibfield  {author} {\bibinfo {author} {\bibfnamefont {M.}~\bibnamefont
  {D'Onofrio}}\ and\ \bibinfo {author} {\bibfnamefont {K.}~\bibnamefont
  {Rummukainen}},\ }\bibfield  {title} {\bibinfo {title} {{Standard model
  cross-over on the lattice}},\ }\href
  {https://doi.org/10.1103/PhysRevD.93.025003} {\bibfield  {journal} {\bibinfo
  {journal} {Phys. Rev. D}\ }\textbf {\bibinfo {volume} {93}},\ \bibinfo
  {pages} {025003} (\bibinfo {year} {2016})},\ \Eprint
  {https://arxiv.org/abs/1508.07161} {arXiv:1508.07161 [hep-ph]} \BibitemShut
  {NoStop}%
\bibitem [{\citenamefont {Laine}\ and\ \citenamefont
  {Rummukainen}(1999)}]{Laine:1998jb}%
  \BibitemOpen
  \bibfield  {author} {\bibinfo {author} {\bibfnamefont {M.}~\bibnamefont
  {Laine}}\ and\ \bibinfo {author} {\bibfnamefont {K.}~\bibnamefont
  {Rummukainen}},\ }\bibfield  {title} {\bibinfo {title} {{What's new with the
  electroweak phase transition?}},\ }\href
  {https://doi.org/10.1016/S0920-5632(99)85017-8} {\bibfield  {journal}
  {\bibinfo  {journal} {Nucl. Phys. B Proc. Suppl.}\ }\textbf {\bibinfo
  {volume} {73}},\ \bibinfo {pages} {180} (\bibinfo {year} {1999})},\ \Eprint
  {https://arxiv.org/abs/hep-lat/9809045} {arXiv:hep-lat/9809045} \BibitemShut
  {NoStop}%
\bibitem [{\citenamefont {Morrissey}\ and\ \citenamefont
  {Ramsey-Musolf}(2012)}]{Morrissey:2012db}%
  \BibitemOpen
  \bibfield  {author} {\bibinfo {author} {\bibfnamefont {D.~E.}\ \bibnamefont
  {Morrissey}}\ and\ \bibinfo {author} {\bibfnamefont {M.~J.}\ \bibnamefont
  {Ramsey-Musolf}},\ }\bibfield  {title} {\bibinfo {title} {{Electroweak
  baryogenesis}},\ }\href {https://doi.org/10.1088/1367-2630/14/12/125003}
  {\bibfield  {journal} {\bibinfo  {journal} {New J. Phys.}\ }\textbf {\bibinfo
  {volume} {14}},\ \bibinfo {pages} {125003} (\bibinfo {year} {2012})},\
  \Eprint {https://arxiv.org/abs/1206.2942} {arXiv:1206.2942 [hep-ph]}
  \BibitemShut {NoStop}%
\bibitem [{\citenamefont {Gould}\ \emph {et~al.}(2022)\citenamefont {Gould},
  \citenamefont {G\"uyer},\ and\ \citenamefont {Rummukainen}}]{Gould:2022ran}%
  \BibitemOpen
  \bibfield  {author} {\bibinfo {author} {\bibfnamefont {O.}~\bibnamefont
  {Gould}}, \bibinfo {author} {\bibfnamefont {S.}~\bibnamefont {G\"uyer}},\
  and\ \bibinfo {author} {\bibfnamefont {K.}~\bibnamefont {Rummukainen}},\
  }\bibfield  {title} {\bibinfo {title} {{First-order electroweak phase
  transitions: A nonperturbative update}},\ }\href
  {https://doi.org/10.1103/PhysRevD.106.114507} {\bibfield  {journal} {\bibinfo
   {journal} {Phys. Rev. D}\ }\textbf {\bibinfo {volume} {106}},\ \bibinfo
  {pages} {114507} (\bibinfo {year} {2022})},\ \Eprint
  {https://arxiv.org/abs/2205.07238} {arXiv:2205.07238 [hep-lat]} \BibitemShut
  {NoStop}%
\bibitem [{\citenamefont {Strassler}\ and\ \citenamefont
  {Zurek}(2007)}]{Strassler:2006im}%
  \BibitemOpen
  \bibfield  {author} {\bibinfo {author} {\bibfnamefont {M.~J.}\ \bibnamefont
  {Strassler}}\ and\ \bibinfo {author} {\bibfnamefont {K.~M.}\ \bibnamefont
  {Zurek}},\ }\bibfield  {title} {\bibinfo {title} {{Echoes of a hidden valley
  at hadron colliders}},\ }\href
  {https://doi.org/10.1016/j.physletb.2007.06.055} {\bibfield  {journal}
  {\bibinfo  {journal} {Phys. Lett. B}\ }\textbf {\bibinfo {volume} {651}},\
  \bibinfo {pages} {374} (\bibinfo {year} {2007})},\ \Eprint
  {https://arxiv.org/abs/hep-ph/0604261} {arXiv:hep-ph/0604261} \BibitemShut
  {NoStop}%
\bibitem [{\citenamefont {Cheung}\ and\ \citenamefont
  {Yuan}(2007)}]{Cheung:2007ut}%
  \BibitemOpen
  \bibfield  {author} {\bibinfo {author} {\bibfnamefont {K.}~\bibnamefont
  {Cheung}}\ and\ \bibinfo {author} {\bibfnamefont {T.-C.}\ \bibnamefont
  {Yuan}},\ }\bibfield  {title} {\bibinfo {title} {{Hidden fermion as
  milli-charged dark matter in Stueckelberg Z- prime model}},\ }\href
  {https://doi.org/10.1088/1126-6708/2007/03/120} {\bibfield  {journal}
  {\bibinfo  {journal} {JHEP}\ }\textbf {\bibinfo {volume} {03}},\ \bibinfo
  {pages} {120}},\ \Eprint {https://arxiv.org/abs/hep-ph/0701107}
  {arXiv:hep-ph/0701107} \BibitemShut {NoStop}%
\bibitem [{\citenamefont {Hambye}(2009)}]{Hambye:2008bq}%
  \BibitemOpen
  \bibfield  {author} {\bibinfo {author} {\bibfnamefont {T.}~\bibnamefont
  {Hambye}},\ }\bibfield  {title} {\bibinfo {title} {{Hidden vector dark
  matter}},\ }\href {https://doi.org/10.1088/1126-6708/2009/01/028} {\bibfield
  {journal} {\bibinfo  {journal} {JHEP}\ }\textbf {\bibinfo {volume} {01}},\
  \bibinfo {pages} {028}},\ \Eprint {https://arxiv.org/abs/0811.0172}
  {arXiv:0811.0172 [hep-ph]} \BibitemShut {NoStop}%
\bibitem [{\citenamefont {Feng}\ \emph {et~al.}(2009)\citenamefont {Feng},
  \citenamefont {Kaplinghat}, \citenamefont {Tu},\ and\ \citenamefont
  {Yu}}]{Feng:2009mn}%
  \BibitemOpen
  \bibfield  {author} {\bibinfo {author} {\bibfnamefont {J.~L.}\ \bibnamefont
  {Feng}}, \bibinfo {author} {\bibfnamefont {M.}~\bibnamefont {Kaplinghat}},
  \bibinfo {author} {\bibfnamefont {H.}~\bibnamefont {Tu}},\ and\ \bibinfo
  {author} {\bibfnamefont {H.-B.}\ \bibnamefont {Yu}},\ }\bibfield  {title}
  {\bibinfo {title} {{Hidden Charged Dark Matter}},\ }\href
  {https://doi.org/10.1088/1475-7516/2009/07/004} {\bibfield  {journal}
  {\bibinfo  {journal} {JCAP}\ }\textbf {\bibinfo {volume} {07}},\ \bibinfo
  {pages} {004}},\ \Eprint {https://arxiv.org/abs/0905.3039} {arXiv:0905.3039
  [hep-ph]} \BibitemShut {NoStop}%
\bibitem [{\citenamefont {Cohen}\ \emph {et~al.}(2010)\citenamefont {Cohen},
  \citenamefont {Phalen}, \citenamefont {Pierce},\ and\ \citenamefont
  {Zurek}}]{Cohen:2010kn}%
  \BibitemOpen
  \bibfield  {author} {\bibinfo {author} {\bibfnamefont {T.}~\bibnamefont
  {Cohen}}, \bibinfo {author} {\bibfnamefont {D.~J.}\ \bibnamefont {Phalen}},
  \bibinfo {author} {\bibfnamefont {A.}~\bibnamefont {Pierce}},\ and\ \bibinfo
  {author} {\bibfnamefont {K.~M.}\ \bibnamefont {Zurek}},\ }\bibfield  {title}
  {\bibinfo {title} {{Asymmetric Dark Matter from a GeV Hidden Sector}},\
  }\href {https://doi.org/10.1103/PhysRevD.82.056001} {\bibfield  {journal}
  {\bibinfo  {journal} {Phys. Rev. D}\ }\textbf {\bibinfo {volume} {82}},\
  \bibinfo {pages} {056001} (\bibinfo {year} {2010})},\ \Eprint
  {https://arxiv.org/abs/1005.1655} {arXiv:1005.1655 [hep-ph]} \BibitemShut
  {NoStop}%
\bibitem [{\citenamefont {Foot}\ and\ \citenamefont
  {Vagnozzi}(2015)}]{Foot:2014uba}%
  \BibitemOpen
  \bibfield  {author} {\bibinfo {author} {\bibfnamefont {R.}~\bibnamefont
  {Foot}}\ and\ \bibinfo {author} {\bibfnamefont {S.}~\bibnamefont
  {Vagnozzi}},\ }\bibfield  {title} {\bibinfo {title} {{Dissipative hidden
  sector dark matter}},\ }\href {https://doi.org/10.1103/PhysRevD.91.023512}
  {\bibfield  {journal} {\bibinfo  {journal} {Phys. Rev. D}\ }\textbf {\bibinfo
  {volume} {91}},\ \bibinfo {pages} {023512} (\bibinfo {year} {2015})},\
  \Eprint {https://arxiv.org/abs/1409.7174} {arXiv:1409.7174 [hep-ph]}
  \BibitemShut {NoStop}%
\bibitem [{\citenamefont {Bertone}\ and\ \citenamefont
  {Hooper}(2018)}]{Bertone:2016nfn}%
  \BibitemOpen
  \bibfield  {author} {\bibinfo {author} {\bibfnamefont {G.}~\bibnamefont
  {Bertone}}\ and\ \bibinfo {author} {\bibfnamefont {D.}~\bibnamefont
  {Hooper}},\ }\bibfield  {title} {\bibinfo {title} {{History of dark
  matter}},\ }\href {https://doi.org/10.1103/RevModPhys.90.045002} {\bibfield
  {journal} {\bibinfo  {journal} {Rev. Mod. Phys.}\ }\textbf {\bibinfo {volume}
  {90}},\ \bibinfo {pages} {045002} (\bibinfo {year} {2018})},\ \Eprint
  {https://arxiv.org/abs/1605.04909} {arXiv:1605.04909 [astro-ph.CO]}
  \BibitemShut {NoStop}%
\bibitem [{\citenamefont {Del~Nobile}\ \emph {et~al.}(2011)\citenamefont
  {Del~Nobile}, \citenamefont {Kouvaris},\ and\ \citenamefont
  {Sannino}}]{DelNobile:2011je}%
  \BibitemOpen
  \bibfield  {author} {\bibinfo {author} {\bibfnamefont {E.}~\bibnamefont
  {Del~Nobile}}, \bibinfo {author} {\bibfnamefont {C.}~\bibnamefont
  {Kouvaris}},\ and\ \bibinfo {author} {\bibfnamefont {F.}~\bibnamefont
  {Sannino}},\ }\bibfield  {title} {\bibinfo {title} {{Interfering Composite
  Asymmetric Dark Matter for DAMA and CoGeNT}},\ }\href
  {https://doi.org/10.1103/PhysRevD.84.027301} {\bibfield  {journal} {\bibinfo
  {journal} {Phys. Rev. D}\ }\textbf {\bibinfo {volume} {84}},\ \bibinfo
  {pages} {027301} (\bibinfo {year} {2011})},\ \Eprint
  {https://arxiv.org/abs/1105.5431} {arXiv:1105.5431 [hep-ph]} \BibitemShut
  {NoStop}%
\bibitem [{\citenamefont {Hietanen}\ \emph {et~al.}(2014)\citenamefont
  {Hietanen}, \citenamefont {Lewis}, \citenamefont {Pica},\ and\ \citenamefont
  {Sannino}}]{Hietanen:2013fya}%
  \BibitemOpen
  \bibfield  {author} {\bibinfo {author} {\bibfnamefont {A.}~\bibnamefont
  {Hietanen}}, \bibinfo {author} {\bibfnamefont {R.}~\bibnamefont {Lewis}},
  \bibinfo {author} {\bibfnamefont {C.}~\bibnamefont {Pica}},\ and\ \bibinfo
  {author} {\bibfnamefont {F.}~\bibnamefont {Sannino}},\ }\bibfield  {title}
  {\bibinfo {title} {{Composite Goldstone Dark Matter: Experimental Predictions
  from the Lattice}},\ }\href {https://doi.org/10.1007/JHEP12(2014)130}
  {\bibfield  {journal} {\bibinfo  {journal} {JHEP}\ }\textbf {\bibinfo
  {volume} {12}},\ \bibinfo {pages} {130}},\ \Eprint
  {https://arxiv.org/abs/1308.4130} {arXiv:1308.4130 [hep-ph]} \BibitemShut
  {NoStop}%
\bibitem [{\citenamefont {Cline}\ \emph {et~al.}(2016)\citenamefont {Cline},
  \citenamefont {Huang},\ and\ \citenamefont {Moore}}]{Cline:2016nab}%
  \BibitemOpen
  \bibfield  {author} {\bibinfo {author} {\bibfnamefont {J.~M.}\ \bibnamefont
  {Cline}}, \bibinfo {author} {\bibfnamefont {W.}~\bibnamefont {Huang}},\ and\
  \bibinfo {author} {\bibfnamefont {G.~D.}\ \bibnamefont {Moore}},\ }\bibfield
  {title} {\bibinfo {title} {{Challenges for models with composite states}},\
  }\href {https://doi.org/10.1103/PhysRevD.94.055029} {\bibfield  {journal}
  {\bibinfo  {journal} {Phys. Rev. D}\ }\textbf {\bibinfo {volume} {94}},\
  \bibinfo {pages} {055029} (\bibinfo {year} {2016})},\ \Eprint
  {https://arxiv.org/abs/1607.07865} {arXiv:1607.07865 [hep-ph]} \BibitemShut
  {NoStop}%
\bibitem [{\citenamefont {Cacciapaglia}\ \emph {et~al.}(2020)\citenamefont
  {Cacciapaglia}, \citenamefont {Pica},\ and\ \citenamefont
  {Sannino}}]{Cacciapaglia:2020kgq}%
  \BibitemOpen
  \bibfield  {author} {\bibinfo {author} {\bibfnamefont {G.}~\bibnamefont
  {Cacciapaglia}}, \bibinfo {author} {\bibfnamefont {C.}~\bibnamefont {Pica}},\
  and\ \bibinfo {author} {\bibfnamefont {F.}~\bibnamefont {Sannino}},\
  }\bibfield  {title} {\bibinfo {title} {{Fundamental Composite Dynamics: A
  Review}},\ }\href {https://doi.org/10.1016/j.physrep.2020.07.002} {\bibfield
  {journal} {\bibinfo  {journal} {Phys. Rept.}\ }\textbf {\bibinfo {volume}
  {877}},\ \bibinfo {pages} {1} (\bibinfo {year} {2020})},\ \Eprint
  {https://arxiv.org/abs/2002.04914} {arXiv:2002.04914 [hep-ph]} \BibitemShut
  {NoStop}%
\bibitem [{\citenamefont {Dondi}\ \emph {et~al.}(2020)\citenamefont {Dondi},
  \citenamefont {Sannino},\ and\ \citenamefont {Smirnov}}]{Dondi:2019olm}%
  \BibitemOpen
  \bibfield  {author} {\bibinfo {author} {\bibfnamefont {N.~A.}\ \bibnamefont
  {Dondi}}, \bibinfo {author} {\bibfnamefont {F.}~\bibnamefont {Sannino}},\
  and\ \bibinfo {author} {\bibfnamefont {J.}~\bibnamefont {Smirnov}},\
  }\bibfield  {title} {\bibinfo {title} {{Thermal history of composite dark
  matter}},\ }\href {https://doi.org/10.1103/PhysRevD.101.103010} {\bibfield
  {journal} {\bibinfo  {journal} {Phys. Rev. D}\ }\textbf {\bibinfo {volume}
  {101}},\ \bibinfo {pages} {103010} (\bibinfo {year} {2020})},\ \Eprint
  {https://arxiv.org/abs/1905.08810} {arXiv:1905.08810 [hep-ph]} \BibitemShut
  {NoStop}%
\bibitem [{\citenamefont {Ge}\ \emph {et~al.}(2019)\citenamefont {Ge},
  \citenamefont {Lawson},\ and\ \citenamefont {Zhitnitsky}}]{Ge:2019voa}%
  \BibitemOpen
  \bibfield  {author} {\bibinfo {author} {\bibfnamefont {S.}~\bibnamefont
  {Ge}}, \bibinfo {author} {\bibfnamefont {K.}~\bibnamefont {Lawson}},\ and\
  \bibinfo {author} {\bibfnamefont {A.}~\bibnamefont {Zhitnitsky}},\ }\bibfield
   {title} {\bibinfo {title} {{Axion quark nugget dark matter model: Size
  distribution and survival pattern}},\ }\href
  {https://doi.org/10.1103/PhysRevD.99.116017} {\bibfield  {journal} {\bibinfo
  {journal} {Phys. Rev. D}\ }\textbf {\bibinfo {volume} {99}},\ \bibinfo
  {pages} {116017} (\bibinfo {year} {2019})},\ \Eprint
  {https://arxiv.org/abs/1903.05090} {arXiv:1903.05090 [hep-ph]} \BibitemShut
  {NoStop}%
\bibitem [{\citenamefont {Beylin}\ \emph {et~al.}(2019)\citenamefont {Beylin},
  \citenamefont {Khlopov}, \citenamefont {Kuksa},\ and\ \citenamefont
  {Volchanskiy}}]{Beylin:2019gtw}%
  \BibitemOpen
  \bibfield  {author} {\bibinfo {author} {\bibfnamefont {V.}~\bibnamefont
  {Beylin}}, \bibinfo {author} {\bibfnamefont {M.~Y.}\ \bibnamefont {Khlopov}},
  \bibinfo {author} {\bibfnamefont {V.}~\bibnamefont {Kuksa}},\ and\ \bibinfo
  {author} {\bibfnamefont {N.}~\bibnamefont {Volchanskiy}},\ }\bibfield
  {title} {\bibinfo {title} {{Hadronic and Hadron-Like Physics of Dark
  Matter}},\ }\href {https://doi.org/10.3390/sym11040587} {\bibfield  {journal}
  {\bibinfo  {journal} {Symmetry}\ }\textbf {\bibinfo {volume} {11}},\ \bibinfo
  {pages} {587} (\bibinfo {year} {2019})},\ \Eprint
  {https://arxiv.org/abs/1904.12013} {arXiv:1904.12013 [hep-ph]} \BibitemShut
  {NoStop}%
\bibitem [{\citenamefont {Yamanaka}\ \emph {et~al.}(2021)\citenamefont
  {Yamanaka}, \citenamefont {Iida}, \citenamefont {Nakamura},\ and\
  \citenamefont {Wakayama}}]{Yamanaka:2019aeq}%
  \BibitemOpen
  \bibfield  {author} {\bibinfo {author} {\bibfnamefont {N.}~\bibnamefont
  {Yamanaka}}, \bibinfo {author} {\bibfnamefont {H.}~\bibnamefont {Iida}},
  \bibinfo {author} {\bibfnamefont {A.}~\bibnamefont {Nakamura}},\ and\
  \bibinfo {author} {\bibfnamefont {M.}~\bibnamefont {Wakayama}},\ }\bibfield
  {title} {\bibinfo {title} {{Dark matter scattering cross section and dynamics
  in dark Yang-Mills theory}},\ }\href
  {https://doi.org/10.1016/j.physletb.2020.136056} {\bibfield  {journal}
  {\bibinfo  {journal} {Phys. Lett. B}\ }\textbf {\bibinfo {volume} {813}},\
  \bibinfo {pages} {136056} (\bibinfo {year} {2021})},\ \Eprint
  {https://arxiv.org/abs/1910.01440} {arXiv:1910.01440 [hep-ph]} \BibitemShut
  {NoStop}%
\bibitem [{\citenamefont {Yamanaka}\ \emph {et~al.}(2020)\citenamefont
  {Yamanaka}, \citenamefont {Iida}, \citenamefont {Nakamura},\ and\
  \citenamefont {Wakayama}}]{Yamanaka:2019yek}%
  \BibitemOpen
  \bibfield  {author} {\bibinfo {author} {\bibfnamefont {N.}~\bibnamefont
  {Yamanaka}}, \bibinfo {author} {\bibfnamefont {H.}~\bibnamefont {Iida}},
  \bibinfo {author} {\bibfnamefont {A.}~\bibnamefont {Nakamura}},\ and\
  \bibinfo {author} {\bibfnamefont {M.}~\bibnamefont {Wakayama}},\ }\bibfield
  {title} {\bibinfo {title} {{Glueball scattering cross section in lattice
  SU(2) Yang-Mills theory}},\ }\href
  {https://doi.org/10.1103/PhysRevD.102.054507} {\bibfield  {journal} {\bibinfo
   {journal} {Phys. Rev. D}\ }\textbf {\bibinfo {volume} {102}},\ \bibinfo
  {pages} {054507} (\bibinfo {year} {2020})},\ \Eprint
  {https://arxiv.org/abs/1910.07756} {arXiv:1910.07756 [hep-lat]} \BibitemShut
  {NoStop}%
\bibitem [{\citenamefont {Cai}\ and\ \citenamefont
  {Cacciapaglia}(2021)}]{Cai:2020njb}%
  \BibitemOpen
  \bibfield  {author} {\bibinfo {author} {\bibfnamefont {H.}~\bibnamefont
  {Cai}}\ and\ \bibinfo {author} {\bibfnamefont {G.}~\bibnamefont
  {Cacciapaglia}},\ }\bibfield  {title} {\bibinfo {title} {{Singlet dark matter
  in the SU(6)/SO(6) composite Higgs model}},\ }\href
  {https://doi.org/10.1103/PhysRevD.103.055002} {\bibfield  {journal} {\bibinfo
   {journal} {Phys. Rev. D}\ }\textbf {\bibinfo {volume} {103}},\ \bibinfo
  {pages} {055002} (\bibinfo {year} {2021})},\ \Eprint
  {https://arxiv.org/abs/2007.04338} {arXiv:2007.04338 [hep-ph]} \BibitemShut
  {NoStop}%
\bibitem [{\citenamefont {Hochberg}\ \emph {et~al.}(2014)\citenamefont
  {Hochberg}, \citenamefont {Kuflik}, \citenamefont {Volansky},\ and\
  \citenamefont {Wacker}}]{Hochberg:2014dra}%
  \BibitemOpen
  \bibfield  {author} {\bibinfo {author} {\bibfnamefont {Y.}~\bibnamefont
  {Hochberg}}, \bibinfo {author} {\bibfnamefont {E.}~\bibnamefont {Kuflik}},
  \bibinfo {author} {\bibfnamefont {T.}~\bibnamefont {Volansky}},\ and\
  \bibinfo {author} {\bibfnamefont {J.~G.}\ \bibnamefont {Wacker}},\ }\bibfield
   {title} {\bibinfo {title} {{Mechanism for Thermal Relic Dark Matter of
  Strongly Interacting Massive Particles}},\ }\href
  {https://doi.org/10.1103/PhysRevLett.113.171301} {\bibfield  {journal}
  {\bibinfo  {journal} {Phys. Rev. Lett.}\ }\textbf {\bibinfo {volume} {113}},\
  \bibinfo {pages} {171301} (\bibinfo {year} {2014})},\ \Eprint
  {https://arxiv.org/abs/1402.5143} {arXiv:1402.5143 [hep-ph]} \BibitemShut
  {NoStop}%
\bibitem [{\citenamefont {Hochberg}\ \emph {et~al.}(2015)\citenamefont
  {Hochberg}, \citenamefont {Kuflik}, \citenamefont {Murayama}, \citenamefont
  {Volansky},\ and\ \citenamefont {Wacker}}]{Hochberg:2014kqa}%
  \BibitemOpen
  \bibfield  {author} {\bibinfo {author} {\bibfnamefont {Y.}~\bibnamefont
  {Hochberg}}, \bibinfo {author} {\bibfnamefont {E.}~\bibnamefont {Kuflik}},
  \bibinfo {author} {\bibfnamefont {H.}~\bibnamefont {Murayama}}, \bibinfo
  {author} {\bibfnamefont {T.}~\bibnamefont {Volansky}},\ and\ \bibinfo
  {author} {\bibfnamefont {J.~G.}\ \bibnamefont {Wacker}},\ }\bibfield  {title}
  {\bibinfo {title} {{Model for Thermal Relic Dark Matter of Strongly
  Interacting Massive Particles}},\ }\href
  {https://doi.org/10.1103/PhysRevLett.115.021301} {\bibfield  {journal}
  {\bibinfo  {journal} {Phys. Rev. Lett.}\ }\textbf {\bibinfo {volume} {115}},\
  \bibinfo {pages} {021301} (\bibinfo {year} {2015})},\ \Eprint
  {https://arxiv.org/abs/1411.3727} {arXiv:1411.3727 [hep-ph]} \BibitemShut
  {NoStop}%
\bibitem [{\citenamefont {Hochberg}\ \emph {et~al.}(2016)\citenamefont
  {Hochberg}, \citenamefont {Kuflik},\ and\ \citenamefont
  {Murayama}}]{Hochberg:2015vrg}%
  \BibitemOpen
  \bibfield  {author} {\bibinfo {author} {\bibfnamefont {Y.}~\bibnamefont
  {Hochberg}}, \bibinfo {author} {\bibfnamefont {E.}~\bibnamefont {Kuflik}},\
  and\ \bibinfo {author} {\bibfnamefont {H.}~\bibnamefont {Murayama}},\
  }\bibfield  {title} {\bibinfo {title} {{SIMP Spectroscopy}},\ }\href
  {https://doi.org/10.1007/JHEP05(2016)090} {\bibfield  {journal} {\bibinfo
  {journal} {JHEP}\ }\textbf {\bibinfo {volume} {05}},\ \bibinfo {pages}
  {090}},\ \Eprint {https://arxiv.org/abs/1512.07917} {arXiv:1512.07917
  [hep-ph]} \BibitemShut {NoStop}%
\bibitem [{\citenamefont {Bernal}\ \emph {et~al.}(2017)\citenamefont {Bernal},
  \citenamefont {Chu},\ and\ \citenamefont {Pradler}}]{Bernal:2017mqb}%
  \BibitemOpen
  \bibfield  {author} {\bibinfo {author} {\bibfnamefont {N.}~\bibnamefont
  {Bernal}}, \bibinfo {author} {\bibfnamefont {X.}~\bibnamefont {Chu}},\ and\
  \bibinfo {author} {\bibfnamefont {J.}~\bibnamefont {Pradler}},\ }\bibfield
  {title} {\bibinfo {title} {{Simply split strongly interacting massive
  particles}},\ }\href {https://doi.org/10.1103/PhysRevD.95.115023} {\bibfield
  {journal} {\bibinfo  {journal} {Phys. Rev. D}\ }\textbf {\bibinfo {volume}
  {95}},\ \bibinfo {pages} {115023} (\bibinfo {year} {2017})},\ \Eprint
  {https://arxiv.org/abs/1702.04906} {arXiv:1702.04906 [hep-ph]} \BibitemShut
  {NoStop}%
\bibitem [{\citenamefont {Berlin}\ \emph {et~al.}(2018)\citenamefont {Berlin},
  \citenamefont {Blinov}, \citenamefont {Gori}, \citenamefont {Schuster},\ and\
  \citenamefont {Toro}}]{Berlin:2018tvf}%
  \BibitemOpen
  \bibfield  {author} {\bibinfo {author} {\bibfnamefont {A.}~\bibnamefont
  {Berlin}}, \bibinfo {author} {\bibfnamefont {N.}~\bibnamefont {Blinov}},
  \bibinfo {author} {\bibfnamefont {S.}~\bibnamefont {Gori}}, \bibinfo {author}
  {\bibfnamefont {P.}~\bibnamefont {Schuster}},\ and\ \bibinfo {author}
  {\bibfnamefont {N.}~\bibnamefont {Toro}},\ }\bibfield  {title} {\bibinfo
  {title} {{Cosmology and Accelerator Tests of Strongly Interacting Dark
  Matter}},\ }\href {https://doi.org/10.1103/PhysRevD.97.055033} {\bibfield
  {journal} {\bibinfo  {journal} {Phys. Rev. D}\ }\textbf {\bibinfo {volume}
  {97}},\ \bibinfo {pages} {055033} (\bibinfo {year} {2018})},\ \Eprint
  {https://arxiv.org/abs/1801.05805} {arXiv:1801.05805 [hep-ph]} \BibitemShut
  {NoStop}%
\bibitem [{\citenamefont {Bernal}\ \emph {et~al.}(2020)\citenamefont {Bernal},
  \citenamefont {Chu}, \citenamefont {Kulkarni},\ and\ \citenamefont
  {Pradler}}]{Bernal:2019uqr}%
  \BibitemOpen
  \bibfield  {author} {\bibinfo {author} {\bibfnamefont {N.}~\bibnamefont
  {Bernal}}, \bibinfo {author} {\bibfnamefont {X.}~\bibnamefont {Chu}},
  \bibinfo {author} {\bibfnamefont {S.}~\bibnamefont {Kulkarni}},\ and\
  \bibinfo {author} {\bibfnamefont {J.}~\bibnamefont {Pradler}},\ }\bibfield
  {title} {\bibinfo {title} {{Self-interacting dark matter without
  prejudice}},\ }\href {https://doi.org/10.1103/PhysRevD.101.055044} {\bibfield
   {journal} {\bibinfo  {journal} {Phys. Rev. D}\ }\textbf {\bibinfo {volume}
  {101}},\ \bibinfo {pages} {055044} (\bibinfo {year} {2020})},\ \Eprint
  {https://arxiv.org/abs/1912.06681} {arXiv:1912.06681 [hep-ph]} \BibitemShut
  {NoStop}%
\bibitem [{\citenamefont {Tsai}\ \emph {et~al.}(2022)\citenamefont {Tsai},
  \citenamefont {McGehee},\ and\ \citenamefont {Murayama}}]{Tsai:2020vpi}%
  \BibitemOpen
  \bibfield  {author} {\bibinfo {author} {\bibfnamefont {Y.-D.}\ \bibnamefont
  {Tsai}}, \bibinfo {author} {\bibfnamefont {R.}~\bibnamefont {McGehee}},\ and\
  \bibinfo {author} {\bibfnamefont {H.}~\bibnamefont {Murayama}},\ }\bibfield
  {title} {\bibinfo {title} {{Resonant Self-Interacting Dark Matter from Dark
  QCD}},\ }\href {https://doi.org/10.1103/PhysRevLett.128.172001} {\bibfield
  {journal} {\bibinfo  {journal} {Phys. Rev. Lett.}\ }\textbf {\bibinfo
  {volume} {128}},\ \bibinfo {pages} {172001} (\bibinfo {year} {2022})},\
  \Eprint {https://arxiv.org/abs/2008.08608} {arXiv:2008.08608 [hep-ph]}
  \BibitemShut {NoStop}%
\bibitem [{\citenamefont {Kondo}\ \emph {et~al.}(2022)\citenamefont {Kondo},
  \citenamefont {McGehee}, \citenamefont {Melia},\ and\ \citenamefont
  {Murayama}}]{Kondo:2022lgg}%
  \BibitemOpen
  \bibfield  {author} {\bibinfo {author} {\bibfnamefont {D.}~\bibnamefont
  {Kondo}}, \bibinfo {author} {\bibfnamefont {R.}~\bibnamefont {McGehee}},
  \bibinfo {author} {\bibfnamefont {T.}~\bibnamefont {Melia}},\ and\ \bibinfo
  {author} {\bibfnamefont {H.}~\bibnamefont {Murayama}},\ }\bibfield  {title}
  {\bibinfo {title} {{Linear sigma dark matter}},\ }\href
  {https://doi.org/10.1007/JHEP09(2022)041} {\bibfield  {journal} {\bibinfo
  {journal} {JHEP}\ }\textbf {\bibinfo {volume} {09}},\ \bibinfo {pages}
  {041}},\ \Eprint {https://arxiv.org/abs/2205.08088} {arXiv:2205.08088
  [hep-ph]} \BibitemShut {NoStop}%
\bibitem [{\citenamefont {Witten}(1984)}]{Witten:1984rs}%
  \BibitemOpen
  \bibfield  {author} {\bibinfo {author} {\bibfnamefont {E.}~\bibnamefont
  {Witten}},\ }\bibfield  {title} {\bibinfo {title} {{Cosmic Separation of
  Phases}},\ }\href {https://doi.org/10.1103/PhysRevD.30.272} {\bibfield
  {journal} {\bibinfo  {journal} {Phys. Rev. D}\ }\textbf {\bibinfo {volume}
  {30}},\ \bibinfo {pages} {272} (\bibinfo {year} {1984})}\BibitemShut
  {NoStop}%
\bibitem [{\citenamefont {Kamionkowski}\ \emph {et~al.}(1994)\citenamefont
  {Kamionkowski}, \citenamefont {Kosowsky},\ and\ \citenamefont
  {Turner}}]{Kamionkowski:1993fg}%
  \BibitemOpen
  \bibfield  {author} {\bibinfo {author} {\bibfnamefont {M.}~\bibnamefont
  {Kamionkowski}}, \bibinfo {author} {\bibfnamefont {A.}~\bibnamefont
  {Kosowsky}},\ and\ \bibinfo {author} {\bibfnamefont {M.~S.}\ \bibnamefont
  {Turner}},\ }\bibfield  {title} {\bibinfo {title} {{Gravitational radiation
  from first order phase transitions}},\ }\href
  {https://doi.org/10.1103/PhysRevD.49.2837} {\bibfield  {journal} {\bibinfo
  {journal} {Phys. Rev. D}\ }\textbf {\bibinfo {volume} {49}},\ \bibinfo
  {pages} {2837} (\bibinfo {year} {1994})},\ \Eprint
  {https://arxiv.org/abs/astro-ph/9310044} {arXiv:astro-ph/9310044}
  \BibitemShut {NoStop}%
\bibitem [{\citenamefont {Allen}(1996)}]{Allen:1996vm}%
  \BibitemOpen
  \bibfield  {author} {\bibinfo {author} {\bibfnamefont {B.}~\bibnamefont
  {Allen}},\ }\bibfield  {title} {\bibinfo {title} {{The Stochastic gravity
  wave background: Sources and detection}},\ }in\ \href@noop {} {\emph
  {\bibinfo {booktitle} {{Les Houches School of Physics: Astrophysical Sources
  of Gravitational Radiation}}}}\ (\bibinfo {year} {1996})\ pp.\ \bibinfo
  {pages} {373--417},\ \Eprint {https://arxiv.org/abs/gr-qc/9604033}
  {arXiv:gr-qc/9604033} \BibitemShut {NoStop}%
\bibitem [{\citenamefont {Schwaller}(2015)}]{Schwaller:2015tja}%
  \BibitemOpen
  \bibfield  {author} {\bibinfo {author} {\bibfnamefont {P.}~\bibnamefont
  {Schwaller}},\ }\bibfield  {title} {\bibinfo {title} {{Gravitational Waves
  from a Dark Phase Transition}},\ }\href
  {https://doi.org/10.1103/PhysRevLett.115.181101} {\bibfield  {journal}
  {\bibinfo  {journal} {Phys. Rev. Lett.}\ }\textbf {\bibinfo {volume} {115}},\
  \bibinfo {pages} {181101} (\bibinfo {year} {2015})},\ \Eprint
  {https://arxiv.org/abs/1504.07263} {arXiv:1504.07263 [hep-ph]} \BibitemShut
  {NoStop}%
\bibitem [{\citenamefont {Croon}\ \emph {et~al.}(2018)\citenamefont {Croon},
  \citenamefont {Sanz},\ and\ \citenamefont {White}}]{Croon:2018erz}%
  \BibitemOpen
  \bibfield  {author} {\bibinfo {author} {\bibfnamefont {D.}~\bibnamefont
  {Croon}}, \bibinfo {author} {\bibfnamefont {V.}~\bibnamefont {Sanz}},\ and\
  \bibinfo {author} {\bibfnamefont {G.}~\bibnamefont {White}},\ }\bibfield
  {title} {\bibinfo {title} {{Model Discrimination in Gravitational Wave
  spectra from Dark Phase Transitions}},\ }\href
  {https://doi.org/10.1007/JHEP08(2018)203} {\bibfield  {journal} {\bibinfo
  {journal} {JHEP}\ }\textbf {\bibinfo {volume} {08}},\ \bibinfo {pages}
  {203}},\ \Eprint {https://arxiv.org/abs/1806.02332} {arXiv:1806.02332
  [hep-ph]} \BibitemShut {NoStop}%
\bibitem [{\citenamefont {Christensen}(2019)}]{Christensen:2018iqi}%
  \BibitemOpen
  \bibfield  {author} {\bibinfo {author} {\bibfnamefont {N.}~\bibnamefont
  {Christensen}},\ }\bibfield  {title} {\bibinfo {title} {{Stochastic
  Gravitational Wave Backgrounds}},\ }\href
  {https://doi.org/10.1088/1361-6633/aae6b5} {\bibfield  {journal} {\bibinfo
  {journal} {Rept. Prog. Phys.}\ }\textbf {\bibinfo {volume} {82}},\ \bibinfo
  {pages} {016903} (\bibinfo {year} {2019})},\ \Eprint
  {https://arxiv.org/abs/1811.08797} {arXiv:1811.08797 [gr-qc]} \BibitemShut
  {NoStop}%
\bibitem [{\citenamefont {Seto}\ \emph {et~al.}(2001)\citenamefont {Seto},
  \citenamefont {Kawamura},\ and\ \citenamefont {Nakamura}}]{Seto:2001qf}%
  \BibitemOpen
  \bibfield  {author} {\bibinfo {author} {\bibfnamefont {N.}~\bibnamefont
  {Seto}}, \bibinfo {author} {\bibfnamefont {S.}~\bibnamefont {Kawamura}},\
  and\ \bibinfo {author} {\bibfnamefont {T.}~\bibnamefont {Nakamura}},\
  }\bibfield  {title} {\bibinfo {title} {{Possibility of direct measurement of
  the acceleration of the universe using 0.1-Hz band laser interferometer
  gravitational wave antenna in space}},\ }\href
  {https://doi.org/10.1103/PhysRevLett.87.221103} {\bibfield  {journal}
  {\bibinfo  {journal} {Phys. Rev. Lett.}\ }\textbf {\bibinfo {volume} {87}},\
  \bibinfo {pages} {221103} (\bibinfo {year} {2001})},\ \Eprint
  {https://arxiv.org/abs/astro-ph/0108011} {arXiv:astro-ph/0108011}
  \BibitemShut {NoStop}%
\bibitem [{\citenamefont {Kawamura}\ \emph {et~al.}(2006)\citenamefont
  {Kawamura} \emph {et~al.}}]{Kawamura:2006up}%
  \BibitemOpen
  \bibfield  {author} {\bibinfo {author} {\bibfnamefont {S.}~\bibnamefont
  {Kawamura}} \emph {et~al.},\ }\bibfield  {title} {\bibinfo {title} {{The
  Japanese space gravitational wave antenna DECIGO}},\ }\href
  {https://doi.org/10.1088/0264-9381/23/8/S17} {\bibfield  {journal} {\bibinfo
  {journal} {Class. Quant. Grav.}\ }\textbf {\bibinfo {volume} {23}},\ \bibinfo
  {pages} {S125} (\bibinfo {year} {2006})}\BibitemShut {NoStop}%
\bibitem [{\citenamefont {Crowder}\ and\ \citenamefont
  {Cornish}(2005)}]{Crowder:2005nr}%
  \BibitemOpen
  \bibfield  {author} {\bibinfo {author} {\bibfnamefont {J.}~\bibnamefont
  {Crowder}}\ and\ \bibinfo {author} {\bibfnamefont {N.~J.}\ \bibnamefont
  {Cornish}},\ }\bibfield  {title} {\bibinfo {title} {{Beyond LISA: Exploring
  future gravitational wave missions}},\ }\href
  {https://doi.org/10.1103/PhysRevD.72.083005} {\bibfield  {journal} {\bibinfo
  {journal} {Phys. Rev. D}\ }\textbf {\bibinfo {volume} {72}},\ \bibinfo
  {pages} {083005} (\bibinfo {year} {2005})},\ \Eprint
  {https://arxiv.org/abs/gr-qc/0506015} {arXiv:gr-qc/0506015} \BibitemShut
  {NoStop}%
\bibitem [{\citenamefont {Corbin}\ and\ \citenamefont
  {Cornish}(2006)}]{Corbin:2005ny}%
  \BibitemOpen
  \bibfield  {author} {\bibinfo {author} {\bibfnamefont {V.}~\bibnamefont
  {Corbin}}\ and\ \bibinfo {author} {\bibfnamefont {N.~J.}\ \bibnamefont
  {Cornish}},\ }\bibfield  {title} {\bibinfo {title} {{Detecting the cosmic
  gravitational wave background with the big bang observer}},\ }\href
  {https://doi.org/10.1088/0264-9381/23/7/014} {\bibfield  {journal} {\bibinfo
  {journal} {Class. Quant. Grav.}\ }\textbf {\bibinfo {volume} {23}},\ \bibinfo
  {pages} {2435} (\bibinfo {year} {2006})},\ \Eprint
  {https://arxiv.org/abs/gr-qc/0512039} {arXiv:gr-qc/0512039} \BibitemShut
  {NoStop}%
\bibitem [{\citenamefont {Harry}\ \emph {et~al.}(2006)\citenamefont {Harry},
  \citenamefont {Fritschel}, \citenamefont {Shaddock}, \citenamefont
  {Folkner},\ and\ \citenamefont {Phinney}}]{Harry:2006fi}%
  \BibitemOpen
  \bibfield  {author} {\bibinfo {author} {\bibfnamefont {G.~M.}\ \bibnamefont
  {Harry}}, \bibinfo {author} {\bibfnamefont {P.}~\bibnamefont {Fritschel}},
  \bibinfo {author} {\bibfnamefont {D.~A.}\ \bibnamefont {Shaddock}}, \bibinfo
  {author} {\bibfnamefont {W.}~\bibnamefont {Folkner}},\ and\ \bibinfo {author}
  {\bibfnamefont {E.~S.}\ \bibnamefont {Phinney}},\ }\bibfield  {title}
  {\bibinfo {title} {{Laser interferometry for the big bang observer}},\ }\href
  {https://doi.org/10.1088/0264-9381/23/15/008} {\bibfield  {journal} {\bibinfo
   {journal} {Class. Quant. Grav.}\ }\textbf {\bibinfo {volume} {23}},\
  \bibinfo {pages} {4887} (\bibinfo {year} {2006})},\ \bibinfo {note}
  {[Erratum: Class.Quant.Grav. 23, 7361 (2006)]}\BibitemShut {NoStop}%
\bibitem [{\citenamefont {Hild}\ \emph {et~al.}(2011)\citenamefont {Hild} \emph
  {et~al.}}]{Hild:2010id}%
  \BibitemOpen
  \bibfield  {author} {\bibinfo {author} {\bibfnamefont {S.}~\bibnamefont
  {Hild}} \emph {et~al.},\ }\bibfield  {title} {\bibinfo {title} {{Sensitivity
  Studies for Third-Generation Gravitational Wave Observatories}},\ }\href
  {https://doi.org/10.1088/0264-9381/28/9/094013} {\bibfield  {journal}
  {\bibinfo  {journal} {Class. Quant. Grav.}\ }\textbf {\bibinfo {volume}
  {28}},\ \bibinfo {pages} {094013} (\bibinfo {year} {2011})},\ \Eprint
  {https://arxiv.org/abs/1012.0908} {arXiv:1012.0908 [gr-qc]} \BibitemShut
  {NoStop}%
\bibitem [{\citenamefont {Yagi}\ and\ \citenamefont
  {Seto}(2011)}]{Yagi:2011wg}%
  \BibitemOpen
  \bibfield  {author} {\bibinfo {author} {\bibfnamefont {K.}~\bibnamefont
  {Yagi}}\ and\ \bibinfo {author} {\bibfnamefont {N.}~\bibnamefont {Seto}},\
  }\bibfield  {title} {\bibinfo {title} {{Detector configuration of DECIGO/BBO
  and identification of cosmological neutron-star binaries}},\ }\href
  {https://doi.org/10.1103/PhysRevD.83.044011} {\bibfield  {journal} {\bibinfo
  {journal} {Phys. Rev. D}\ }\textbf {\bibinfo {volume} {83}},\ \bibinfo
  {pages} {044011} (\bibinfo {year} {2011})},\ \bibinfo {note} {[Erratum:
  Phys.Rev.D 95, 109901 (2017)]},\ \Eprint {https://arxiv.org/abs/1101.3940}
  {arXiv:1101.3940 [astro-ph.CO]} \BibitemShut {NoStop}%
\bibitem [{\citenamefont {Sathyaprakash}\ \emph {et~al.}(2012)\citenamefont
  {Sathyaprakash} \emph {et~al.}}]{Sathyaprakash:2012jk}%
  \BibitemOpen
  \bibfield  {author} {\bibinfo {author} {\bibfnamefont {B.}~\bibnamefont
  {Sathyaprakash}} \emph {et~al.},\ }\bibfield  {title} {\bibinfo {title}
  {{Scientific Objectives of Einstein Telescope}},\ }\href
  {https://doi.org/10.1088/0264-9381/29/12/124013} {\bibfield  {journal}
  {\bibinfo  {journal} {Class. Quant. Grav.}\ }\textbf {\bibinfo {volume}
  {29}},\ \bibinfo {pages} {124013} (\bibinfo {year} {2012})},\ \bibinfo {note}
  {[Erratum: Class.Quant.Grav. 30, 079501 (2013)]},\ \Eprint
  {https://arxiv.org/abs/1206.0331} {arXiv:1206.0331 [gr-qc]} \BibitemShut
  {NoStop}%
\bibitem [{\citenamefont {Thrane}\ and\ \citenamefont
  {Romano}(2013)}]{Thrane:2013oya}%
  \BibitemOpen
  \bibfield  {author} {\bibinfo {author} {\bibfnamefont {E.}~\bibnamefont
  {Thrane}}\ and\ \bibinfo {author} {\bibfnamefont {J.~D.}\ \bibnamefont
  {Romano}},\ }\bibfield  {title} {\bibinfo {title} {{Sensitivity curves for
  searches for gravitational-wave backgrounds}},\ }\href
  {https://doi.org/10.1103/PhysRevD.88.124032} {\bibfield  {journal} {\bibinfo
  {journal} {Phys. Rev. D}\ }\textbf {\bibinfo {volume} {88}},\ \bibinfo
  {pages} {124032} (\bibinfo {year} {2013})},\ \Eprint
  {https://arxiv.org/abs/1310.5300} {arXiv:1310.5300 [astro-ph.IM]}
  \BibitemShut {NoStop}%
\bibitem [{\citenamefont {Caprini}\ \emph {et~al.}(2016)\citenamefont {Caprini}
  \emph {et~al.}}]{Caprini:2015zlo}%
  \BibitemOpen
  \bibfield  {author} {\bibinfo {author} {\bibfnamefont {C.}~\bibnamefont
  {Caprini}} \emph {et~al.},\ }\bibfield  {title} {\bibinfo {title} {{Science
  with the space-based interferometer eLISA. II: Gravitational waves from
  cosmological phase transitions}},\ }\href
  {https://doi.org/10.1088/1475-7516/2016/04/001} {\bibfield  {journal}
  {\bibinfo  {journal} {JCAP}\ }\textbf {\bibinfo {volume} {04}},\ \bibinfo
  {pages} {001}},\ \Eprint {https://arxiv.org/abs/1512.06239} {arXiv:1512.06239
  [astro-ph.CO]} \BibitemShut {NoStop}%
\bibitem [{\citenamefont {Amaro-Seoane}\ \emph {et~al.}(2017)\citenamefont
  {Amaro-Seoane} \emph {et~al.}}]{LISA:2017pwj}%
  \BibitemOpen
  \bibfield  {author} {\bibinfo {author} {\bibfnamefont {P.}~\bibnamefont
  {Amaro-Seoane}} \emph {et~al.} (\bibinfo {collaboration} {LISA}),\ }\bibfield
   {title} {\bibinfo {title} {{Laser Interferometer Space Antenna}}\ }(\bibinfo
  {year} {2017})\ \Eprint {https://arxiv.org/abs/1702.00786} {arXiv:1702.00786
  [astro-ph.IM]} \BibitemShut {NoStop}%
\bibitem [{\citenamefont {Abbott}\ \emph {et~al.}(2017)\citenamefont {Abbott}
  \emph {et~al.}}]{LIGOScientific:2016wof}%
  \BibitemOpen
  \bibfield  {author} {\bibinfo {author} {\bibfnamefont {B.~P.}\ \bibnamefont
  {Abbott}} \emph {et~al.} (\bibinfo {collaboration} {LIGO Scientific}),\
  }\bibfield  {title} {\bibinfo {title} {{Exploring the Sensitivity of Next
  Generation Gravitational Wave Detectors}},\ }\href
  {https://doi.org/10.1088/1361-6382/aa51f4} {\bibfield  {journal} {\bibinfo
  {journal} {Class. Quant. Grav.}\ }\textbf {\bibinfo {volume} {34}},\ \bibinfo
  {pages} {044001} (\bibinfo {year} {2017})},\ \Eprint
  {https://arxiv.org/abs/1607.08697} {arXiv:1607.08697 [astro-ph.IM]}
  \BibitemShut {NoStop}%
\bibitem [{\citenamefont {Isoyama}\ \emph {et~al.}(2018)\citenamefont
  {Isoyama}, \citenamefont {Nakano},\ and\ \citenamefont
  {Nakamura}}]{Isoyama:2018rjb}%
  \BibitemOpen
  \bibfield  {author} {\bibinfo {author} {\bibfnamefont {S.}~\bibnamefont
  {Isoyama}}, \bibinfo {author} {\bibfnamefont {H.}~\bibnamefont {Nakano}},\
  and\ \bibinfo {author} {\bibfnamefont {T.}~\bibnamefont {Nakamura}},\
  }\bibfield  {title} {\bibinfo {title} {{Multiband Gravitational-Wave
  Astronomy: Observing binary inspirals with a decihertz detector, B-DECIGO}},\
  }\href {https://doi.org/10.1093/ptep/pty078} {\bibfield  {journal} {\bibinfo
  {journal} {PTEP}\ }\textbf {\bibinfo {volume} {2018}},\ \bibinfo {pages}
  {073E01} (\bibinfo {year} {2018})},\ \Eprint
  {https://arxiv.org/abs/1802.06977} {arXiv:1802.06977 [gr-qc]} \BibitemShut
  {NoStop}%
\bibitem [{\citenamefont {Baker}\ \emph {et~al.}(2019)\citenamefont {Baker}
  \emph {et~al.}}]{Baker:2019nia}%
  \BibitemOpen
  \bibfield  {author} {\bibinfo {author} {\bibfnamefont {J.}~\bibnamefont
  {Baker}} \emph {et~al.},\ }\href@noop {} {\bibinfo {title} {{The Laser
  Interferometer Space Antenna: Unveiling the Millihertz Gravitational Wave
  Sky}}} (\bibinfo {year} {2019}),\ \Eprint {https://arxiv.org/abs/1907.06482}
  {arXiv:1907.06482 [astro-ph.IM]} \BibitemShut {NoStop}%
\bibitem [{\citenamefont {Brdar}\ \emph {et~al.}(2019)\citenamefont {Brdar},
  \citenamefont {Helmboldt},\ and\ \citenamefont {Kubo}}]{Brdar:2018num}%
  \BibitemOpen
  \bibfield  {author} {\bibinfo {author} {\bibfnamefont {V.}~\bibnamefont
  {Brdar}}, \bibinfo {author} {\bibfnamefont {A.~J.}\ \bibnamefont
  {Helmboldt}},\ and\ \bibinfo {author} {\bibfnamefont {J.}~\bibnamefont
  {Kubo}},\ }\bibfield  {title} {\bibinfo {title} {{Gravitational Waves from
  First-Order Phase Transitions: LIGO as a Window to Unexplored Seesaw
  Scales}},\ }\href {https://doi.org/10.1088/1475-7516/2019/02/021} {\bibfield
  {journal} {\bibinfo  {journal} {JCAP}\ }\textbf {\bibinfo {volume} {02}},\
  \bibinfo {pages} {021}},\ \Eprint {https://arxiv.org/abs/1810.12306}
  {arXiv:1810.12306 [hep-ph]} \BibitemShut {NoStop}%
\bibitem [{\citenamefont {Reitze}\ \emph {et~al.}(2019)\citenamefont {Reitze}
  \emph {et~al.}}]{Reitze:2019iox}%
  \BibitemOpen
  \bibfield  {author} {\bibinfo {author} {\bibfnamefont {D.}~\bibnamefont
  {Reitze}} \emph {et~al.},\ }\bibfield  {title} {\bibinfo {title} {{Cosmic
  Explorer: The U.S. Contribution to Gravitational-Wave Astronomy beyond
  LIGO}},\ }\href@noop {} {\bibfield  {journal} {\bibinfo  {journal} {Bull. Am.
  Astron. Soc.}\ }\textbf {\bibinfo {volume} {51}},\ \bibinfo {pages} {035}
  (\bibinfo {year} {2019})},\ \Eprint {https://arxiv.org/abs/1907.04833}
  {arXiv:1907.04833 [astro-ph.IM]} \BibitemShut {NoStop}%
\bibitem [{\citenamefont {Caprini}\ \emph {et~al.}(2020)\citenamefont {Caprini}
  \emph {et~al.}}]{Caprini:2019egz}%
  \BibitemOpen
  \bibfield  {author} {\bibinfo {author} {\bibfnamefont {C.}~\bibnamefont
  {Caprini}} \emph {et~al.},\ }\bibfield  {title} {\bibinfo {title} {{Detecting
  gravitational waves from cosmological phase transitions with LISA: an
  update}},\ }\href {https://doi.org/10.1088/1475-7516/2020/03/024} {\bibfield
  {journal} {\bibinfo  {journal} {JCAP}\ }\textbf {\bibinfo {volume} {03}},\
  \bibinfo {pages} {024}},\ \Eprint {https://arxiv.org/abs/1910.13125}
  {arXiv:1910.13125 [astro-ph.CO]} \BibitemShut {NoStop}%
\bibitem [{\citenamefont {Maggiore}\ \emph {et~al.}(2020)\citenamefont
  {Maggiore} \emph {et~al.}}]{Maggiore:2019uih}%
  \BibitemOpen
  \bibfield  {author} {\bibinfo {author} {\bibfnamefont {M.}~\bibnamefont
  {Maggiore}} \emph {et~al.},\ }\bibfield  {title} {\bibinfo {title} {{Science
  Case for the Einstein Telescope}},\ }\href
  {https://doi.org/10.1088/1475-7516/2020/03/050} {\bibfield  {journal}
  {\bibinfo  {journal} {JCAP}\ }\textbf {\bibinfo {volume} {03}},\ \bibinfo
  {pages} {050}},\ \Eprint {https://arxiv.org/abs/1912.02622} {arXiv:1912.02622
  [astro-ph.CO]} \BibitemShut {NoStop}%
\bibitem [{\citenamefont {Huang}\ \emph {et~al.}(2021)\citenamefont {Huang},
  \citenamefont {Reichert}, \citenamefont {Sannino},\ and\ \citenamefont
  {Wang}}]{Huang:2020crf}%
  \BibitemOpen
  \bibfield  {author} {\bibinfo {author} {\bibfnamefont {W.-C.}\ \bibnamefont
  {Huang}}, \bibinfo {author} {\bibfnamefont {M.}~\bibnamefont {Reichert}},
  \bibinfo {author} {\bibfnamefont {F.}~\bibnamefont {Sannino}},\ and\ \bibinfo
  {author} {\bibfnamefont {Z.-W.}\ \bibnamefont {Wang}},\ }\bibfield  {title}
  {\bibinfo {title} {{Testing the dark SU(N) Yang-Mills theory confined
  landscape: From the lattice to gravitational waves}},\ }\href
  {https://doi.org/10.1103/PhysRevD.104.035005} {\bibfield  {journal} {\bibinfo
   {journal} {Phys. Rev. D}\ }\textbf {\bibinfo {volume} {104}},\ \bibinfo
  {pages} {035005} (\bibinfo {year} {2021})},\ \Eprint
  {https://arxiv.org/abs/2012.11614} {arXiv:2012.11614 [hep-ph]} \BibitemShut
  {NoStop}%
\bibitem [{\citenamefont {Halverson}\ \emph {et~al.}(2021)\citenamefont
  {Halverson}, \citenamefont {Long}, \citenamefont {Maiti}, \citenamefont
  {Nelson},\ and\ \citenamefont {Salinas}}]{Halverson:2020xpg}%
  \BibitemOpen
  \bibfield  {author} {\bibinfo {author} {\bibfnamefont {J.}~\bibnamefont
  {Halverson}}, \bibinfo {author} {\bibfnamefont {C.}~\bibnamefont {Long}},
  \bibinfo {author} {\bibfnamefont {A.}~\bibnamefont {Maiti}}, \bibinfo
  {author} {\bibfnamefont {B.}~\bibnamefont {Nelson}},\ and\ \bibinfo {author}
  {\bibfnamefont {G.}~\bibnamefont {Salinas}},\ }\bibfield  {title} {\bibinfo
  {title} {{Gravitational waves from dark Yang-Mills sectors}},\ }\href
  {https://doi.org/10.1007/JHEP05(2021)154} {\bibfield  {journal} {\bibinfo
  {journal} {JHEP}\ }\textbf {\bibinfo {volume} {05}},\ \bibinfo {pages}
  {154}},\ \Eprint {https://arxiv.org/abs/2012.04071} {arXiv:2012.04071
  [hep-ph]} \BibitemShut {NoStop}%
\bibitem [{\citenamefont {Kang}\ \emph {et~al.}(2021)\citenamefont {Kang},
  \citenamefont {Zhu},\ and\ \citenamefont {Matsuzaki}}]{Kang:2021epo}%
  \BibitemOpen
  \bibfield  {author} {\bibinfo {author} {\bibfnamefont {Z.}~\bibnamefont
  {Kang}}, \bibinfo {author} {\bibfnamefont {J.}~\bibnamefont {Zhu}},\ and\
  \bibinfo {author} {\bibfnamefont {S.}~\bibnamefont {Matsuzaki}},\ }\bibfield
  {title} {\bibinfo {title} {{Dark confinement-deconfinement phase transition:
  a roadmap from Polyakov loop models to gravitational waves}},\ }\href
  {https://doi.org/10.1007/JHEP09(2021)060} {\bibfield  {journal} {\bibinfo
  {journal} {JHEP}\ }\textbf {\bibinfo {volume} {09}},\ \bibinfo {pages}
  {060}},\ \Eprint {https://arxiv.org/abs/2101.03795} {arXiv:2101.03795
  [hep-ph]} \BibitemShut {NoStop}%
\bibitem [{\citenamefont {Reichert}\ \emph {et~al.}(2022)\citenamefont
  {Reichert}, \citenamefont {Sannino}, \citenamefont {Wang},\ and\
  \citenamefont {Zhang}}]{Reichert:2021cvs}%
  \BibitemOpen
  \bibfield  {author} {\bibinfo {author} {\bibfnamefont {M.}~\bibnamefont
  {Reichert}}, \bibinfo {author} {\bibfnamefont {F.}~\bibnamefont {Sannino}},
  \bibinfo {author} {\bibfnamefont {Z.-W.}\ \bibnamefont {Wang}},\ and\
  \bibinfo {author} {\bibfnamefont {C.}~\bibnamefont {Zhang}},\ }\bibfield
  {title} {\bibinfo {title} {{Dark confinement and chiral phase transitions:
  gravitational waves vs matter representations}},\ }\href
  {https://doi.org/10.1007/JHEP01(2022)003} {\bibfield  {journal} {\bibinfo
  {journal} {JHEP}\ }\textbf {\bibinfo {volume} {01}},\ \bibinfo {pages}
  {003}},\ \Eprint {https://arxiv.org/abs/2109.11552} {arXiv:2109.11552
  [hep-ph]} \BibitemShut {NoStop}%
\bibitem [{\citenamefont {Reichert}\ and\ \citenamefont
  {Wang}(2022)}]{Reichert:2022naa}%
  \BibitemOpen
  \bibfield  {author} {\bibinfo {author} {\bibfnamefont {M.}~\bibnamefont
  {Reichert}}\ and\ \bibinfo {author} {\bibfnamefont {Z.-W.}\ \bibnamefont
  {Wang}},\ }\bibfield  {title} {\bibinfo {title} {{Gravitational Waves from
  dark composite dynamics}},\ }\href
  {https://doi.org/10.1051/epjconf/202227408003} {\bibfield  {journal}
  {\bibinfo  {journal} {EPJ Web Conf.}\ }\textbf {\bibinfo {volume} {274}},\
  \bibinfo {pages} {08003} (\bibinfo {year} {2022})},\ \Eprint
  {https://arxiv.org/abs/2211.08877} {arXiv:2211.08877 [hep-ph]} \BibitemShut
  {NoStop}%
\bibitem [{\citenamefont {Pisarski}(2000)}]{Pisarski:2000eq}%
  \BibitemOpen
  \bibfield  {author} {\bibinfo {author} {\bibfnamefont {R.~D.}\ \bibnamefont
  {Pisarski}},\ }\bibfield  {title} {\bibinfo {title} {{Quark gluon plasma as a
  condensate of SU(3) Wilson lines}},\ }\href
  {https://doi.org/10.1103/PhysRevD.62.111501} {\bibfield  {journal} {\bibinfo
  {journal} {Phys. Rev. D}\ }\textbf {\bibinfo {volume} {62}},\ \bibinfo
  {pages} {111501} (\bibinfo {year} {2000})},\ \Eprint
  {https://arxiv.org/abs/hep-ph/0006205} {arXiv:hep-ph/0006205} \BibitemShut
  {NoStop}%
\bibitem [{\citenamefont {Pisarski}(2002{\natexlab{a}})}]{Pisarski:2001pe}%
  \BibitemOpen
  \bibfield  {author} {\bibinfo {author} {\bibfnamefont {R.~D.}\ \bibnamefont
  {Pisarski}},\ }\bibfield  {title} {\bibinfo {title} {{Tests of the Polyakov
  loops model}},\ }\href {https://doi.org/10.1016/S0375-9474(02)00699-1}
  {\bibfield  {journal} {\bibinfo  {journal} {Nucl. Phys. A}\ }\textbf
  {\bibinfo {volume} {702}},\ \bibinfo {pages} {151} (\bibinfo {year}
  {2002}{\natexlab{a}})},\ \Eprint {https://arxiv.org/abs/hep-ph/0112037}
  {arXiv:hep-ph/0112037} \BibitemShut {NoStop}%
\bibitem [{\citenamefont {Pisarski}(2002{\natexlab{b}})}]{Pisarski:2002ji}%
  \BibitemOpen
  \bibfield  {author} {\bibinfo {author} {\bibfnamefont {R.~D.}\ \bibnamefont
  {Pisarski}},\ }\bibfield  {title} {\bibinfo {title} {{Notes on the
  deconfining phase transition}},\ }in\ \href@noop {} {\emph {\bibinfo
  {booktitle} {{Cargese Summer School on QCD Perspectives on Hot and Dense
  Matter}}}}\ (\bibinfo {year} {2002})\ pp.\ \bibinfo {pages} {353--384},\
  \Eprint {https://arxiv.org/abs/hep-ph/0203271} {arXiv:hep-ph/0203271}
  \BibitemShut {NoStop}%
\bibitem [{\citenamefont {Sannino}(2002)}]{Sannino:2002wb}%
  \BibitemOpen
  \bibfield  {author} {\bibinfo {author} {\bibfnamefont {F.}~\bibnamefont
  {Sannino}},\ }\bibfield  {title} {\bibinfo {title} {{Polyakov loops versus
  hadronic states}},\ }\href {https://doi.org/10.1103/PhysRevD.66.034013}
  {\bibfield  {journal} {\bibinfo  {journal} {Phys. Rev. D}\ }\textbf {\bibinfo
  {volume} {66}},\ \bibinfo {pages} {034013} (\bibinfo {year} {2002})},\
  \Eprint {https://arxiv.org/abs/hep-ph/0204174} {arXiv:hep-ph/0204174}
  \BibitemShut {NoStop}%
\bibitem [{\citenamefont {Ratti}\ \emph {et~al.}(2006)\citenamefont {Ratti},
  \citenamefont {Thaler},\ and\ \citenamefont {Weise}}]{Ratti:2005jh}%
  \BibitemOpen
  \bibfield  {author} {\bibinfo {author} {\bibfnamefont {C.}~\bibnamefont
  {Ratti}}, \bibinfo {author} {\bibfnamefont {M.~A.}\ \bibnamefont {Thaler}},\
  and\ \bibinfo {author} {\bibfnamefont {W.}~\bibnamefont {Weise}},\ }\bibfield
   {title} {\bibinfo {title} {{Phases of QCD: Lattice thermodynamics and a
  field theoretical model}},\ }\href
  {https://doi.org/10.1103/PhysRevD.73.014019} {\bibfield  {journal} {\bibinfo
  {journal} {Phys. Rev. D}\ }\textbf {\bibinfo {volume} {73}},\ \bibinfo
  {pages} {014019} (\bibinfo {year} {2006})},\ \Eprint
  {https://arxiv.org/abs/hep-ph/0506234} {arXiv:hep-ph/0506234} \BibitemShut
  {NoStop}%
\bibitem [{\citenamefont {Fukushima}\ and\ \citenamefont
  {Sasaki}(2013)}]{Fukushima:2013rx}%
  \BibitemOpen
  \bibfield  {author} {\bibinfo {author} {\bibfnamefont {K.}~\bibnamefont
  {Fukushima}}\ and\ \bibinfo {author} {\bibfnamefont {C.}~\bibnamefont
  {Sasaki}},\ }\bibfield  {title} {\bibinfo {title} {{The phase diagram of
  nuclear and quark matter at high baryon density}},\ }\href
  {https://doi.org/10.1016/j.ppnp.2013.05.003} {\bibfield  {journal} {\bibinfo
  {journal} {Prog. Part. Nucl. Phys.}\ }\textbf {\bibinfo {volume} {72}},\
  \bibinfo {pages} {99} (\bibinfo {year} {2013})},\ \Eprint
  {https://arxiv.org/abs/1301.6377} {arXiv:1301.6377 [hep-ph]} \BibitemShut
  {NoStop}%
\bibitem [{\citenamefont {Fukushima}\ and\ \citenamefont
  {Skokov}(2017)}]{Fukushima:2017csk}%
  \BibitemOpen
  \bibfield  {author} {\bibinfo {author} {\bibfnamefont {K.}~\bibnamefont
  {Fukushima}}\ and\ \bibinfo {author} {\bibfnamefont {V.}~\bibnamefont
  {Skokov}},\ }\bibfield  {title} {\bibinfo {title} {{Polyakov loop modeling
  for hot QCD}},\ }\href {https://doi.org/10.1016/j.ppnp.2017.05.002}
  {\bibfield  {journal} {\bibinfo  {journal} {Prog. Part. Nucl. Phys.}\
  }\textbf {\bibinfo {volume} {96}},\ \bibinfo {pages} {154} (\bibinfo {year}
  {2017})},\ \Eprint {https://arxiv.org/abs/1705.00718} {arXiv:1705.00718
  [hep-ph]} \BibitemShut {NoStop}%
\bibitem [{\citenamefont {Lo}\ \emph {et~al.}(2013)\citenamefont {Lo},
  \citenamefont {Friman}, \citenamefont {Kaczmarek}, \citenamefont {Redlich},\
  and\ \citenamefont {Sasaki}}]{Lo:2013hla}%
  \BibitemOpen
  \bibfield  {author} {\bibinfo {author} {\bibfnamefont {P.~M.}\ \bibnamefont
  {Lo}}, \bibinfo {author} {\bibfnamefont {B.}~\bibnamefont {Friman}}, \bibinfo
  {author} {\bibfnamefont {O.}~\bibnamefont {Kaczmarek}}, \bibinfo {author}
  {\bibfnamefont {K.}~\bibnamefont {Redlich}},\ and\ \bibinfo {author}
  {\bibfnamefont {C.}~\bibnamefont {Sasaki}},\ }\bibfield  {title} {\bibinfo
  {title} {{Polyakov loop fluctuations in SU(3) lattice gauge theory and an
  effective gluon potential}},\ }\href
  {https://doi.org/10.1103/PhysRevD.88.074502} {\bibfield  {journal} {\bibinfo
  {journal} {Phys. Rev. D}\ }\textbf {\bibinfo {volume} {88}},\ \bibinfo
  {pages} {074502} (\bibinfo {year} {2013})},\ \Eprint
  {https://arxiv.org/abs/1307.5958} {arXiv:1307.5958 [hep-lat]} \BibitemShut
  {NoStop}%
\bibitem [{\citenamefont {Hansen}\ \emph {et~al.}(2020)\citenamefont {Hansen},
  \citenamefont {Stiele},\ and\ \citenamefont {Costa}}]{Hansen:2019lnf}%
  \BibitemOpen
  \bibfield  {author} {\bibinfo {author} {\bibfnamefont {H.}~\bibnamefont
  {Hansen}}, \bibinfo {author} {\bibfnamefont {R.}~\bibnamefont {Stiele}},\
  and\ \bibinfo {author} {\bibfnamefont {P.}~\bibnamefont {Costa}},\ }\bibfield
   {title} {\bibinfo {title} {{Quark and Polyakov-loop correlations in
  effective models at zero and nonvanishing density}},\ }\href
  {https://doi.org/10.1103/PhysRevD.101.094001} {\bibfield  {journal} {\bibinfo
   {journal} {Phys. Rev. D}\ }\textbf {\bibinfo {volume} {101}},\ \bibinfo
  {pages} {094001} (\bibinfo {year} {2020})},\ \Eprint
  {https://arxiv.org/abs/1904.08965} {arXiv:1904.08965 [hep-ph]} \BibitemShut
  {NoStop}%
\bibitem [{\citenamefont {Meisinger}\ \emph {et~al.}(2002)\citenamefont
  {Meisinger}, \citenamefont {Miller},\ and\ \citenamefont
  {Ogilvie}}]{Meisinger:2001cq}%
  \BibitemOpen
  \bibfield  {author} {\bibinfo {author} {\bibfnamefont {P.~N.}\ \bibnamefont
  {Meisinger}}, \bibinfo {author} {\bibfnamefont {T.~R.}\ \bibnamefont
  {Miller}},\ and\ \bibinfo {author} {\bibfnamefont {M.~C.}\ \bibnamefont
  {Ogilvie}},\ }\bibfield  {title} {\bibinfo {title} {{Phenomenological
  equations of state for the quark gluon plasma}},\ }\href
  {https://doi.org/10.1103/PhysRevD.65.034009} {\bibfield  {journal} {\bibinfo
  {journal} {Phys. Rev. D}\ }\textbf {\bibinfo {volume} {65}},\ \bibinfo
  {pages} {034009} (\bibinfo {year} {2002})},\ \Eprint
  {https://arxiv.org/abs/hep-ph/0108009} {arXiv:hep-ph/0108009} \BibitemShut
  {NoStop}%
\bibitem [{\citenamefont {Dumitru}\ \emph {et~al.}(2011)\citenamefont
  {Dumitru}, \citenamefont {Guo}, \citenamefont {Hidaka}, \citenamefont
  {Altes},\ and\ \citenamefont {Pisarski}}]{Dumitru:2010mj}%
  \BibitemOpen
  \bibfield  {author} {\bibinfo {author} {\bibfnamefont {A.}~\bibnamefont
  {Dumitru}}, \bibinfo {author} {\bibfnamefont {Y.}~\bibnamefont {Guo}},
  \bibinfo {author} {\bibfnamefont {Y.}~\bibnamefont {Hidaka}}, \bibinfo
  {author} {\bibfnamefont {C.~P.~K.}\ \bibnamefont {Altes}},\ and\ \bibinfo
  {author} {\bibfnamefont {R.~D.}\ \bibnamefont {Pisarski}},\ }\bibfield
  {title} {\bibinfo {title} {{How Wide is the Transition to Deconfinement?}},\
  }\href {https://doi.org/10.1103/PhysRevD.83.034022} {\bibfield  {journal}
  {\bibinfo  {journal} {Phys. Rev. D}\ }\textbf {\bibinfo {volume} {83}},\
  \bibinfo {pages} {034022} (\bibinfo {year} {2011})},\ \Eprint
  {https://arxiv.org/abs/1011.3820} {arXiv:1011.3820 [hep-ph]} \BibitemShut
  {NoStop}%
\bibitem [{\citenamefont {Dumitru}\ \emph {et~al.}(2012)\citenamefont
  {Dumitru}, \citenamefont {Guo}, \citenamefont {Hidaka}, \citenamefont
  {Altes},\ and\ \citenamefont {Pisarski}}]{Dumitru:2012fw}%
  \BibitemOpen
  \bibfield  {author} {\bibinfo {author} {\bibfnamefont {A.}~\bibnamefont
  {Dumitru}}, \bibinfo {author} {\bibfnamefont {Y.}~\bibnamefont {Guo}},
  \bibinfo {author} {\bibfnamefont {Y.}~\bibnamefont {Hidaka}}, \bibinfo
  {author} {\bibfnamefont {C.~P.~K.}\ \bibnamefont {Altes}},\ and\ \bibinfo
  {author} {\bibfnamefont {R.~D.}\ \bibnamefont {Pisarski}},\ }\bibfield
  {title} {\bibinfo {title} {{Effective Matrix Model for Deconfinement in Pure
  Gauge Theories}},\ }\href {https://doi.org/10.1103/PhysRevD.86.105017}
  {\bibfield  {journal} {\bibinfo  {journal} {Phys. Rev. D}\ }\textbf {\bibinfo
  {volume} {86}},\ \bibinfo {pages} {105017} (\bibinfo {year} {2012})},\
  \Eprint {https://arxiv.org/abs/1205.0137} {arXiv:1205.0137 [hep-ph]}
  \BibitemShut {NoStop}%
\bibitem [{\citenamefont {Kondo}(2015)}]{Kondo:2015noa}%
  \BibitemOpen
  \bibfield  {author} {\bibinfo {author} {\bibfnamefont {K.-I.}\ \bibnamefont
  {Kondo}},\ }\href@noop {} {\bibinfo {title} {{Confinement--deconfinement
  phase transition and gauge-invariant gluonic mass in Yang-Mills theory}}}
  (\bibinfo {year} {2015}),\ \Eprint {https://arxiv.org/abs/1508.02656}
  {arXiv:1508.02656 [hep-th]} \BibitemShut {NoStop}%
\bibitem [{\citenamefont {Pisarski}\ and\ \citenamefont
  {Skokov}(2016)}]{Pisarski:2016ixt}%
  \BibitemOpen
  \bibfield  {author} {\bibinfo {author} {\bibfnamefont {R.~D.}\ \bibnamefont
  {Pisarski}}\ and\ \bibinfo {author} {\bibfnamefont {V.~V.}\ \bibnamefont
  {Skokov}},\ }\bibfield  {title} {\bibinfo {title} {{Chiral matrix model of
  the semi-QGP in QCD}},\ }\href {https://doi.org/10.1103/PhysRevD.94.034015}
  {\bibfield  {journal} {\bibinfo  {journal} {Phys. Rev. D}\ }\textbf {\bibinfo
  {volume} {94}},\ \bibinfo {pages} {034015} (\bibinfo {year} {2016})},\
  \Eprint {https://arxiv.org/abs/1604.00022} {arXiv:1604.00022 [hep-ph]}
  \BibitemShut {NoStop}%
\bibitem [{\citenamefont {Nishimura}\ \emph {et~al.}(2018)\citenamefont
  {Nishimura}, \citenamefont {Pisarski},\ and\ \citenamefont
  {Skokov}}]{Nishimura:2017crr}%
  \BibitemOpen
  \bibfield  {author} {\bibinfo {author} {\bibfnamefont {H.}~\bibnamefont
  {Nishimura}}, \bibinfo {author} {\bibfnamefont {R.~D.}\ \bibnamefont
  {Pisarski}},\ and\ \bibinfo {author} {\bibfnamefont {V.~V.}\ \bibnamefont
  {Skokov}},\ }\bibfield  {title} {\bibinfo {title} {{Finite-temperature phase
  transitions of third and higher order in gauge theories at large $N$}},\
  }\href {https://doi.org/10.1103/PhysRevD.97.036014} {\bibfield  {journal}
  {\bibinfo  {journal} {Phys. Rev. D}\ }\textbf {\bibinfo {volume} {97}},\
  \bibinfo {pages} {036014} (\bibinfo {year} {2018})},\ \Eprint
  {https://arxiv.org/abs/1712.04465} {arXiv:1712.04465 [hep-th]} \BibitemShut
  {NoStop}%
\bibitem [{\citenamefont {Guo}\ and\ \citenamefont {Du}(2019)}]{Guo:2018scp}%
  \BibitemOpen
  \bibfield  {author} {\bibinfo {author} {\bibfnamefont {Y.}~\bibnamefont
  {Guo}}\ and\ \bibinfo {author} {\bibfnamefont {Q.}~\bibnamefont {Du}},\
  }\bibfield  {title} {\bibinfo {title} {{Two-loop perturbative corrections to
  the constrained effective potential in thermal QCD}},\ }\href
  {https://doi.org/10.1007/JHEP05(2019)042} {\bibfield  {journal} {\bibinfo
  {journal} {JHEP}\ }\textbf {\bibinfo {volume} {05}},\ \bibinfo {pages}
  {042}},\ \Eprint {https://arxiv.org/abs/1810.13090} {arXiv:1810.13090
  [hep-ph]} \BibitemShut {NoStop}%
\bibitem [{\citenamefont {Korthals~Altes}\ \emph {et~al.}(2020)\citenamefont
  {Korthals~Altes}, \citenamefont {Nishimura}, \citenamefont {Pisarski},\ and\
  \citenamefont {Skokov}}]{KorthalsAltes:2020ryu}%
  \BibitemOpen
  \bibfield  {author} {\bibinfo {author} {\bibfnamefont {C.~P.}\ \bibnamefont
  {Korthals~Altes}}, \bibinfo {author} {\bibfnamefont {H.}~\bibnamefont
  {Nishimura}}, \bibinfo {author} {\bibfnamefont {R.~D.}\ \bibnamefont
  {Pisarski}},\ and\ \bibinfo {author} {\bibfnamefont {V.~V.}\ \bibnamefont
  {Skokov}},\ }\bibfield  {title} {\bibinfo {title} {{Free energy of a
  Holonomous Plasma}},\ }\href {https://doi.org/10.1103/PhysRevD.101.094025}
  {\bibfield  {journal} {\bibinfo  {journal} {Phys. Rev. D}\ }\textbf {\bibinfo
  {volume} {101}},\ \bibinfo {pages} {094025} (\bibinfo {year} {2020})},\
  \Eprint {https://arxiv.org/abs/2002.00968} {arXiv:2002.00968 [hep-ph]}
  \BibitemShut {NoStop}%
\bibitem [{\citenamefont {Hidaka}\ and\ \citenamefont
  {Pisarski}(2021)}]{Hidaka:2020vna}%
  \BibitemOpen
  \bibfield  {author} {\bibinfo {author} {\bibfnamefont {Y.}~\bibnamefont
  {Hidaka}}\ and\ \bibinfo {author} {\bibfnamefont {R.~D.}\ \bibnamefont
  {Pisarski}},\ }\bibfield  {title} {\bibinfo {title} {{Effective models of a
  semi-quark-gluon plasma}},\ }\href
  {https://doi.org/10.1103/PhysRevD.104.074036} {\bibfield  {journal} {\bibinfo
   {journal} {Phys. Rev. D}\ }\textbf {\bibinfo {volume} {104}},\ \bibinfo
  {pages} {074036} (\bibinfo {year} {2021})},\ \Eprint
  {https://arxiv.org/abs/2009.03903} {arXiv:2009.03903 [hep-ph]} \BibitemShut
  {NoStop}%
\bibitem [{\citenamefont {Lucini}\ \emph {et~al.}(2002)\citenamefont {Lucini},
  \citenamefont {Teper},\ and\ \citenamefont {Wenger}}]{Lucini:2002ku}%
  \BibitemOpen
  \bibfield  {author} {\bibinfo {author} {\bibfnamefont {B.}~\bibnamefont
  {Lucini}}, \bibinfo {author} {\bibfnamefont {M.}~\bibnamefont {Teper}},\ and\
  \bibinfo {author} {\bibfnamefont {U.}~\bibnamefont {Wenger}},\ }\bibfield
  {title} {\bibinfo {title} {{The Deconfinement transition in SU(N) gauge
  theories}},\ }\href {https://doi.org/10.1016/S0370-2693(02)02556-X}
  {\bibfield  {journal} {\bibinfo  {journal} {Phys. Lett. B}\ }\textbf
  {\bibinfo {volume} {545}},\ \bibinfo {pages} {197} (\bibinfo {year}
  {2002})},\ \Eprint {https://arxiv.org/abs/hep-lat/0206029}
  {arXiv:hep-lat/0206029} \BibitemShut {NoStop}%
\bibitem [{\citenamefont {Lucini}\ \emph {et~al.}(2004)\citenamefont {Lucini},
  \citenamefont {Teper},\ and\ \citenamefont {Wenger}}]{Lucini:2003zr}%
  \BibitemOpen
  \bibfield  {author} {\bibinfo {author} {\bibfnamefont {B.}~\bibnamefont
  {Lucini}}, \bibinfo {author} {\bibfnamefont {M.}~\bibnamefont {Teper}},\ and\
  \bibinfo {author} {\bibfnamefont {U.}~\bibnamefont {Wenger}},\ }\bibfield
  {title} {\bibinfo {title} {{The High temperature phase transition in SU(N)
  gauge theories}},\ }\href {https://doi.org/10.1088/1126-6708/2004/01/061}
  {\bibfield  {journal} {\bibinfo  {journal} {JHEP}\ }\textbf {\bibinfo
  {volume} {01}},\ \bibinfo {pages} {061}},\ \Eprint
  {https://arxiv.org/abs/hep-lat/0307017} {arXiv:hep-lat/0307017} \BibitemShut
  {NoStop}%
\bibitem [{\citenamefont {Lucini}\ \emph {et~al.}(2005)\citenamefont {Lucini},
  \citenamefont {Teper},\ and\ \citenamefont {Wenger}}]{Lucini:2005vg}%
  \BibitemOpen
  \bibfield  {author} {\bibinfo {author} {\bibfnamefont {B.}~\bibnamefont
  {Lucini}}, \bibinfo {author} {\bibfnamefont {M.}~\bibnamefont {Teper}},\ and\
  \bibinfo {author} {\bibfnamefont {U.}~\bibnamefont {Wenger}},\ }\bibfield
  {title} {\bibinfo {title} {{Properties of the deconfining phase transition in
  SU(N) gauge theories}},\ }\href
  {https://doi.org/10.1088/1126-6708/2005/02/033} {\bibfield  {journal}
  {\bibinfo  {journal} {JHEP}\ }\textbf {\bibinfo {volume} {02}},\ \bibinfo
  {pages} {033}},\ \Eprint {https://arxiv.org/abs/hep-lat/0502003}
  {arXiv:hep-lat/0502003} \BibitemShut {NoStop}%
\bibitem [{\citenamefont {Panero}(2009)}]{Panero:2009tv}%
  \BibitemOpen
  \bibfield  {author} {\bibinfo {author} {\bibfnamefont {M.}~\bibnamefont
  {Panero}},\ }\bibfield  {title} {\bibinfo {title} {{Thermodynamics of the QCD
  plasma and the large-N limit}},\ }\href
  {https://doi.org/10.1103/PhysRevLett.103.232001} {\bibfield  {journal}
  {\bibinfo  {journal} {Phys. Rev. Lett.}\ }\textbf {\bibinfo {volume} {103}},\
  \bibinfo {pages} {232001} (\bibinfo {year} {2009})},\ \Eprint
  {https://arxiv.org/abs/0907.3719} {arXiv:0907.3719 [hep-lat]} \BibitemShut
  {NoStop}%
\bibitem [{\citenamefont {Datta}\ and\ \citenamefont
  {Gupta}(2010)}]{Datta:2010sq}%
  \BibitemOpen
  \bibfield  {author} {\bibinfo {author} {\bibfnamefont {S.}~\bibnamefont
  {Datta}}\ and\ \bibinfo {author} {\bibfnamefont {S.}~\bibnamefont {Gupta}},\
  }\bibfield  {title} {\bibinfo {title} {{Continuum Thermodynamics of the
  Gluo$N_c$ Plasma}},\ }\href {https://doi.org/10.1103/PhysRevD.82.114505}
  {\bibfield  {journal} {\bibinfo  {journal} {Phys. Rev. D}\ }\textbf {\bibinfo
  {volume} {82}},\ \bibinfo {pages} {114505} (\bibinfo {year} {2010})},\
  \Eprint {https://arxiv.org/abs/1006.0938} {arXiv:1006.0938 [hep-lat]}
  \BibitemShut {NoStop}%
\bibitem [{\citenamefont {Lucini}\ \emph {et~al.}(2012)\citenamefont {Lucini},
  \citenamefont {Rago},\ and\ \citenamefont {Rinaldi}}]{Lucini:2012wq}%
  \BibitemOpen
  \bibfield  {author} {\bibinfo {author} {\bibfnamefont {B.}~\bibnamefont
  {Lucini}}, \bibinfo {author} {\bibfnamefont {A.}~\bibnamefont {Rago}},\ and\
  \bibinfo {author} {\bibfnamefont {E.}~\bibnamefont {Rinaldi}},\ }\bibfield
  {title} {\bibinfo {title} {{SU($N_c$) gauge theories at deconfinement}},\
  }\href {https://doi.org/10.1016/j.physletb.2012.04.070} {\bibfield  {journal}
  {\bibinfo  {journal} {Phys. Lett. B}\ }\textbf {\bibinfo {volume} {712}},\
  \bibinfo {pages} {279} (\bibinfo {year} {2012})},\ \Eprint
  {https://arxiv.org/abs/1202.6684} {arXiv:1202.6684 [hep-lat]} \BibitemShut
  {NoStop}%
\bibitem [{\citenamefont {Holland}\ \emph {et~al.}(2004)\citenamefont
  {Holland}, \citenamefont {Pepe},\ and\ \citenamefont
  {Wiese}}]{Holland:2003kg}%
  \BibitemOpen
  \bibfield  {author} {\bibinfo {author} {\bibfnamefont {K.}~\bibnamefont
  {Holland}}, \bibinfo {author} {\bibfnamefont {M.}~\bibnamefont {Pepe}},\ and\
  \bibinfo {author} {\bibfnamefont {U.~J.}\ \bibnamefont {Wiese}},\ }\bibfield
  {title} {\bibinfo {title} {{The Deconfinement phase transition of Sp(2) and
  Sp(3) Yang-Mills theories in (2+1)-dimensions and (3+1)-dimensions}},\ }\href
  {https://doi.org/10.1016/j.nuclphysb.2004.06.026} {\bibfield  {journal}
  {\bibinfo  {journal} {Nucl. Phys. B}\ }\textbf {\bibinfo {volume} {694}},\
  \bibinfo {pages} {35} (\bibinfo {year} {2004})},\ \Eprint
  {https://arxiv.org/abs/hep-lat/0312022} {arXiv:hep-lat/0312022} \BibitemShut
  {NoStop}%
\bibitem [{\citenamefont {Pepe}(2006)}]{Pepe:2005sz}%
  \BibitemOpen
  \bibfield  {author} {\bibinfo {author} {\bibfnamefont {M.}~\bibnamefont
  {Pepe}},\ }\bibfield  {title} {\bibinfo {title} {{Confinement and the center
  of the gauge group}},\ }\href
  {https://doi.org/10.1016/j.nuclphysbps.2006.01.045} {\bibfield  {journal}
  {\bibinfo  {journal} {PoS}\ }\textbf {\bibinfo {volume} {LAT2005}},\ \bibinfo
  {pages} {017} (\bibinfo {year} {2006})},\ \Eprint
  {https://arxiv.org/abs/hep-lat/0510013} {arXiv:hep-lat/0510013} \BibitemShut
  {NoStop}%
\bibitem [{\citenamefont {Pepe}\ and\ \citenamefont
  {Wiese}(2007)}]{Pepe:2006er}%
  \BibitemOpen
  \bibfield  {author} {\bibinfo {author} {\bibfnamefont {M.}~\bibnamefont
  {Pepe}}\ and\ \bibinfo {author} {\bibfnamefont {U.~J.}\ \bibnamefont
  {Wiese}},\ }\bibfield  {title} {\bibinfo {title} {{Exceptional Deconfinement
  in G(2) Gauge Theory}},\ }\href
  {https://doi.org/10.1016/j.nuclphysb.2006.12.024} {\bibfield  {journal}
  {\bibinfo  {journal} {Nucl. Phys. B}\ }\textbf {\bibinfo {volume} {768}},\
  \bibinfo {pages} {21} (\bibinfo {year} {2007})},\ \Eprint
  {https://arxiv.org/abs/hep-lat/0610076} {arXiv:hep-lat/0610076} \BibitemShut
  {NoStop}%
\bibitem [{\citenamefont {Cossu}\ \emph {et~al.}(2007)\citenamefont {Cossu},
  \citenamefont {D'Elia}, \citenamefont {Di~Giacomo}, \citenamefont {Lucini},\
  and\ \citenamefont {Pica}}]{Cossu:2007dk}%
  \BibitemOpen
  \bibfield  {author} {\bibinfo {author} {\bibfnamefont {G.}~\bibnamefont
  {Cossu}}, \bibinfo {author} {\bibfnamefont {M.}~\bibnamefont {D'Elia}},
  \bibinfo {author} {\bibfnamefont {A.}~\bibnamefont {Di~Giacomo}}, \bibinfo
  {author} {\bibfnamefont {B.}~\bibnamefont {Lucini}},\ and\ \bibinfo {author}
  {\bibfnamefont {C.}~\bibnamefont {Pica}},\ }\bibfield  {title} {\bibinfo
  {title} {{G(2) gauge theory at finite temperature}},\ }\href
  {https://doi.org/10.1088/1126-6708/2007/10/100} {\bibfield  {journal}
  {\bibinfo  {journal} {JHEP}\ }\textbf {\bibinfo {volume} {10}},\ \bibinfo
  {pages} {100}},\ \Eprint {https://arxiv.org/abs/0709.0669} {arXiv:0709.0669
  [hep-lat]} \BibitemShut {NoStop}%
\bibitem [{\citenamefont {Bruno}\ \emph {et~al.}(2015)\citenamefont {Bruno},
  \citenamefont {Caselle}, \citenamefont {Panero},\ and\ \citenamefont
  {Pellegrini}}]{Bruno:2014rxa}%
  \BibitemOpen
  \bibfield  {author} {\bibinfo {author} {\bibfnamefont {M.}~\bibnamefont
  {Bruno}}, \bibinfo {author} {\bibfnamefont {M.}~\bibnamefont {Caselle}},
  \bibinfo {author} {\bibfnamefont {M.}~\bibnamefont {Panero}},\ and\ \bibinfo
  {author} {\bibfnamefont {R.}~\bibnamefont {Pellegrini}},\ }\bibfield  {title}
  {\bibinfo {title} {{Exceptional thermodynamics: the equation of state of
  G$_{2}$ gauge theory}},\ }\href {https://doi.org/10.1007/JHEP03(2015)057}
  {\bibfield  {journal} {\bibinfo  {journal} {JHEP}\ }\textbf {\bibinfo
  {volume} {03}},\ \bibinfo {pages} {057}},\ \Eprint
  {https://arxiv.org/abs/1409.8305} {arXiv:1409.8305 [hep-lat]} \BibitemShut
  {NoStop}%
\bibitem [{\citenamefont {Appelquist}\ \emph
  {et~al.}(2015{\natexlab{a}})\citenamefont {Appelquist} \emph
  {et~al.}}]{Appelquist:2015yfa}%
  \BibitemOpen
  \bibfield  {author} {\bibinfo {author} {\bibfnamefont {T.}~\bibnamefont
  {Appelquist}} \emph {et~al.},\ }\bibfield  {title} {\bibinfo {title}
  {{Stealth Dark Matter: Dark scalar baryons through the Higgs portal}},\
  }\href {https://doi.org/10.1103/PhysRevD.92.075030} {\bibfield  {journal}
  {\bibinfo  {journal} {Phys. Rev. D}\ }\textbf {\bibinfo {volume} {92}},\
  \bibinfo {pages} {075030} (\bibinfo {year} {2015}{\natexlab{a}})},\ \Eprint
  {https://arxiv.org/abs/1503.04203} {arXiv:1503.04203 [hep-ph]} \BibitemShut
  {NoStop}%
\bibitem [{\citenamefont {Appelquist}\ \emph
  {et~al.}(2015{\natexlab{b}})\citenamefont {Appelquist} \emph
  {et~al.}}]{Appelquist:2015zfa}%
  \BibitemOpen
  \bibfield  {author} {\bibinfo {author} {\bibfnamefont {T.}~\bibnamefont
  {Appelquist}} \emph {et~al.},\ }\bibfield  {title} {\bibinfo {title}
  {{Detecting Stealth Dark Matter Directly through Electromagnetic
  Polarizability}},\ }\href {https://doi.org/10.1103/PhysRevLett.115.171803}
  {\bibfield  {journal} {\bibinfo  {journal} {Phys. Rev. Lett.}\ }\textbf
  {\bibinfo {volume} {115}},\ \bibinfo {pages} {171803} (\bibinfo {year}
  {2015}{\natexlab{b}})},\ \Eprint {https://arxiv.org/abs/1503.04205}
  {arXiv:1503.04205 [hep-ph]} \BibitemShut {NoStop}%
\bibitem [{\citenamefont {Brower}\ \emph {et~al.}(2021)\citenamefont {Brower}
  \emph {et~al.}}]{LatticeStrongDynamics:2020jwi}%
  \BibitemOpen
  \bibfield  {author} {\bibinfo {author} {\bibfnamefont {R.~C.}\ \bibnamefont
  {Brower}} \emph {et~al.} (\bibinfo {collaboration} {Lattice Strong
  Dynamics}),\ }\bibfield  {title} {\bibinfo {title} {{Stealth dark matter
  confinement transition and gravitational waves}},\ }\href
  {https://doi.org/10.1103/PhysRevD.103.014505} {\bibfield  {journal} {\bibinfo
   {journal} {Phys. Rev. D}\ }\textbf {\bibinfo {volume} {103}},\ \bibinfo
  {pages} {014505} (\bibinfo {year} {2021})},\ \Eprint
  {https://arxiv.org/abs/2006.16429} {arXiv:2006.16429 [hep-lat]} \BibitemShut
  {NoStop}%
\bibitem [{\citenamefont {Maas}\ and\ \citenamefont
  {Zierler}(2022)}]{Maas:2021gbf}%
  \BibitemOpen
  \bibfield  {author} {\bibinfo {author} {\bibfnamefont {A.}~\bibnamefont
  {Maas}}\ and\ \bibinfo {author} {\bibfnamefont {F.}~\bibnamefont {Zierler}},\
  }\bibfield  {title} {\bibinfo {title} {{Strong isospin breaking in
  \ensuremath{\boldsymbol{\mathit{S}}}\ensuremath{\boldsymbol{\mathit{p}}}(4)
  gauge theory}},\ }\href {https://doi.org/10.22323/1.396.0130} {\bibfield
  {journal} {\bibinfo  {journal} {PoS}\ }\textbf {\bibinfo {volume}
  {LATTICE2021}},\ \bibinfo {pages} {130} (\bibinfo {year} {2022})},\ \Eprint
  {https://arxiv.org/abs/2109.14377} {arXiv:2109.14377 [hep-lat]} \BibitemShut
  {NoStop}%
\bibitem [{\citenamefont {Zierler}\ and\ \citenamefont
  {Maas}(2021)}]{Zierler:2021cfa}%
  \BibitemOpen
  \bibfield  {author} {\bibinfo {author} {\bibfnamefont {F.}~\bibnamefont
  {Zierler}}\ and\ \bibinfo {author} {\bibfnamefont {A.}~\bibnamefont {Maas}},\
  }\bibfield  {title} {\bibinfo {title} {{$Sp(4)$ SIMP Dark Matter on the
  Lattice}},\ }\href {https://doi.org/10.22323/1.397.0162} {\bibfield
  {journal} {\bibinfo  {journal} {PoS}\ }\textbf {\bibinfo {volume}
  {LHCP2021}},\ \bibinfo {pages} {162} (\bibinfo {year} {2021})}\BibitemShut
  {NoStop}%
\bibitem [{\citenamefont {Kulkarni}\ \emph {et~al.}(2023)\citenamefont
  {Kulkarni}, \citenamefont {Maas}, \citenamefont {Mee}, \citenamefont
  {Nikolic}, \citenamefont {Pradler},\ and\ \citenamefont
  {Zierler}}]{Kulkarni:2022bvh}%
  \BibitemOpen
  \bibfield  {author} {\bibinfo {author} {\bibfnamefont {S.}~\bibnamefont
  {Kulkarni}}, \bibinfo {author} {\bibfnamefont {A.}~\bibnamefont {Maas}},
  \bibinfo {author} {\bibfnamefont {S.}~\bibnamefont {Mee}}, \bibinfo {author}
  {\bibfnamefont {M.}~\bibnamefont {Nikolic}}, \bibinfo {author} {\bibfnamefont
  {J.}~\bibnamefont {Pradler}},\ and\ \bibinfo {author} {\bibfnamefont
  {F.}~\bibnamefont {Zierler}},\ }\bibfield  {title} {\bibinfo {title}
  {{Low-energy effective description of dark $Sp(4)$ theories}},\ }\href
  {https://doi.org/10.21468/SciPostPhys.14.3.044} {\bibfield  {journal}
  {\bibinfo  {journal} {SciPost Phys.}\ }\textbf {\bibinfo {volume} {14}},\
  \bibinfo {pages} {044} (\bibinfo {year} {2023})},\ \Eprint
  {https://arxiv.org/abs/2202.05191} {arXiv:2202.05191 [hep-ph]} \BibitemShut
  {NoStop}%
\bibitem [{\citenamefont {Maldacena}(1998)}]{Maldacena:1997re}%
  \BibitemOpen
  \bibfield  {author} {\bibinfo {author} {\bibfnamefont {J.~M.}\ \bibnamefont
  {Maldacena}},\ }\bibfield  {title} {\bibinfo {title} {{The Large N limit of
  superconformal field theories and supergravity}},\ }\href
  {https://doi.org/10.1023/A:1026654312961} {\bibfield  {journal} {\bibinfo
  {journal} {Adv. Theor. Math. Phys.}\ }\textbf {\bibinfo {volume} {2}},\
  \bibinfo {pages} {231} (\bibinfo {year} {1998})},\ \Eprint
  {https://arxiv.org/abs/hep-th/9711200} {arXiv:hep-th/9711200} \BibitemShut
  {NoStop}%
\bibitem [{\citenamefont {Gubser}\ \emph {et~al.}(1998)\citenamefont {Gubser},
  \citenamefont {Klebanov},\ and\ \citenamefont {Polyakov}}]{Gubser:1998bc}%
  \BibitemOpen
  \bibfield  {author} {\bibinfo {author} {\bibfnamefont {S.~S.}\ \bibnamefont
  {Gubser}}, \bibinfo {author} {\bibfnamefont {I.~R.}\ \bibnamefont
  {Klebanov}},\ and\ \bibinfo {author} {\bibfnamefont {A.~M.}\ \bibnamefont
  {Polyakov}},\ }\bibfield  {title} {\bibinfo {title} {{Gauge theory
  correlators from noncritical string theory}},\ }\href
  {https://doi.org/10.1016/S0370-2693(98)00377-3} {\bibfield  {journal}
  {\bibinfo  {journal} {Phys. Lett. B}\ }\textbf {\bibinfo {volume} {428}},\
  \bibinfo {pages} {105} (\bibinfo {year} {1998})},\ \Eprint
  {https://arxiv.org/abs/hep-th/9802109} {arXiv:hep-th/9802109} \BibitemShut
  {NoStop}%
\bibitem [{\citenamefont {Witten}(1998{\natexlab{a}})}]{Witten:1998qj}%
  \BibitemOpen
  \bibfield  {author} {\bibinfo {author} {\bibfnamefont {E.}~\bibnamefont
  {Witten}},\ }\bibfield  {title} {\bibinfo {title} {{Anti-de Sitter space and
  holography}},\ }\href {https://doi.org/10.4310/ATMP.1998.v2.n2.a2} {\bibfield
   {journal} {\bibinfo  {journal} {Adv. Theor. Math. Phys.}\ }\textbf {\bibinfo
  {volume} {2}},\ \bibinfo {pages} {253} (\bibinfo {year}
  {1998}{\natexlab{a}})},\ \Eprint {https://arxiv.org/abs/hep-th/9802150}
  {arXiv:hep-th/9802150} \BibitemShut {NoStop}%
\bibitem [{\citenamefont {Aharony}\ \emph {et~al.}(2000)\citenamefont
  {Aharony}, \citenamefont {Gubser}, \citenamefont {Maldacena}, \citenamefont
  {Ooguri},\ and\ \citenamefont {Oz}}]{Aharony:1999ti}%
  \BibitemOpen
  \bibfield  {author} {\bibinfo {author} {\bibfnamefont {O.}~\bibnamefont
  {Aharony}}, \bibinfo {author} {\bibfnamefont {S.~S.}\ \bibnamefont {Gubser}},
  \bibinfo {author} {\bibfnamefont {J.~M.}\ \bibnamefont {Maldacena}}, \bibinfo
  {author} {\bibfnamefont {H.}~\bibnamefont {Ooguri}},\ and\ \bibinfo {author}
  {\bibfnamefont {Y.}~\bibnamefont {Oz}},\ }\bibfield  {title} {\bibinfo
  {title} {{Large N field theories, string theory and gravity}},\ }\href
  {https://doi.org/10.1016/S0370-1573(99)00083-6} {\bibfield  {journal}
  {\bibinfo  {journal} {Phys. Rept.}\ }\textbf {\bibinfo {volume} {323}},\
  \bibinfo {pages} {183} (\bibinfo {year} {2000})},\ \Eprint
  {https://arxiv.org/abs/hep-th/9905111} {arXiv:hep-th/9905111} \BibitemShut
  {NoStop}%
\bibitem [{\citenamefont {Witten}(1998{\natexlab{b}})}]{Witten:1998zw}%
  \BibitemOpen
  \bibfield  {author} {\bibinfo {author} {\bibfnamefont {E.}~\bibnamefont
  {Witten}},\ }\bibfield  {title} {\bibinfo {title} {{Anti-de Sitter space,
  thermal phase transition, and confinement in gauge theories}},\ }\href
  {https://doi.org/10.4310/ATMP.1998.v2.n3.a3} {\bibfield  {journal} {\bibinfo
  {journal} {Adv. Theor. Math. Phys.}\ }\textbf {\bibinfo {volume} {2}},\
  \bibinfo {pages} {505} (\bibinfo {year} {1998}{\natexlab{b}})},\ \Eprint
  {https://arxiv.org/abs/hep-th/9803131} {arXiv:hep-th/9803131} \BibitemShut
  {NoStop}%
\bibitem [{\citenamefont {Klebanov}\ and\ \citenamefont
  {Strassler}(2000)}]{Klebanov:2000hb}%
  \BibitemOpen
  \bibfield  {author} {\bibinfo {author} {\bibfnamefont {I.~R.}\ \bibnamefont
  {Klebanov}}\ and\ \bibinfo {author} {\bibfnamefont {M.~J.}\ \bibnamefont
  {Strassler}},\ }\bibfield  {title} {\bibinfo {title} {{Supergravity and a
  confining gauge theory: Duality cascades and chi SB resolution of naked
  singularities}},\ }\href {https://doi.org/10.1088/1126-6708/2000/08/052}
  {\bibfield  {journal} {\bibinfo  {journal} {JHEP}\ }\textbf {\bibinfo
  {volume} {08}},\ \bibinfo {pages} {052}},\ \Eprint
  {https://arxiv.org/abs/hep-th/0007191} {arXiv:hep-th/0007191} \BibitemShut
  {NoStop}%
\bibitem [{\citenamefont {Maldacena}\ and\ \citenamefont
  {Nunez}(2001)}]{Maldacena:2000yy}%
  \BibitemOpen
  \bibfield  {author} {\bibinfo {author} {\bibfnamefont {J.~M.}\ \bibnamefont
  {Maldacena}}\ and\ \bibinfo {author} {\bibfnamefont {C.}~\bibnamefont
  {Nunez}},\ }\bibfield  {title} {\bibinfo {title} {{Towards the large N limit
  of pure N=1 superYang-Mills}},\ }\href
  {https://doi.org/10.1103/PhysRevLett.86.588} {\bibfield  {journal} {\bibinfo
  {journal} {Phys. Rev. Lett.}\ }\textbf {\bibinfo {volume} {86}},\ \bibinfo
  {pages} {588} (\bibinfo {year} {2001})},\ \Eprint
  {https://arxiv.org/abs/hep-th/0008001} {arXiv:hep-th/0008001} \BibitemShut
  {NoStop}%
\bibitem [{\citenamefont {Chamseddine}\ and\ \citenamefont
  {Volkov}(1997)}]{Chamseddine:1997nm}%
  \BibitemOpen
  \bibfield  {author} {\bibinfo {author} {\bibfnamefont {A.~H.}\ \bibnamefont
  {Chamseddine}}\ and\ \bibinfo {author} {\bibfnamefont {M.~S.}\ \bibnamefont
  {Volkov}},\ }\bibfield  {title} {\bibinfo {title} {{NonAbelian BPS monopoles
  in N=4 gauged supergravity}},\ }\href
  {https://doi.org/10.1103/PhysRevLett.79.3343} {\bibfield  {journal} {\bibinfo
   {journal} {Phys. Rev. Lett.}\ }\textbf {\bibinfo {volume} {79}},\ \bibinfo
  {pages} {3343} (\bibinfo {year} {1997})},\ \Eprint
  {https://arxiv.org/abs/hep-th/9707176} {arXiv:hep-th/9707176} \BibitemShut
  {NoStop}%
\bibitem [{\citenamefont {Butti}\ \emph {et~al.}(2005)\citenamefont {Butti},
  \citenamefont {Grana}, \citenamefont {Minasian}, \citenamefont {Petrini},\
  and\ \citenamefont {Zaffaroni}}]{Butti:2004pk}%
  \BibitemOpen
  \bibfield  {author} {\bibinfo {author} {\bibfnamefont {A.}~\bibnamefont
  {Butti}}, \bibinfo {author} {\bibfnamefont {M.}~\bibnamefont {Grana}},
  \bibinfo {author} {\bibfnamefont {R.}~\bibnamefont {Minasian}}, \bibinfo
  {author} {\bibfnamefont {M.}~\bibnamefont {Petrini}},\ and\ \bibinfo {author}
  {\bibfnamefont {A.}~\bibnamefont {Zaffaroni}},\ }\bibfield  {title} {\bibinfo
  {title} {{The Baryonic branch of Klebanov-Strassler solution: A
  supersymmetric family of SU(3) structure backgrounds}},\ }\href
  {https://doi.org/10.1088/1126-6708/2005/03/069} {\bibfield  {journal}
  {\bibinfo  {journal} {JHEP}\ }\textbf {\bibinfo {volume} {03}},\ \bibinfo
  {pages} {069}},\ \Eprint {https://arxiv.org/abs/hep-th/0412187}
  {arXiv:hep-th/0412187} \BibitemShut {NoStop}%
\bibitem [{\citenamefont {Brower}\ \emph {et~al.}(2000)\citenamefont {Brower},
  \citenamefont {Mathur},\ and\ \citenamefont {Tan}}]{Brower:2000rp}%
  \BibitemOpen
  \bibfield  {author} {\bibinfo {author} {\bibfnamefont {R.~C.}\ \bibnamefont
  {Brower}}, \bibinfo {author} {\bibfnamefont {S.~D.}\ \bibnamefont {Mathur}},\
  and\ \bibinfo {author} {\bibfnamefont {C.-I.}\ \bibnamefont {Tan}},\
  }\bibfield  {title} {\bibinfo {title} {{Glueball spectrum for QCD from AdS
  supergravity duality}},\ }\href
  {https://doi.org/10.1016/S0550-3213(00)00435-1} {\bibfield  {journal}
  {\bibinfo  {journal} {Nucl. Phys. B}\ }\textbf {\bibinfo {volume} {587}},\
  \bibinfo {pages} {249} (\bibinfo {year} {2000})},\ \Eprint
  {https://arxiv.org/abs/hep-th/0003115} {arXiv:hep-th/0003115} \BibitemShut
  {NoStop}%
\bibitem [{\citenamefont {Karch}\ and\ \citenamefont
  {Katz}(2002)}]{Karch:2002sh}%
  \BibitemOpen
  \bibfield  {author} {\bibinfo {author} {\bibfnamefont {A.}~\bibnamefont
  {Karch}}\ and\ \bibinfo {author} {\bibfnamefont {E.}~\bibnamefont {Katz}},\
  }\bibfield  {title} {\bibinfo {title} {{Adding flavor to AdS / CFT}},\ }\href
  {https://doi.org/10.1088/1126-6708/2002/06/043} {\bibfield  {journal}
  {\bibinfo  {journal} {JHEP}\ }\textbf {\bibinfo {volume} {06}},\ \bibinfo
  {pages} {043}},\ \Eprint {https://arxiv.org/abs/hep-th/0205236}
  {arXiv:hep-th/0205236} \BibitemShut {NoStop}%
\bibitem [{\citenamefont {Kruczenski}\ \emph {et~al.}(2003)\citenamefont
  {Kruczenski}, \citenamefont {Mateos}, \citenamefont {Myers},\ and\
  \citenamefont {Winters}}]{Kruczenski:2003be}%
  \BibitemOpen
  \bibfield  {author} {\bibinfo {author} {\bibfnamefont {M.}~\bibnamefont
  {Kruczenski}}, \bibinfo {author} {\bibfnamefont {D.}~\bibnamefont {Mateos}},
  \bibinfo {author} {\bibfnamefont {R.~C.}\ \bibnamefont {Myers}},\ and\
  \bibinfo {author} {\bibfnamefont {D.~J.}\ \bibnamefont {Winters}},\
  }\bibfield  {title} {\bibinfo {title} {{Meson spectroscopy in AdS / CFT with
  flavor}},\ }\href {https://doi.org/10.1088/1126-6708/2003/07/049} {\bibfield
  {journal} {\bibinfo  {journal} {JHEP}\ }\textbf {\bibinfo {volume} {07}},\
  \bibinfo {pages} {049}},\ \Eprint {https://arxiv.org/abs/hep-th/0304032}
  {arXiv:hep-th/0304032} \BibitemShut {NoStop}%
\bibitem [{\citenamefont {Sakai}\ and\ \citenamefont
  {Sugimoto}(2005{\natexlab{a}})}]{Sakai:2004cn}%
  \BibitemOpen
  \bibfield  {author} {\bibinfo {author} {\bibfnamefont {T.}~\bibnamefont
  {Sakai}}\ and\ \bibinfo {author} {\bibfnamefont {S.}~\bibnamefont
  {Sugimoto}},\ }\bibfield  {title} {\bibinfo {title} {{Low energy hadron
  physics in holographic QCD}},\ }\href {https://doi.org/10.1143/PTP.113.843}
  {\bibfield  {journal} {\bibinfo  {journal} {Prog. Theor. Phys.}\ }\textbf
  {\bibinfo {volume} {113}},\ \bibinfo {pages} {843} (\bibinfo {year}
  {2005}{\natexlab{a}})},\ \Eprint {https://arxiv.org/abs/hep-th/0412141}
  {arXiv:hep-th/0412141} \BibitemShut {NoStop}%
\bibitem [{\citenamefont {Sakai}\ and\ \citenamefont
  {Sugimoto}(2005{\natexlab{b}})}]{Sakai:2005yt}%
  \BibitemOpen
  \bibfield  {author} {\bibinfo {author} {\bibfnamefont {T.}~\bibnamefont
  {Sakai}}\ and\ \bibinfo {author} {\bibfnamefont {S.}~\bibnamefont
  {Sugimoto}},\ }\bibfield  {title} {\bibinfo {title} {{More on a holographic
  dual of QCD}},\ }\href {https://doi.org/10.1143/PTP.114.1083} {\bibfield
  {journal} {\bibinfo  {journal} {Prog. Theor. Phys.}\ }\textbf {\bibinfo
  {volume} {114}},\ \bibinfo {pages} {1083} (\bibinfo {year}
  {2005}{\natexlab{b}})},\ \Eprint {https://arxiv.org/abs/hep-th/0507073}
  {arXiv:hep-th/0507073} \BibitemShut {NoStop}%
\bibitem [{\citenamefont {Bigazzi}\ \emph {et~al.}(2020)\citenamefont
  {Bigazzi}, \citenamefont {Caddeo}, \citenamefont {Cotrone},\ and\
  \citenamefont {Paredes}}]{Bigazzi:2020phm}%
  \BibitemOpen
  \bibfield  {author} {\bibinfo {author} {\bibfnamefont {F.}~\bibnamefont
  {Bigazzi}}, \bibinfo {author} {\bibfnamefont {A.}~\bibnamefont {Caddeo}},
  \bibinfo {author} {\bibfnamefont {A.~L.}\ \bibnamefont {Cotrone}},\ and\
  \bibinfo {author} {\bibfnamefont {A.}~\bibnamefont {Paredes}},\ }\bibfield
  {title} {\bibinfo {title} {{Fate of false vacua in holographic first-order
  phase transitions}},\ }\href {https://doi.org/10.1007/JHEP12(2020)200}
  {\bibfield  {journal} {\bibinfo  {journal} {JHEP}\ }\textbf {\bibinfo
  {volume} {12}},\ \bibinfo {pages} {200}},\ \Eprint
  {https://arxiv.org/abs/2008.02579} {arXiv:2008.02579 [hep-th]} \BibitemShut
  {NoStop}%
\bibitem [{\citenamefont {Ares}\ \emph {et~al.}(2020)\citenamefont {Ares},
  \citenamefont {Hindmarsh}, \citenamefont {Hoyos},\ and\ \citenamefont
  {Jokela}}]{Ares:2020lbt}%
  \BibitemOpen
  \bibfield  {author} {\bibinfo {author} {\bibfnamefont {F.~R.}\ \bibnamefont
  {Ares}}, \bibinfo {author} {\bibfnamefont {M.}~\bibnamefont {Hindmarsh}},
  \bibinfo {author} {\bibfnamefont {C.}~\bibnamefont {Hoyos}},\ and\ \bibinfo
  {author} {\bibfnamefont {N.}~\bibnamefont {Jokela}},\ }\bibfield  {title}
  {\bibinfo {title} {{Gravitational waves from a holographic phase
  transition}},\ }\href {https://doi.org/10.1007/JHEP04(2021)100} {\bibfield
  {journal} {\bibinfo  {journal} {JHEP}\ }\textbf {\bibinfo {volume} {21}},\
  \bibinfo {pages} {100}},\ \Eprint {https://arxiv.org/abs/2011.12878}
  {arXiv:2011.12878 [hep-th]} \BibitemShut {NoStop}%
\bibitem [{\citenamefont {Bea}\ \emph {et~al.}(2021)\citenamefont {Bea},
  \citenamefont {Casalderrey-Solana}, \citenamefont {Giannakopoulos},
  \citenamefont {Mateos}, \citenamefont {Sanchez-Garitaonandia},\ and\
  \citenamefont {Zilh\~ao}}]{Bea:2021zsu}%
  \BibitemOpen
  \bibfield  {author} {\bibinfo {author} {\bibfnamefont {Y.}~\bibnamefont
  {Bea}}, \bibinfo {author} {\bibfnamefont {J.}~\bibnamefont
  {Casalderrey-Solana}}, \bibinfo {author} {\bibfnamefont {T.}~\bibnamefont
  {Giannakopoulos}}, \bibinfo {author} {\bibfnamefont {D.}~\bibnamefont
  {Mateos}}, \bibinfo {author} {\bibfnamefont {M.}~\bibnamefont
  {Sanchez-Garitaonandia}},\ and\ \bibinfo {author} {\bibfnamefont
  {M.}~\bibnamefont {Zilh\~ao}},\ }\bibfield  {title} {\bibinfo {title}
  {{Bubble wall velocity from holography}},\ }\href
  {https://doi.org/10.1103/PhysRevD.104.L121903} {\bibfield  {journal}
  {\bibinfo  {journal} {Phys. Rev. D}\ }\textbf {\bibinfo {volume} {104}},\
  \bibinfo {pages} {L121903} (\bibinfo {year} {2021})},\ \Eprint
  {https://arxiv.org/abs/2104.05708} {arXiv:2104.05708 [hep-th]} \BibitemShut
  {NoStop}%
\bibitem [{\citenamefont {Bigazzi}\ \emph {et~al.}(2021)\citenamefont
  {Bigazzi}, \citenamefont {Caddeo}, \citenamefont {Canneti},\ and\
  \citenamefont {Cotrone}}]{Bigazzi:2021ucw}%
  \BibitemOpen
  \bibfield  {author} {\bibinfo {author} {\bibfnamefont {F.}~\bibnamefont
  {Bigazzi}}, \bibinfo {author} {\bibfnamefont {A.}~\bibnamefont {Caddeo}},
  \bibinfo {author} {\bibfnamefont {T.}~\bibnamefont {Canneti}},\ and\ \bibinfo
  {author} {\bibfnamefont {A.~L.}\ \bibnamefont {Cotrone}},\ }\bibfield
  {title} {\bibinfo {title} {{Bubble wall velocity at strong coupling}},\
  }\href {https://doi.org/10.1007/JHEP08(2021)090} {\bibfield  {journal}
  {\bibinfo  {journal} {JHEP}\ }\textbf {\bibinfo {volume} {08}},\ \bibinfo
  {pages} {090}},\ \Eprint {https://arxiv.org/abs/2104.12817} {arXiv:2104.12817
  [hep-ph]} \BibitemShut {NoStop}%
\bibitem [{\citenamefont {Henriksson}(2022)}]{Henriksson:2021zei}%
  \BibitemOpen
  \bibfield  {author} {\bibinfo {author} {\bibfnamefont {O.}~\bibnamefont
  {Henriksson}},\ }\bibfield  {title} {\bibinfo {title} {{Black brane
  evaporation through D-brane bubble nucleation}},\ }\href
  {https://doi.org/10.1103/PhysRevD.105.L041901} {\bibfield  {journal}
  {\bibinfo  {journal} {Phys. Rev. D}\ }\textbf {\bibinfo {volume} {105}},\
  \bibinfo {pages} {L041901} (\bibinfo {year} {2022})},\ \Eprint
  {https://arxiv.org/abs/2106.13254} {arXiv:2106.13254 [hep-th]} \BibitemShut
  {NoStop}%
\bibitem [{\citenamefont {Ares}\ \emph
  {et~al.}(2022{\natexlab{a}})\citenamefont {Ares}, \citenamefont {Henriksson},
  \citenamefont {Hindmarsh}, \citenamefont {Hoyos},\ and\ \citenamefont
  {Jokela}}]{Ares:2021ntv}%
  \BibitemOpen
  \bibfield  {author} {\bibinfo {author} {\bibfnamefont {F.~R.}\ \bibnamefont
  {Ares}}, \bibinfo {author} {\bibfnamefont {O.}~\bibnamefont {Henriksson}},
  \bibinfo {author} {\bibfnamefont {M.}~\bibnamefont {Hindmarsh}}, \bibinfo
  {author} {\bibfnamefont {C.}~\bibnamefont {Hoyos}},\ and\ \bibinfo {author}
  {\bibfnamefont {N.}~\bibnamefont {Jokela}},\ }\bibfield  {title} {\bibinfo
  {title} {{Effective actions and bubble nucleation from holography}},\ }\href
  {https://doi.org/10.1103/PhysRevD.105.066020} {\bibfield  {journal} {\bibinfo
   {journal} {Phys. Rev. D}\ }\textbf {\bibinfo {volume} {105}},\ \bibinfo
  {pages} {066020} (\bibinfo {year} {2022}{\natexlab{a}})},\ \Eprint
  {https://arxiv.org/abs/2109.13784} {arXiv:2109.13784 [hep-th]} \BibitemShut
  {NoStop}%
\bibitem [{\citenamefont {Ares}\ \emph
  {et~al.}(2022{\natexlab{b}})\citenamefont {Ares}, \citenamefont {Henriksson},
  \citenamefont {Hindmarsh}, \citenamefont {Hoyos},\ and\ \citenamefont
  {Jokela}}]{Ares:2021nap}%
  \BibitemOpen
  \bibfield  {author} {\bibinfo {author} {\bibfnamefont {F.~R.}\ \bibnamefont
  {Ares}}, \bibinfo {author} {\bibfnamefont {O.}~\bibnamefont {Henriksson}},
  \bibinfo {author} {\bibfnamefont {M.}~\bibnamefont {Hindmarsh}}, \bibinfo
  {author} {\bibfnamefont {C.}~\bibnamefont {Hoyos}},\ and\ \bibinfo {author}
  {\bibfnamefont {N.}~\bibnamefont {Jokela}},\ }\bibfield  {title} {\bibinfo
  {title} {{Gravitational Waves at Strong Coupling from an Effective Action}},\
  }\href {https://doi.org/10.1103/PhysRevLett.128.131101} {\bibfield  {journal}
  {\bibinfo  {journal} {Phys. Rev. Lett.}\ }\textbf {\bibinfo {volume} {128}},\
  \bibinfo {pages} {131101} (\bibinfo {year} {2022}{\natexlab{b}})},\ \Eprint
  {https://arxiv.org/abs/2110.14442} {arXiv:2110.14442 [hep-th]} \BibitemShut
  {NoStop}%
\bibitem [{\citenamefont {Morgante}\ \emph {et~al.}(2023)\citenamefont
  {Morgante}, \citenamefont {Ramberg},\ and\ \citenamefont
  {Schwaller}}]{Morgante:2022zvc}%
  \BibitemOpen
  \bibfield  {author} {\bibinfo {author} {\bibfnamefont {E.}~\bibnamefont
  {Morgante}}, \bibinfo {author} {\bibfnamefont {N.}~\bibnamefont {Ramberg}},\
  and\ \bibinfo {author} {\bibfnamefont {P.}~\bibnamefont {Schwaller}},\
  }\bibfield  {title} {\bibinfo {title} {{Gravitational waves from dark SU(3)
  Yang-Mills theory}},\ }\href {https://doi.org/10.1103/PhysRevD.107.036010}
  {\bibfield  {journal} {\bibinfo  {journal} {Phys. Rev. D}\ }\textbf {\bibinfo
  {volume} {107}},\ \bibinfo {pages} {036010} (\bibinfo {year} {2023})},\
  \Eprint {https://arxiv.org/abs/2210.11821} {arXiv:2210.11821 [hep-ph]}
  \BibitemShut {NoStop}%
\bibitem [{\citenamefont {Borsanyi}\ \emph
  {et~al.}(2022{\natexlab{a}})\citenamefont {Borsanyi}, \citenamefont {R.},
  \citenamefont {Fodor}, \citenamefont {Godzieba}, \citenamefont {Parotto},\
  and\ \citenamefont {Sexty}}]{borsanyi:2022xml}%
  \BibitemOpen
  \bibfield  {author} {\bibinfo {author} {\bibfnamefont {S.}~\bibnamefont
  {Borsanyi}}, \bibinfo {author} {\bibfnamefont {K.}~\bibnamefont {R.}},
  \bibinfo {author} {\bibfnamefont {Z.}~\bibnamefont {Fodor}}, \bibinfo
  {author} {\bibfnamefont {D.~A.}\ \bibnamefont {Godzieba}}, \bibinfo {author}
  {\bibfnamefont {P.}~\bibnamefont {Parotto}},\ and\ \bibinfo {author}
  {\bibfnamefont {D.}~\bibnamefont {Sexty}},\ }\bibfield  {title} {\bibinfo
  {title} {{Precision study of the continuum SU(3) Yang-Mills theory: How to
  use parallel tempering to improve on supercritical slowing down for first
  order phase transitions}},\ }\href
  {https://doi.org/10.1103/PhysRevD.105.074513} {\bibfield  {journal} {\bibinfo
   {journal} {Phys. Rev. D}\ }\textbf {\bibinfo {volume} {105}},\ \bibinfo
  {pages} {074513} (\bibinfo {year} {2022}{\natexlab{a}})},\ \Eprint
  {https://arxiv.org/abs/2202.05234} {arXiv:2202.05234 [hep-lat]} \BibitemShut
  {NoStop}%
\bibitem [{\citenamefont {Svetitsky}\ and\ \citenamefont
  {Yaffe}(1982)}]{Svetitsky:1982gs}%
  \BibitemOpen
  \bibfield  {author} {\bibinfo {author} {\bibfnamefont {B.}~\bibnamefont
  {Svetitsky}}\ and\ \bibinfo {author} {\bibfnamefont {L.~G.}\ \bibnamefont
  {Yaffe}},\ }\bibfield  {title} {\bibinfo {title} {{Critical Behavior at
  Finite Temperature Confinement Transitions}},\ }\href
  {https://doi.org/10.1016/0550-3213(82)90172-9} {\bibfield  {journal}
  {\bibinfo  {journal} {Nucl. Phys. B}\ }\textbf {\bibinfo {volume} {210}},\
  \bibinfo {pages} {423} (\bibinfo {year} {1982})}\BibitemShut {NoStop}%
\bibitem [{\citenamefont {Yaffe}\ and\ \citenamefont
  {Svetitsky}(1982)}]{Yaffe:1982qf}%
  \BibitemOpen
  \bibfield  {author} {\bibinfo {author} {\bibfnamefont {L.~G.}\ \bibnamefont
  {Yaffe}}\ and\ \bibinfo {author} {\bibfnamefont {B.}~\bibnamefont
  {Svetitsky}},\ }\bibfield  {title} {\bibinfo {title} {{First Order Phase
  Transition in the SU(3) Gauge Theory at Finite Temperature}},\ }\href
  {https://doi.org/10.1103/PhysRevD.26.963} {\bibfield  {journal} {\bibinfo
  {journal} {Phys. Rev. D}\ }\textbf {\bibinfo {volume} {26}},\ \bibinfo
  {pages} {963} (\bibinfo {year} {1982})}\BibitemShut {NoStop}%
\bibitem [{\citenamefont {Saito}\ \emph {et~al.}(2011)\citenamefont {Saito},
  \citenamefont {Ejiri}, \citenamefont {Aoki}, \citenamefont {Hatsuda},
  \citenamefont {Kanaya}, \citenamefont {Maezawa}, \citenamefont {Ohno},\ and\
  \citenamefont {Umeda}}]{Saito:2011fs}%
  \BibitemOpen
  \bibfield  {author} {\bibinfo {author} {\bibfnamefont {H.}~\bibnamefont
  {Saito}}, \bibinfo {author} {\bibfnamefont {S.}~\bibnamefont {Ejiri}},
  \bibinfo {author} {\bibfnamefont {S.}~\bibnamefont {Aoki}}, \bibinfo {author}
  {\bibfnamefont {T.}~\bibnamefont {Hatsuda}}, \bibinfo {author} {\bibfnamefont
  {K.}~\bibnamefont {Kanaya}}, \bibinfo {author} {\bibfnamefont
  {Y.}~\bibnamefont {Maezawa}}, \bibinfo {author} {\bibfnamefont
  {H.}~\bibnamefont {Ohno}},\ and\ \bibinfo {author} {\bibfnamefont
  {T.}~\bibnamefont {Umeda}} (\bibinfo {collaboration} {WHOT-QCD}),\ }\bibfield
   {title} {\bibinfo {title} {{Phase structure of finite temperature QCD in the
  heavy quark region}},\ }\href {https://doi.org/10.1103/PhysRevD.85.079902}
  {\bibfield  {journal} {\bibinfo  {journal} {Phys. Rev. D}\ }\textbf {\bibinfo
  {volume} {84}},\ \bibinfo {pages} {054502} (\bibinfo {year} {2011})},\
  \bibinfo {note} {[Erratum: Phys.Rev.D 85, 079902 (2012)]},\ \Eprint
  {https://arxiv.org/abs/1106.0974} {arXiv:1106.0974 [hep-lat]} \BibitemShut
  {NoStop}%
\bibitem [{\citenamefont {Ejiri}\ \emph {et~al.}(2020)\citenamefont {Ejiri},
  \citenamefont {Itagaki}, \citenamefont {Iwami}, \citenamefont {Kanaya},
  \citenamefont {Kitazawa}, \citenamefont {Kiyohara}, \citenamefont
  {Shirogane},\ and\ \citenamefont {Umeda}}]{Ejiri:2019csa}%
  \BibitemOpen
  \bibfield  {author} {\bibinfo {author} {\bibfnamefont {S.}~\bibnamefont
  {Ejiri}}, \bibinfo {author} {\bibfnamefont {S.}~\bibnamefont {Itagaki}},
  \bibinfo {author} {\bibfnamefont {R.}~\bibnamefont {Iwami}}, \bibinfo
  {author} {\bibfnamefont {K.}~\bibnamefont {Kanaya}}, \bibinfo {author}
  {\bibfnamefont {M.}~\bibnamefont {Kitazawa}}, \bibinfo {author}
  {\bibfnamefont {A.}~\bibnamefont {Kiyohara}}, \bibinfo {author}
  {\bibfnamefont {M.}~\bibnamefont {Shirogane}},\ and\ \bibinfo {author}
  {\bibfnamefont {T.}~\bibnamefont {Umeda}} (\bibinfo {collaboration}
  {WHOT-QCD}),\ }\bibfield  {title} {\bibinfo {title} {{End point of the
  first-order phase transition of QCD in the heavy quark region by reweighting
  from quenched QCD}},\ }\href {https://doi.org/10.1103/PhysRevD.101.054505}
  {\bibfield  {journal} {\bibinfo  {journal} {Phys. Rev. D}\ }\textbf {\bibinfo
  {volume} {101}},\ \bibinfo {pages} {054505} (\bibinfo {year} {2020})},\
  \Eprint {https://arxiv.org/abs/1912.10500} {arXiv:1912.10500 [hep-lat]}
  \BibitemShut {NoStop}%
\bibitem [{\citenamefont {Kiyohara}\ \emph {et~al.}(2021)\citenamefont
  {Kiyohara}, \citenamefont {Kitazawa}, \citenamefont {Ejiri},\ and\
  \citenamefont {Kanaya}}]{Kiyohara:2021smr}%
  \BibitemOpen
  \bibfield  {author} {\bibinfo {author} {\bibfnamefont {A.}~\bibnamefont
  {Kiyohara}}, \bibinfo {author} {\bibfnamefont {M.}~\bibnamefont {Kitazawa}},
  \bibinfo {author} {\bibfnamefont {S.}~\bibnamefont {Ejiri}},\ and\ \bibinfo
  {author} {\bibfnamefont {K.}~\bibnamefont {Kanaya}},\ }\bibfield  {title}
  {\bibinfo {title} {{Finite-size scaling around the critical point in the
  heavy quark region of QCD}},\ }\href
  {https://doi.org/10.1103/PhysRevD.104.114509} {\bibfield  {journal} {\bibinfo
   {journal} {Phys. Rev. D}\ }\textbf {\bibinfo {volume} {104}},\ \bibinfo
  {pages} {114509} (\bibinfo {year} {2021})},\ \Eprint
  {https://arxiv.org/abs/2108.00118} {arXiv:2108.00118 [hep-lat]} \BibitemShut
  {NoStop}%
\bibitem [{\citenamefont {Fromm}\ \emph {et~al.}(2012)\citenamefont {Fromm},
  \citenamefont {Langelage}, \citenamefont {Lottini},\ and\ \citenamefont
  {Philipsen}}]{Fromm:2011qi}%
  \BibitemOpen
  \bibfield  {author} {\bibinfo {author} {\bibfnamefont {M.}~\bibnamefont
  {Fromm}}, \bibinfo {author} {\bibfnamefont {J.}~\bibnamefont {Langelage}},
  \bibinfo {author} {\bibfnamefont {S.}~\bibnamefont {Lottini}},\ and\ \bibinfo
  {author} {\bibfnamefont {O.}~\bibnamefont {Philipsen}},\ }\bibfield  {title}
  {\bibinfo {title} {{The QCD deconfinement transition for heavy quarks and all
  baryon chemical potentials}},\ }\href
  {https://doi.org/10.1007/JHEP01(2012)042} {\bibfield  {journal} {\bibinfo
  {journal} {JHEP}\ }\textbf {\bibinfo {volume} {01}},\ \bibinfo {pages}
  {042}},\ \Eprint {https://arxiv.org/abs/1111.4953} {arXiv:1111.4953
  [hep-lat]} \BibitemShut {NoStop}%
\bibitem [{\citenamefont {Cuteri}\ \emph {et~al.}(2021)\citenamefont {Cuteri},
  \citenamefont {Philipsen}, \citenamefont {Sch\"on},\ and\ \citenamefont
  {Sciarra}}]{Cuteri:2020yke}%
  \BibitemOpen
  \bibfield  {author} {\bibinfo {author} {\bibfnamefont {F.}~\bibnamefont
  {Cuteri}}, \bibinfo {author} {\bibfnamefont {O.}~\bibnamefont {Philipsen}},
  \bibinfo {author} {\bibfnamefont {A.}~\bibnamefont {Sch\"on}},\ and\ \bibinfo
  {author} {\bibfnamefont {A.}~\bibnamefont {Sciarra}},\ }\bibfield  {title}
  {\bibinfo {title} {{Deconfinement critical point of lattice QCD with $N_f$=2
  Wilson fermions}},\ }\href {https://doi.org/10.1103/PhysRevD.103.014513}
  {\bibfield  {journal} {\bibinfo  {journal} {Phys. Rev. D}\ }\textbf {\bibinfo
  {volume} {103}},\ \bibinfo {pages} {014513} (\bibinfo {year} {2021})},\
  \Eprint {https://arxiv.org/abs/2009.14033} {arXiv:2009.14033 [hep-lat]}
  \BibitemShut {NoStop}%
\bibitem [{\citenamefont {Borsanyi}\ \emph
  {et~al.}(2022{\natexlab{b}})\citenamefont {Borsanyi}, \citenamefont {Fodor},
  \citenamefont {Guenther}, \citenamefont {Kara}, \citenamefont {Parotto},
  \citenamefont {Pasztor},\ and\ \citenamefont {Sexty}}]{Borsanyi:2021yoz}%
  \BibitemOpen
  \bibfield  {author} {\bibinfo {author} {\bibfnamefont {S.}~\bibnamefont
  {Borsanyi}}, \bibinfo {author} {\bibfnamefont {Z.}~\bibnamefont {Fodor}},
  \bibinfo {author} {\bibfnamefont {J.~N.}\ \bibnamefont {Guenther}}, \bibinfo
  {author} {\bibfnamefont {R.}~\bibnamefont {Kara}}, \bibinfo {author}
  {\bibfnamefont {P.}~\bibnamefont {Parotto}}, \bibinfo {author} {\bibfnamefont
  {A.}~\bibnamefont {Pasztor}},\ and\ \bibinfo {author} {\bibfnamefont
  {D.}~\bibnamefont {Sexty}},\ }\bibfield  {title} {\bibinfo {title} {{The
  upper right corner of the Columbia plot with staggered fermions}},\ }\href
  {https://doi.org/10.22323/1.396.0496} {\bibfield  {journal} {\bibinfo
  {journal} {PoS}\ }\textbf {\bibinfo {volume} {LATTICE2021}},\ \bibinfo
  {pages} {496} (\bibinfo {year} {2022}{\natexlab{b}})},\ \Eprint
  {https://arxiv.org/abs/2112.04192} {arXiv:2112.04192 [hep-lat]} \BibitemShut
  {NoStop}%
\bibitem [{\citenamefont {Aarts}\ \emph {et~al.}(2023)\citenamefont {Aarts}
  \emph {et~al.}}]{Aarts:2023vsf}%
  \BibitemOpen
  \bibfield  {author} {\bibinfo {author} {\bibfnamefont {G.}~\bibnamefont
  {Aarts}} \emph {et~al.},\ }\bibfield  {title} {\bibinfo {title} {{Phase
  Transitions in Particle Physics - Results and Perspectives from Lattice
  Quantum Chromo-Dynamics}},\ }in\ \href@noop {} {\emph {\bibinfo {booktitle}
  {{Phase Transitions in Particle Physics}: {Results and Perspectives from
  Lattice Quantum Chromo-Dynamics}}}}\ (\bibinfo {year} {2023})\ \Eprint
  {https://arxiv.org/abs/2301.04382} {arXiv:2301.04382 [hep-lat]} \BibitemShut
  {NoStop}%
\bibitem [{\citenamefont {Kajantie}\ \emph {et~al.}(1981)\citenamefont
  {Kajantie}, \citenamefont {Montonen},\ and\ \citenamefont
  {Pietarinen}}]{Kajantie:1981wh}%
  \BibitemOpen
  \bibfield  {author} {\bibinfo {author} {\bibfnamefont {K.}~\bibnamefont
  {Kajantie}}, \bibinfo {author} {\bibfnamefont {C.}~\bibnamefont {Montonen}},\
  and\ \bibinfo {author} {\bibfnamefont {E.}~\bibnamefont {Pietarinen}},\
  }\bibfield  {title} {\bibinfo {title} {{Phase Transition of SU(3) Gauge
  Theory at Finite Temperature}},\ }\href {https://doi.org/10.1007/BF01410665}
  {\bibfield  {journal} {\bibinfo  {journal} {Z. Phys. C}\ }\textbf {\bibinfo
  {volume} {9}},\ \bibinfo {pages} {253} (\bibinfo {year} {1981})}\BibitemShut
  {NoStop}%
\bibitem [{\citenamefont {Celik}\ \emph {et~al.}(1983)\citenamefont {Celik},
  \citenamefont {Engels},\ and\ \citenamefont {Satz}}]{Celik:1983wz}%
  \BibitemOpen
  \bibfield  {author} {\bibinfo {author} {\bibfnamefont {T.}~\bibnamefont
  {Celik}}, \bibinfo {author} {\bibfnamefont {J.}~\bibnamefont {Engels}},\ and\
  \bibinfo {author} {\bibfnamefont {H.}~\bibnamefont {Satz}},\ }\bibfield
  {title} {\bibinfo {title} {{The Order of the Deconfinement Transition in
  SU(3) Yang-Mills Theory}},\ }\href
  {https://doi.org/10.1016/0370-2693(83)91314-X} {\bibfield  {journal}
  {\bibinfo  {journal} {Phys. Lett. B}\ }\textbf {\bibinfo {volume} {125}},\
  \bibinfo {pages} {411} (\bibinfo {year} {1983})}\BibitemShut {NoStop}%
\bibitem [{\citenamefont {Kogut}\ \emph {et~al.}(1983)\citenamefont {Kogut},
  \citenamefont {Matsuoka}, \citenamefont {Stone}, \citenamefont {Wyld},
  \citenamefont {Shenker}, \citenamefont {Shigemitsu},\ and\ \citenamefont
  {Sinclair}}]{Kogut:1983mn}%
  \BibitemOpen
  \bibfield  {author} {\bibinfo {author} {\bibfnamefont {J.~B.}\ \bibnamefont
  {Kogut}}, \bibinfo {author} {\bibfnamefont {H.}~\bibnamefont {Matsuoka}},
  \bibinfo {author} {\bibfnamefont {M.}~\bibnamefont {Stone}}, \bibinfo
  {author} {\bibfnamefont {H.~W.}\ \bibnamefont {Wyld}}, \bibinfo {author}
  {\bibfnamefont {S.~H.}\ \bibnamefont {Shenker}}, \bibinfo {author}
  {\bibfnamefont {J.}~\bibnamefont {Shigemitsu}},\ and\ \bibinfo {author}
  {\bibfnamefont {D.~K.}\ \bibnamefont {Sinclair}},\ }\bibfield  {title}
  {\bibinfo {title} {{Quark and Gluon Latent Heats at the Deconfinement Phase
  Transition in SU(3) Gauge Theory}},\ }\href
  {https://doi.org/10.1103/PhysRevLett.51.869} {\bibfield  {journal} {\bibinfo
  {journal} {Phys. Rev. Lett.}\ }\textbf {\bibinfo {volume} {51}},\ \bibinfo
  {pages} {869} (\bibinfo {year} {1983})}\BibitemShut {NoStop}%
\bibitem [{\citenamefont {Svetitsky}\ and\ \citenamefont
  {Fucito}(1983)}]{Svetitsky:1983bq}%
  \BibitemOpen
  \bibfield  {author} {\bibinfo {author} {\bibfnamefont {B.}~\bibnamefont
  {Svetitsky}}\ and\ \bibinfo {author} {\bibfnamefont {F.}~\bibnamefont
  {Fucito}},\ }\bibfield  {title} {\bibinfo {title} {{Latent Heat of the SU(3)
  Gauge Theory}},\ }\href {https://doi.org/10.1016/0370-2693(83)91112-7}
  {\bibfield  {journal} {\bibinfo  {journal} {Phys. Lett. B}\ }\textbf
  {\bibinfo {volume} {131}},\ \bibinfo {pages} {165} (\bibinfo {year}
  {1983})}\BibitemShut {NoStop}%
\bibitem [{\citenamefont {Gottlieb}\ \emph {et~al.}(1985)\citenamefont
  {Gottlieb}, \citenamefont {Kuti}, \citenamefont {Toussaint}, \citenamefont
  {Kennedy}, \citenamefont {Meyer}, \citenamefont {Pendleton},\ and\
  \citenamefont {Sugar}}]{Gottlieb:1985ug}%
  \BibitemOpen
  \bibfield  {author} {\bibinfo {author} {\bibfnamefont {S.~A.}\ \bibnamefont
  {Gottlieb}}, \bibinfo {author} {\bibfnamefont {J.}~\bibnamefont {Kuti}},
  \bibinfo {author} {\bibfnamefont {D.}~\bibnamefont {Toussaint}}, \bibinfo
  {author} {\bibfnamefont {A.~D.}\ \bibnamefont {Kennedy}}, \bibinfo {author}
  {\bibfnamefont {S.}~\bibnamefont {Meyer}}, \bibinfo {author} {\bibfnamefont
  {B.~J.}\ \bibnamefont {Pendleton}},\ and\ \bibinfo {author} {\bibfnamefont
  {R.~L.}\ \bibnamefont {Sugar}},\ }\bibfield  {title} {\bibinfo {title} {{The
  Deconfining Phase Transition and the Continuum Limit of Lattice Quantum
  Chromodynamics}},\ }\href {https://doi.org/10.1103/PhysRevLett.55.1958}
  {\bibfield  {journal} {\bibinfo  {journal} {Phys. Rev. Lett.}\ }\textbf
  {\bibinfo {volume} {55}},\ \bibinfo {pages} {1958} (\bibinfo {year}
  {1985})}\BibitemShut {NoStop}%
\bibitem [{\citenamefont {Brown}\ \emph {et~al.}(1988)\citenamefont {Brown},
  \citenamefont {Christ}, \citenamefont {Deng}, \citenamefont {Gao},\ and\
  \citenamefont {Woch}}]{Brown:1988qe}%
  \BibitemOpen
  \bibfield  {author} {\bibinfo {author} {\bibfnamefont {F.~R.}\ \bibnamefont
  {Brown}}, \bibinfo {author} {\bibfnamefont {N.~H.}\ \bibnamefont {Christ}},
  \bibinfo {author} {\bibfnamefont {Y.~F.}\ \bibnamefont {Deng}}, \bibinfo
  {author} {\bibfnamefont {M.~S.}\ \bibnamefont {Gao}},\ and\ \bibinfo {author}
  {\bibfnamefont {T.~J.}\ \bibnamefont {Woch}},\ }\bibfield  {title} {\bibinfo
  {title} {{Nature of the Deconfining Phase Transition in SU(3) Lattice Gauge
  Theory}},\ }\href {https://doi.org/10.1103/PhysRevLett.61.2058} {\bibfield
  {journal} {\bibinfo  {journal} {Phys. Rev. Lett.}\ }\textbf {\bibinfo
  {volume} {61}},\ \bibinfo {pages} {2058} (\bibinfo {year}
  {1988})}\BibitemShut {NoStop}%
\bibitem [{\citenamefont {Fukugita}\ \emph {et~al.}(1989)\citenamefont
  {Fukugita}, \citenamefont {Okawa},\ and\ \citenamefont
  {Ukawa}}]{Fukugita:1989yb}%
  \BibitemOpen
  \bibfield  {author} {\bibinfo {author} {\bibfnamefont {M.}~\bibnamefont
  {Fukugita}}, \bibinfo {author} {\bibfnamefont {M.}~\bibnamefont {Okawa}},\
  and\ \bibinfo {author} {\bibfnamefont {A.}~\bibnamefont {Ukawa}},\ }\bibfield
   {title} {\bibinfo {title} {{Order of the Deconfining Phase Transition in
  SU(3) Lattice Gauge Theory}},\ }\href
  {https://doi.org/10.1103/PhysRevLett.63.1768} {\bibfield  {journal} {\bibinfo
   {journal} {Phys. Rev. Lett.}\ }\textbf {\bibinfo {volume} {63}},\ \bibinfo
  {pages} {1768} (\bibinfo {year} {1989})}\BibitemShut {NoStop}%
\bibitem [{\citenamefont {Bacilieri}\ \emph {et~al.}(1989)\citenamefont
  {Bacilieri} \emph {et~al.}}]{Bacilieri:1989ir}%
  \BibitemOpen
  \bibfield  {author} {\bibinfo {author} {\bibfnamefont {P.}~\bibnamefont
  {Bacilieri}} \emph {et~al.},\ }\bibfield  {title} {\bibinfo {title} {{A New
  Computation of the Correlation Length Near the Deconfining Transition in
  SU(3)}},\ }\href {https://doi.org/10.1016/0370-2693(89)91241-0} {\bibfield
  {journal} {\bibinfo  {journal} {Phys. Lett. B}\ }\textbf {\bibinfo {volume}
  {224}},\ \bibinfo {pages} {333} (\bibinfo {year} {1989})}\BibitemShut
  {NoStop}%
\bibitem [{\citenamefont {Alves}\ \emph {et~al.}(1990)\citenamefont {Alves},
  \citenamefont {Berg},\ and\ \citenamefont {Sanielevici}}]{Alves:1990pn}%
  \BibitemOpen
  \bibfield  {author} {\bibinfo {author} {\bibfnamefont {N.~A.}\ \bibnamefont
  {Alves}}, \bibinfo {author} {\bibfnamefont {B.~A.}\ \bibnamefont {Berg}},\
  and\ \bibinfo {author} {\bibfnamefont {S.}~\bibnamefont {Sanielevici}},\
  }\bibfield  {title} {\bibinfo {title} {{Binder Energy Cumulant for SU(3)
  Lattice Gauge Theory}},\ }\href
  {https://doi.org/10.1016/0370-2693(90)91869-D} {\bibfield  {journal}
  {\bibinfo  {journal} {Phys. Lett. B}\ }\textbf {\bibinfo {volume} {241}},\
  \bibinfo {pages} {557} (\bibinfo {year} {1990})}\BibitemShut {NoStop}%
\bibitem [{\citenamefont {Boyd}\ \emph {et~al.}(1995)\citenamefont {Boyd},
  \citenamefont {Engels}, \citenamefont {Karsch}, \citenamefont {Laermann},
  \citenamefont {Legeland}, \citenamefont {Lutgemeier},\ and\ \citenamefont
  {Petersson}}]{Boyd:1995zg}%
  \BibitemOpen
  \bibfield  {author} {\bibinfo {author} {\bibfnamefont {G.}~\bibnamefont
  {Boyd}}, \bibinfo {author} {\bibfnamefont {J.}~\bibnamefont {Engels}},
  \bibinfo {author} {\bibfnamefont {F.}~\bibnamefont {Karsch}}, \bibinfo
  {author} {\bibfnamefont {E.}~\bibnamefont {Laermann}}, \bibinfo {author}
  {\bibfnamefont {C.}~\bibnamefont {Legeland}}, \bibinfo {author}
  {\bibfnamefont {M.}~\bibnamefont {Lutgemeier}},\ and\ \bibinfo {author}
  {\bibfnamefont {B.}~\bibnamefont {Petersson}},\ }\bibfield  {title} {\bibinfo
  {title} {{Equation of state for the SU(3) gauge theory}},\ }\href
  {https://doi.org/10.1103/PhysRevLett.75.4169} {\bibfield  {journal} {\bibinfo
   {journal} {Phys. Rev. Lett.}\ }\textbf {\bibinfo {volume} {75}},\ \bibinfo
  {pages} {4169} (\bibinfo {year} {1995})},\ \Eprint
  {https://arxiv.org/abs/hep-lat/9506025} {arXiv:hep-lat/9506025} \BibitemShut
  {NoStop}%
\bibitem [{\citenamefont {Boyd}\ \emph {et~al.}(1996)\citenamefont {Boyd},
  \citenamefont {Engels}, \citenamefont {Karsch}, \citenamefont {Laermann},
  \citenamefont {Legeland}, \citenamefont {Lutgemeier},\ and\ \citenamefont
  {Petersson}}]{Boyd:1996bx}%
  \BibitemOpen
  \bibfield  {author} {\bibinfo {author} {\bibfnamefont {G.}~\bibnamefont
  {Boyd}}, \bibinfo {author} {\bibfnamefont {J.}~\bibnamefont {Engels}},
  \bibinfo {author} {\bibfnamefont {F.}~\bibnamefont {Karsch}}, \bibinfo
  {author} {\bibfnamefont {E.}~\bibnamefont {Laermann}}, \bibinfo {author}
  {\bibfnamefont {C.}~\bibnamefont {Legeland}}, \bibinfo {author}
  {\bibfnamefont {M.}~\bibnamefont {Lutgemeier}},\ and\ \bibinfo {author}
  {\bibfnamefont {B.}~\bibnamefont {Petersson}},\ }\bibfield  {title} {\bibinfo
  {title} {{Thermodynamics of SU(3) lattice gauge theory}},\ }\href
  {https://doi.org/10.1016/0550-3213(96)00170-8} {\bibfield  {journal}
  {\bibinfo  {journal} {Nucl. Phys. B}\ }\textbf {\bibinfo {volume} {469}},\
  \bibinfo {pages} {419} (\bibinfo {year} {1996})},\ \Eprint
  {https://arxiv.org/abs/hep-lat/9602007} {arXiv:hep-lat/9602007} \BibitemShut
  {NoStop}%
\bibitem [{\citenamefont {Borsanyi}\ \emph {et~al.}(2012)\citenamefont
  {Borsanyi}, \citenamefont {Endrodi}, \citenamefont {Fodor}, \citenamefont
  {Katz},\ and\ \citenamefont {Szabo}}]{Borsanyi:2012ve}%
  \BibitemOpen
  \bibfield  {author} {\bibinfo {author} {\bibfnamefont {S.}~\bibnamefont
  {Borsanyi}}, \bibinfo {author} {\bibfnamefont {G.}~\bibnamefont {Endrodi}},
  \bibinfo {author} {\bibfnamefont {Z.}~\bibnamefont {Fodor}}, \bibinfo
  {author} {\bibfnamefont {S.~D.}\ \bibnamefont {Katz}},\ and\ \bibinfo
  {author} {\bibfnamefont {K.~K.}\ \bibnamefont {Szabo}},\ }\bibfield  {title}
  {\bibinfo {title} {{Precision SU(3) lattice thermodynamics for a large
  temperature range}},\ }\href {https://doi.org/10.1007/JHEP07(2012)056}
  {\bibfield  {journal} {\bibinfo  {journal} {JHEP}\ }\textbf {\bibinfo
  {volume} {07}},\ \bibinfo {pages} {056}},\ \Eprint
  {https://arxiv.org/abs/1204.6184} {arXiv:1204.6184 [hep-lat]} \BibitemShut
  {NoStop}%
\bibitem [{\citenamefont {Shirogane}\ \emph {et~al.}(2016)\citenamefont
  {Shirogane}, \citenamefont {Ejiri}, \citenamefont {Iwami}, \citenamefont
  {Kanaya},\ and\ \citenamefont {Kitazawa}}]{Shirogane:2016zbf}%
  \BibitemOpen
  \bibfield  {author} {\bibinfo {author} {\bibfnamefont {M.}~\bibnamefont
  {Shirogane}}, \bibinfo {author} {\bibfnamefont {S.}~\bibnamefont {Ejiri}},
  \bibinfo {author} {\bibfnamefont {R.}~\bibnamefont {Iwami}}, \bibinfo
  {author} {\bibfnamefont {K.}~\bibnamefont {Kanaya}},\ and\ \bibinfo {author}
  {\bibfnamefont {M.}~\bibnamefont {Kitazawa}},\ }\bibfield  {title} {\bibinfo
  {title} {{Latent heat at the first order phase transition point of SU(3)
  gauge theory}},\ }\href {https://doi.org/10.1103/PhysRevD.94.014506}
  {\bibfield  {journal} {\bibinfo  {journal} {Phys. Rev. D}\ }\textbf {\bibinfo
  {volume} {94}},\ \bibinfo {pages} {014506} (\bibinfo {year} {2016})},\
  \Eprint {https://arxiv.org/abs/1605.02997} {arXiv:1605.02997 [hep-lat]}
  \BibitemShut {NoStop}%
\bibitem [{\citenamefont {Bennett}\ \emph {et~al.}(2018)\citenamefont
  {Bennett}, \citenamefont {Hong}, \citenamefont {Lee}, \citenamefont {Lin},
  \citenamefont {Lucini}, \citenamefont {Piai},\ and\ \citenamefont
  {Vadacchino}}]{Bennett:2017kga}%
  \BibitemOpen
  \bibfield  {author} {\bibinfo {author} {\bibfnamefont {E.}~\bibnamefont
  {Bennett}}, \bibinfo {author} {\bibfnamefont {D.~K.}\ \bibnamefont {Hong}},
  \bibinfo {author} {\bibfnamefont {J.-W.}\ \bibnamefont {Lee}}, \bibinfo
  {author} {\bibfnamefont {C.~J.~D.}\ \bibnamefont {Lin}}, \bibinfo {author}
  {\bibfnamefont {B.}~\bibnamefont {Lucini}}, \bibinfo {author} {\bibfnamefont
  {M.}~\bibnamefont {Piai}},\ and\ \bibinfo {author} {\bibfnamefont
  {D.}~\bibnamefont {Vadacchino}},\ }\bibfield  {title} {\bibinfo {title}
  {{Sp(4) gauge theory on the lattice: towards SU(4)/Sp(4) composite Higgs (and
  beyond)}},\ }\href {https://doi.org/10.1007/JHEP03(2018)185} {\bibfield
  {journal} {\bibinfo  {journal} {JHEP}\ }\textbf {\bibinfo {volume} {03}},\
  \bibinfo {pages} {185}},\ \Eprint {https://arxiv.org/abs/1712.04220}
  {arXiv:1712.04220 [hep-lat]} \BibitemShut {NoStop}%
\bibitem [{\citenamefont {Bennett}\ \emph {et~al.}(2019)\citenamefont
  {Bennett}, \citenamefont {Hong}, \citenamefont {Lee}, \citenamefont {Lin},
  \citenamefont {Lucini}, \citenamefont {Piai},\ and\ \citenamefont
  {Vadacchino}}]{Bennett:2019jzz}%
  \BibitemOpen
  \bibfield  {author} {\bibinfo {author} {\bibfnamefont {E.}~\bibnamefont
  {Bennett}}, \bibinfo {author} {\bibfnamefont {D.~K.}\ \bibnamefont {Hong}},
  \bibinfo {author} {\bibfnamefont {J.-W.}\ \bibnamefont {Lee}}, \bibinfo
  {author} {\bibfnamefont {C.~J.~D.}\ \bibnamefont {Lin}}, \bibinfo {author}
  {\bibfnamefont {B.}~\bibnamefont {Lucini}}, \bibinfo {author} {\bibfnamefont
  {M.}~\bibnamefont {Piai}},\ and\ \bibinfo {author} {\bibfnamefont
  {D.}~\bibnamefont {Vadacchino}},\ }\bibfield  {title} {\bibinfo {title}
  {{Sp(4) gauge theories on the lattice: $N_f=2$ dynamical fundamental
  fermions}},\ }\href {https://doi.org/10.1007/JHEP12(2019)053} {\bibfield
  {journal} {\bibinfo  {journal} {JHEP}\ }\textbf {\bibinfo {volume} {12}},\
  \bibinfo {pages} {053}},\ \Eprint {https://arxiv.org/abs/1909.12662}
  {arXiv:1909.12662 [hep-lat]} \BibitemShut {NoStop}%
\bibitem [{\citenamefont {Bennett}\ \emph
  {et~al.}(2020{\natexlab{a}})\citenamefont {Bennett}, \citenamefont {Hong},
  \citenamefont {Lee}, \citenamefont {Lin}, \citenamefont {Lucini},
  \citenamefont {Mesiti}, \citenamefont {Piai}, \citenamefont {Rantaharju},\
  and\ \citenamefont {Vadacchino}}]{Bennett:2019cxd}%
  \BibitemOpen
  \bibfield  {author} {\bibinfo {author} {\bibfnamefont {E.}~\bibnamefont
  {Bennett}}, \bibinfo {author} {\bibfnamefont {D.~K.}\ \bibnamefont {Hong}},
  \bibinfo {author} {\bibfnamefont {J.-W.}\ \bibnamefont {Lee}}, \bibinfo
  {author} {\bibfnamefont {C.-J.~D.}\ \bibnamefont {Lin}}, \bibinfo {author}
  {\bibfnamefont {B.}~\bibnamefont {Lucini}}, \bibinfo {author} {\bibfnamefont
  {M.}~\bibnamefont {Mesiti}}, \bibinfo {author} {\bibfnamefont
  {M.}~\bibnamefont {Piai}}, \bibinfo {author} {\bibfnamefont {J.}~\bibnamefont
  {Rantaharju}},\ and\ \bibinfo {author} {\bibfnamefont {D.}~\bibnamefont
  {Vadacchino}},\ }\bibfield  {title} {\bibinfo {title} {{$Sp(4)$ gauge
  theories on the lattice: quenched fundamental and antisymmetric fermions}},\
  }\href {https://doi.org/10.1103/PhysRevD.101.074516} {\bibfield  {journal}
  {\bibinfo  {journal} {Phys. Rev. D}\ }\textbf {\bibinfo {volume} {101}},\
  \bibinfo {pages} {074516} (\bibinfo {year} {2020}{\natexlab{a}})},\ \Eprint
  {https://arxiv.org/abs/1912.06505} {arXiv:1912.06505 [hep-lat]} \BibitemShut
  {NoStop}%
\bibitem [{\citenamefont {Bennett}\ \emph
  {et~al.}(2020{\natexlab{b}})\citenamefont {Bennett}, \citenamefont
  {Holligan}, \citenamefont {Hong}, \citenamefont {Lee}, \citenamefont {Lin},
  \citenamefont {Lucini}, \citenamefont {Piai},\ and\ \citenamefont
  {Vadacchino}}]{Bennett:2020hqd}%
  \BibitemOpen
  \bibfield  {author} {\bibinfo {author} {\bibfnamefont {E.}~\bibnamefont
  {Bennett}}, \bibinfo {author} {\bibfnamefont {J.}~\bibnamefont {Holligan}},
  \bibinfo {author} {\bibfnamefont {D.~K.}\ \bibnamefont {Hong}}, \bibinfo
  {author} {\bibfnamefont {J.-W.}\ \bibnamefont {Lee}}, \bibinfo {author}
  {\bibfnamefont {C.~J.~D.}\ \bibnamefont {Lin}}, \bibinfo {author}
  {\bibfnamefont {B.}~\bibnamefont {Lucini}}, \bibinfo {author} {\bibfnamefont
  {M.}~\bibnamefont {Piai}},\ and\ \bibinfo {author} {\bibfnamefont
  {D.}~\bibnamefont {Vadacchino}},\ }\bibfield  {title} {\bibinfo {title}
  {{Color dependence of tensor and scalar glueball masses in Yang-Mills
  theories}},\ }\href {https://doi.org/10.1103/PhysRevD.102.011501} {\bibfield
  {journal} {\bibinfo  {journal} {Phys. Rev. D}\ }\textbf {\bibinfo {volume}
  {102}},\ \bibinfo {pages} {011501} (\bibinfo {year} {2020}{\natexlab{b}})},\
  \Eprint {https://arxiv.org/abs/2004.11063} {arXiv:2004.11063 [hep-lat]}
  \BibitemShut {NoStop}%
\bibitem [{\citenamefont {Bennett}\ \emph {et~al.}(2021)\citenamefont
  {Bennett}, \citenamefont {Holligan}, \citenamefont {Hong}, \citenamefont
  {Lee}, \citenamefont {Lin}, \citenamefont {Lucini}, \citenamefont {Piai},\
  and\ \citenamefont {Vadacchino}}]{Bennett:2020qtj}%
  \BibitemOpen
  \bibfield  {author} {\bibinfo {author} {\bibfnamefont {E.}~\bibnamefont
  {Bennett}}, \bibinfo {author} {\bibfnamefont {J.}~\bibnamefont {Holligan}},
  \bibinfo {author} {\bibfnamefont {D.~K.}\ \bibnamefont {Hong}}, \bibinfo
  {author} {\bibfnamefont {J.-W.}\ \bibnamefont {Lee}}, \bibinfo {author}
  {\bibfnamefont {C.~J.~D.}\ \bibnamefont {Lin}}, \bibinfo {author}
  {\bibfnamefont {B.}~\bibnamefont {Lucini}}, \bibinfo {author} {\bibfnamefont
  {M.}~\bibnamefont {Piai}},\ and\ \bibinfo {author} {\bibfnamefont
  {D.}~\bibnamefont {Vadacchino}},\ }\bibfield  {title} {\bibinfo {title}
  {{Glueballs and strings in $Sp(2N)$ Yang-Mills theories}},\ }\href
  {https://doi.org/10.1103/PhysRevD.103.054509} {\bibfield  {journal} {\bibinfo
   {journal} {Phys. Rev. D}\ }\textbf {\bibinfo {volume} {103}},\ \bibinfo
  {pages} {054509} (\bibinfo {year} {2021})},\ \Eprint
  {https://arxiv.org/abs/2010.15781} {arXiv:2010.15781 [hep-lat]} \BibitemShut
  {NoStop}%
\bibitem [{\citenamefont {Bennett}\ \emph
  {et~al.}(2022{\natexlab{a}})\citenamefont {Bennett}, \citenamefont {Hong},
  \citenamefont {Hsiao}, \citenamefont {Lee}, \citenamefont {Lin},
  \citenamefont {Lucini}, \citenamefont {Mesiti}, \citenamefont {Piai},\ and\
  \citenamefont {Vadacchino}}]{Bennett:2022yfa}%
  \BibitemOpen
  \bibfield  {author} {\bibinfo {author} {\bibfnamefont {E.}~\bibnamefont
  {Bennett}}, \bibinfo {author} {\bibfnamefont {D.~K.}\ \bibnamefont {Hong}},
  \bibinfo {author} {\bibfnamefont {H.}~\bibnamefont {Hsiao}}, \bibinfo
  {author} {\bibfnamefont {J.-W.}\ \bibnamefont {Lee}}, \bibinfo {author}
  {\bibfnamefont {C.~J.~D.}\ \bibnamefont {Lin}}, \bibinfo {author}
  {\bibfnamefont {B.}~\bibnamefont {Lucini}}, \bibinfo {author} {\bibfnamefont
  {M.}~\bibnamefont {Mesiti}}, \bibinfo {author} {\bibfnamefont
  {M.}~\bibnamefont {Piai}},\ and\ \bibinfo {author} {\bibfnamefont
  {D.}~\bibnamefont {Vadacchino}},\ }\bibfield  {title} {\bibinfo {title}
  {{Lattice studies of the Sp(4) gauge theory with two fundamental and three
  antisymmetric Dirac fermions}},\ }\href
  {https://doi.org/10.1103/PhysRevD.106.014501} {\bibfield  {journal} {\bibinfo
   {journal} {Phys. Rev. D}\ }\textbf {\bibinfo {volume} {106}},\ \bibinfo
  {pages} {014501} (\bibinfo {year} {2022}{\natexlab{a}})},\ \Eprint
  {https://arxiv.org/abs/2202.05516} {arXiv:2202.05516 [hep-lat]} \BibitemShut
  {NoStop}%
\bibitem [{\citenamefont {Bennett}\ \emph
  {et~al.}(2022{\natexlab{b}})\citenamefont {Bennett}, \citenamefont {Hong},
  \citenamefont {Lee}, \citenamefont {Lin}, \citenamefont {Lucini},
  \citenamefont {Piai},\ and\ \citenamefont {Vadacchino}}]{Bennett:2022ftz}%
  \BibitemOpen
  \bibfield  {author} {\bibinfo {author} {\bibfnamefont {E.}~\bibnamefont
  {Bennett}}, \bibinfo {author} {\bibfnamefont {D.~K.}\ \bibnamefont {Hong}},
  \bibinfo {author} {\bibfnamefont {J.-W.}\ \bibnamefont {Lee}}, \bibinfo
  {author} {\bibfnamefont {C.~J.~D.}\ \bibnamefont {Lin}}, \bibinfo {author}
  {\bibfnamefont {B.}~\bibnamefont {Lucini}}, \bibinfo {author} {\bibfnamefont
  {M.}~\bibnamefont {Piai}},\ and\ \bibinfo {author} {\bibfnamefont
  {D.}~\bibnamefont {Vadacchino}},\ }\bibfield  {title} {\bibinfo {title}
  {{Sp(2N) Yang-Mills theories on the lattice: Scale setting and topology}},\
  }\href {https://doi.org/10.1103/PhysRevD.106.094503} {\bibfield  {journal}
  {\bibinfo  {journal} {Phys. Rev. D}\ }\textbf {\bibinfo {volume} {106}},\
  \bibinfo {pages} {094503} (\bibinfo {year} {2022}{\natexlab{b}})},\ \Eprint
  {https://arxiv.org/abs/2205.09364} {arXiv:2205.09364 [hep-lat]} \BibitemShut
  {NoStop}%
\bibitem [{\citenamefont {Bennett}\ \emph
  {et~al.}(2022{\natexlab{c}})\citenamefont {Bennett}, \citenamefont {Hong},
  \citenamefont {Lee}, \citenamefont {Lin}, \citenamefont {Lucini},
  \citenamefont {Piai},\ and\ \citenamefont {Vadacchino}}]{Bennett:2022gdz}%
  \BibitemOpen
  \bibfield  {author} {\bibinfo {author} {\bibfnamefont {E.}~\bibnamefont
  {Bennett}}, \bibinfo {author} {\bibfnamefont {D.~K.}\ \bibnamefont {Hong}},
  \bibinfo {author} {\bibfnamefont {J.-W.}\ \bibnamefont {Lee}}, \bibinfo
  {author} {\bibfnamefont {C.~J.~D.}\ \bibnamefont {Lin}}, \bibinfo {author}
  {\bibfnamefont {B.}~\bibnamefont {Lucini}}, \bibinfo {author} {\bibfnamefont
  {M.}~\bibnamefont {Piai}},\ and\ \bibinfo {author} {\bibfnamefont
  {D.}~\bibnamefont {Vadacchino}},\ }\bibfield  {title} {\bibinfo {title}
  {{Color dependence of the topological susceptibility in Yang-Mills
  theories}},\ }\href {https://doi.org/10.1016/j.physletb.2022.137504}
  {\bibfield  {journal} {\bibinfo  {journal} {Phys. Lett. B}\ }\textbf
  {\bibinfo {volume} {835}},\ \bibinfo {pages} {137504} (\bibinfo {year}
  {2022}{\natexlab{c}})},\ \Eprint {https://arxiv.org/abs/2205.09254}
  {arXiv:2205.09254 [hep-lat]} \BibitemShut {NoStop}%
\bibitem [{\citenamefont {Bennett}\ \emph
  {et~al.}(2023{\natexlab{a}})\citenamefont {Bennett}, \citenamefont
  {Holligan}, \citenamefont {Hong}, \citenamefont {Hsiao}, \citenamefont {Lee},
  \citenamefont {Lin}, \citenamefont {Lucini}, \citenamefont {Mesiti},
  \citenamefont {Piai},\ and\ \citenamefont {Vadacchino}}]{Bennett:2023wjw}%
  \BibitemOpen
  \bibfield  {author} {\bibinfo {author} {\bibfnamefont {E.}~\bibnamefont
  {Bennett}}, \bibinfo {author} {\bibfnamefont {J.}~\bibnamefont {Holligan}},
  \bibinfo {author} {\bibfnamefont {D.~K.}\ \bibnamefont {Hong}}, \bibinfo
  {author} {\bibfnamefont {H.}~\bibnamefont {Hsiao}}, \bibinfo {author}
  {\bibfnamefont {J.-W.}\ \bibnamefont {Lee}}, \bibinfo {author} {\bibfnamefont
  {C.~J.~D.}\ \bibnamefont {Lin}}, \bibinfo {author} {\bibfnamefont
  {B.}~\bibnamefont {Lucini}}, \bibinfo {author} {\bibfnamefont
  {M.}~\bibnamefont {Mesiti}}, \bibinfo {author} {\bibfnamefont
  {M.}~\bibnamefont {Piai}},\ and\ \bibinfo {author} {\bibfnamefont
  {D.}~\bibnamefont {Vadacchino}},\ }\bibfield  {title} {\bibinfo {title}
  {{$Sp(2N)$ Lattice Gauge Theories and Extensions of the Standard Model of
  Particle Physics}}\ }(\bibinfo {year} {2023})\ \Eprint
  {https://arxiv.org/abs/2304.01070} {arXiv:2304.01070 [hep-lat]} \BibitemShut
  {NoStop}%
\bibitem [{\citenamefont {Bennett}\ \emph
  {et~al.}(2023{\natexlab{b}})\citenamefont {Bennett}, \citenamefont {Hsiao},
  \citenamefont {Lee}, \citenamefont {Lucini}, \citenamefont {Maas},
  \citenamefont {Piai},\ and\ \citenamefont {Zierler}}]{Bennett:2023rsl}%
  \BibitemOpen
  \bibfield  {author} {\bibinfo {author} {\bibfnamefont {E.}~\bibnamefont
  {Bennett}}, \bibinfo {author} {\bibfnamefont {H.}~\bibnamefont {Hsiao}},
  \bibinfo {author} {\bibfnamefont {J.-W.}\ \bibnamefont {Lee}}, \bibinfo
  {author} {\bibfnamefont {B.}~\bibnamefont {Lucini}}, \bibinfo {author}
  {\bibfnamefont {A.}~\bibnamefont {Maas}}, \bibinfo {author} {\bibfnamefont
  {M.}~\bibnamefont {Piai}},\ and\ \bibinfo {author} {\bibfnamefont
  {F.}~\bibnamefont {Zierler}},\ }\href@noop {} {\bibinfo {title} {{Singlets in
  gauge theories with fundamental matter}}} (\bibinfo {year}
  {2023}{\natexlab{b}}),\ \Eprint {https://arxiv.org/abs/2304.07191}
  {arXiv:2304.07191 [hep-lat]} \BibitemShut {NoStop}%
\bibitem [{\citenamefont {Langfeld}\ \emph {et~al.}(2012)\citenamefont
  {Langfeld}, \citenamefont {Lucini},\ and\ \citenamefont
  {Rago}}]{Langfeld:2012ah}%
  \BibitemOpen
  \bibfield  {author} {\bibinfo {author} {\bibfnamefont {K.}~\bibnamefont
  {Langfeld}}, \bibinfo {author} {\bibfnamefont {B.}~\bibnamefont {Lucini}},\
  and\ \bibinfo {author} {\bibfnamefont {A.}~\bibnamefont {Rago}},\ }\bibfield
  {title} {\bibinfo {title} {{The density of states in gauge theories}},\
  }\href {https://doi.org/10.1103/PhysRevLett.109.111601} {\bibfield  {journal}
  {\bibinfo  {journal} {Phys. Rev. Lett.}\ }\textbf {\bibinfo {volume} {109}},\
  \bibinfo {pages} {111601} (\bibinfo {year} {2012})},\ \Eprint
  {https://arxiv.org/abs/1204.3243} {arXiv:1204.3243 [hep-lat]} \BibitemShut
  {NoStop}%
\bibitem [{\citenamefont {Langfeld}\ and\ \citenamefont
  {Pawlowski}(2013)}]{Langfeld:2013xbf}%
  \BibitemOpen
  \bibfield  {author} {\bibinfo {author} {\bibfnamefont {K.}~\bibnamefont
  {Langfeld}}\ and\ \bibinfo {author} {\bibfnamefont {J.~M.}\ \bibnamefont
  {Pawlowski}},\ }\bibfield  {title} {\bibinfo {title} {{Two-color QCD with
  heavy quarks at finite densities}},\ }\href
  {https://doi.org/10.1103/PhysRevD.88.071502} {\bibfield  {journal} {\bibinfo
  {journal} {Phys. Rev. D}\ }\textbf {\bibinfo {volume} {88}},\ \bibinfo
  {pages} {071502} (\bibinfo {year} {2013})},\ \Eprint
  {https://arxiv.org/abs/1307.0455} {arXiv:1307.0455 [hep-lat]} \BibitemShut
  {NoStop}%
\bibitem [{\citenamefont {Langfeld}\ \emph {et~al.}(2016)\citenamefont
  {Langfeld}, \citenamefont {Lucini}, \citenamefont {Pellegrini},\ and\
  \citenamefont {Rago}}]{Langfeld:2015fua}%
  \BibitemOpen
  \bibfield  {author} {\bibinfo {author} {\bibfnamefont {K.}~\bibnamefont
  {Langfeld}}, \bibinfo {author} {\bibfnamefont {B.}~\bibnamefont {Lucini}},
  \bibinfo {author} {\bibfnamefont {R.}~\bibnamefont {Pellegrini}},\ and\
  \bibinfo {author} {\bibfnamefont {A.}~\bibnamefont {Rago}},\ }\bibfield
  {title} {\bibinfo {title} {{An efficient algorithm for numerical computations
  of continuous densities of states}},\ }\href
  {https://doi.org/10.1140/epjc/s10052-016-4142-5} {\bibfield  {journal}
  {\bibinfo  {journal} {Eur. Phys. J. C}\ }\textbf {\bibinfo {volume} {76}},\
  \bibinfo {pages} {306} (\bibinfo {year} {2016})},\ \Eprint
  {https://arxiv.org/abs/1509.08391} {arXiv:1509.08391 [hep-lat]} \BibitemShut
  {NoStop}%
\bibitem [{\citenamefont {Cossu}\ \emph {et~al.}(2021)\citenamefont {Cossu},
  \citenamefont {Lancaster}, \citenamefont {Lucini}, \citenamefont
  {Pellegrini},\ and\ \citenamefont {Rago}}]{Cossu:2021bgn}%
  \BibitemOpen
  \bibfield  {author} {\bibinfo {author} {\bibfnamefont {G.}~\bibnamefont
  {Cossu}}, \bibinfo {author} {\bibfnamefont {D.}~\bibnamefont {Lancaster}},
  \bibinfo {author} {\bibfnamefont {B.}~\bibnamefont {Lucini}}, \bibinfo
  {author} {\bibfnamefont {R.}~\bibnamefont {Pellegrini}},\ and\ \bibinfo
  {author} {\bibfnamefont {A.}~\bibnamefont {Rago}},\ }\bibfield  {title}
  {\bibinfo {title} {{Ergodic sampling of the topological charge using the
  density of states}},\ }\href
  {https://doi.org/10.1140/epjc/s10052-021-09161-1} {\bibfield  {journal}
  {\bibinfo  {journal} {Eur. Phys. J. C}\ }\textbf {\bibinfo {volume} {81}},\
  \bibinfo {pages} {375} (\bibinfo {year} {2021})},\ \Eprint
  {https://arxiv.org/abs/2102.03630} {arXiv:2102.03630 [hep-lat]} \BibitemShut
  {NoStop}%
\bibitem [{\citenamefont {Springer}\ and\ \citenamefont
  {Schaich}(2022{\natexlab{a}})}]{Springer:2021liy}%
  \BibitemOpen
  \bibfield  {author} {\bibinfo {author} {\bibfnamefont {F.}~\bibnamefont
  {Springer}}\ and\ \bibinfo {author} {\bibfnamefont {D.}~\bibnamefont
  {Schaich}},\ }\bibfield  {title} {\bibinfo {title} {{Density of states for
  gravitational waves}},\ }\href {https://doi.org/10.22323/1.396.0043}
  {\bibfield  {journal} {\bibinfo  {journal} {PoS}\ }\textbf {\bibinfo {volume}
  {LATTICE2021}},\ \bibinfo {pages} {043} (\bibinfo {year}
  {2022}{\natexlab{a}})},\ \Eprint {https://arxiv.org/abs/2112.11868}
  {arXiv:2112.11868 [hep-lat]} \BibitemShut {NoStop}%
\bibitem [{\citenamefont {Springer}\ and\ \citenamefont
  {Schaich}(2022{\natexlab{b}})}]{Springer:2022qos}%
  \BibitemOpen
  \bibfield  {author} {\bibinfo {author} {\bibfnamefont {F.}~\bibnamefont
  {Springer}}\ and\ \bibinfo {author} {\bibfnamefont {D.}~\bibnamefont
  {Schaich}},\ }\bibfield  {title} {\bibinfo {title} {{Progress applying
  density of states for gravitational waves}},\ }\href
  {https://doi.org/10.1051/epjconf/202227408008} {\bibfield  {journal}
  {\bibinfo  {journal} {EPJ Web Conf.}\ }\textbf {\bibinfo {volume} {274}},\
  \bibinfo {pages} {08008} (\bibinfo {year} {2022}{\natexlab{b}})},\ \Eprint
  {https://arxiv.org/abs/2212.09199} {arXiv:2212.09199 [hep-lat]} \BibitemShut
  {NoStop}%
\bibitem [{\citenamefont {Springer}\ and\ \citenamefont
  {Schaich}(2023)}]{Springer:2023wok}%
  \BibitemOpen
  \bibfield  {author} {\bibinfo {author} {\bibfnamefont {F.}~\bibnamefont
  {Springer}}\ and\ \bibinfo {author} {\bibfnamefont {D.}~\bibnamefont
  {Schaich}} (\bibinfo {collaboration} {Lattice Strong Dynamics (LSD)}),\
  }\bibfield  {title} {\bibinfo {title} {{Advances in using density of states
  for large-N Yang\textendash{}Mills}},\ }\href
  {https://doi.org/10.22323/1.430.0223} {\bibfield  {journal} {\bibinfo
  {journal} {PoS}\ }\textbf {\bibinfo {volume} {LATTICE2022}},\ \bibinfo
  {pages} {223} (\bibinfo {year} {2023})}\BibitemShut {NoStop}%
\bibitem [{\citenamefont {Mason}\ \emph {et~al.}(2022)\citenamefont {Mason},
  \citenamefont {Lucini}, \citenamefont {Piai}, \citenamefont {Rinaldi},\ and\
  \citenamefont {Vadacchino}}]{Mason:2022trc}%
  \BibitemOpen
  \bibfield  {author} {\bibinfo {author} {\bibfnamefont {D.}~\bibnamefont
  {Mason}}, \bibinfo {author} {\bibfnamefont {B.}~\bibnamefont {Lucini}},
  \bibinfo {author} {\bibfnamefont {M.}~\bibnamefont {Piai}}, \bibinfo {author}
  {\bibfnamefont {E.}~\bibnamefont {Rinaldi}},\ and\ \bibinfo {author}
  {\bibfnamefont {D.}~\bibnamefont {Vadacchino}},\ }\bibfield  {title}
  {\bibinfo {title} {{The density of states method in Yang-Mills theories and
  first order phase transitions}},\ }\href
  {https://doi.org/10.1051/epjconf/202227408007} {\bibfield  {journal}
  {\bibinfo  {journal} {EPJ Web Conf.}\ }\textbf {\bibinfo {volume} {274}},\
  \bibinfo {pages} {08007} (\bibinfo {year} {2022})},\ \Eprint
  {https://arxiv.org/abs/2211.10373} {arXiv:2211.10373 [hep-lat]} \BibitemShut
  {NoStop}%
\bibitem [{\citenamefont {Mason}\ \emph {et~al.}(2023)\citenamefont {Mason},
  \citenamefont {Lucini}, \citenamefont {Piai}, \citenamefont {Rinaldi},\ and\
  \citenamefont {Vadacchino}}]{Mason:2022aka}%
  \BibitemOpen
  \bibfield  {author} {\bibinfo {author} {\bibfnamefont {D.}~\bibnamefont
  {Mason}}, \bibinfo {author} {\bibfnamefont {B.}~\bibnamefont {Lucini}},
  \bibinfo {author} {\bibfnamefont {M.}~\bibnamefont {Piai}}, \bibinfo {author}
  {\bibfnamefont {E.}~\bibnamefont {Rinaldi}},\ and\ \bibinfo {author}
  {\bibfnamefont {D.}~\bibnamefont {Vadacchino}},\ }\bibfield  {title}
  {\bibinfo {title} {{The density of state method for first-order phase
  transitions in Yang-Mills theories}},\ }\href
  {https://doi.org/10.22323/1.430.0216} {\bibfield  {journal} {\bibinfo
  {journal} {PoS}\ }\textbf {\bibinfo {volume} {LATTICE2022}},\ \bibinfo
  {pages} {216} (\bibinfo {year} {2023})},\ \Eprint
  {https://arxiv.org/abs/2212.01074} {arXiv:2212.01074 [hep-lat]} \BibitemShut
  {NoStop}%
\bibitem [{\citenamefont {Lucini}\ \emph {et~al.}(2016)\citenamefont {Lucini},
  \citenamefont {Fall},\ and\ \citenamefont {Langfeld}}]{Lucini:2016fid}%
  \BibitemOpen
  \bibfield  {author} {\bibinfo {author} {\bibfnamefont {B.}~\bibnamefont
  {Lucini}}, \bibinfo {author} {\bibfnamefont {W.}~\bibnamefont {Fall}},\ and\
  \bibinfo {author} {\bibfnamefont {K.}~\bibnamefont {Langfeld}},\ }\bibfield
  {title} {\bibinfo {title} {{Overcoming strong metastabilities with the LLR
  method}},\ }\href {https://doi.org/10.22323/1.256.0275} {\bibfield  {journal}
  {\bibinfo  {journal} {PoS}\ }\textbf {\bibinfo {volume} {LATTICE2016}},\
  \bibinfo {pages} {275} (\bibinfo {year} {2016})},\ \Eprint
  {https://arxiv.org/abs/1611.00019} {arXiv:1611.00019 [hep-lat]} \BibitemShut
  {NoStop}%
\bibitem [{\citenamefont {Robbins}\ and\ \citenamefont
  {Monro}(1951)}]{Robbins&Monro:1951}%
  \BibitemOpen
  \bibfield  {author} {\bibinfo {author} {\bibfnamefont {H.}~\bibnamefont
  {Robbins}}\ and\ \bibinfo {author} {\bibfnamefont {S.}~\bibnamefont
  {Monro}},\ }\bibfield  {title} {\bibinfo {title} {A stochastic approximation
  method},\ }\href@noop {} {\bibfield  {journal} {\bibinfo  {journal} {Annals
  of Mathematical Statistics}\ }\textbf {\bibinfo {volume} {22}},\ \bibinfo
  {pages} {400} (\bibinfo {year} {1951})}\BibitemShut {NoStop}%
\bibitem [{\citenamefont {Karsch}(1982)}]{Karsch:1982ve}%
  \BibitemOpen
  \bibfield  {author} {\bibinfo {author} {\bibfnamefont {F.}~\bibnamefont
  {Karsch}},\ }\bibfield  {title} {\bibinfo {title} {{SU(N) Gauge Theory
  Couplings on Asymmetric Lattices}},\ }\href
  {https://doi.org/10.1016/0550-3213(82)90390-X} {\bibfield  {journal}
  {\bibinfo  {journal} {Nucl. Phys. B}\ }\textbf {\bibinfo {volume} {205}},\
  \bibinfo {pages} {285} (\bibinfo {year} {1982})}\BibitemShut {NoStop}%
\bibitem [{\citenamefont {Lucini}\ \emph
  {et~al.}(2023{\natexlab{a}})\citenamefont {Lucini}, \citenamefont {Mason},
  \citenamefont {Piai}, \citenamefont {Rinaldi},\ and\ \citenamefont
  {Vadacchino}}]{LMPRV}%
  \BibitemOpen
  \bibfield  {author} {\bibinfo {author} {\bibfnamefont {B.}~\bibnamefont
  {Lucini}}, \bibinfo {author} {\bibfnamefont {D.}~\bibnamefont {Mason}},
  \bibinfo {author} {\bibfnamefont {M.}~\bibnamefont {Piai}}, \bibinfo {author}
  {\bibfnamefont {E.}~\bibnamefont {Rinaldi}},\ and\ \bibinfo {author}
  {\bibfnamefont {D.}~\bibnamefont {Vadacchino}},\ }\href
  {https://doi.org/10.5281/zenodo.8124749} {\bibinfo {title} {{First-order
  phase transitions in Yang-Mills theories and the density of state
  method---data and analysis code release}}} (\bibinfo {year}
  {2023}{\natexlab{a}}),\ \Eprint
  {https://arxiv.org/abs/10.5281/zenodo.8124749} {DOI:10.5281/zenodo.8124749}
  \BibitemShut {NoStop}%
\bibitem [{\citenamefont {Lucini}\ \emph
  {et~al.}(2023{\natexlab{b}})\citenamefont {Lucini}, \citenamefont {Mason},
  \citenamefont {Piai}, \citenamefont {Rinaldi},\ and\ \citenamefont
  {Vadacchino}}]{LMPRV2}%
  \BibitemOpen
  \bibfield  {author} {\bibinfo {author} {\bibfnamefont {B.}~\bibnamefont
  {Lucini}}, \bibinfo {author} {\bibfnamefont {D.}~\bibnamefont {Mason}},
  \bibinfo {author} {\bibfnamefont {M.}~\bibnamefont {Piai}}, \bibinfo {author}
  {\bibfnamefont {E.}~\bibnamefont {Rinaldi}},\ and\ \bibinfo {author}
  {\bibfnamefont {D.}~\bibnamefont {Vadacchino}},\ }\href
  {https://doi.org/10.5281/zenodo.8134756} {\bibinfo {title} {{First-order
  phase transitions in Yang-Mills theories and the density of state method ---
  HiRep LLR Code v1.0.0}}} (\bibinfo {year} {2023}{\natexlab{b}}),\ \Eprint
  {https://arxiv.org/abs/10.5281/zenodo.8134756} {DOI:10.5281/zenodo.8134756}
  \BibitemShut {NoStop}%
\bibitem [{\citenamefont {Creutz}(1980)}]{Creutz:1980zw}%
  \BibitemOpen
  \bibfield  {author} {\bibinfo {author} {\bibfnamefont {M.}~\bibnamefont
  {Creutz}},\ }\bibfield  {title} {\bibinfo {title} {{Monte Carlo Study of
  Quantized SU(2) Gauge Theory}},\ }\href
  {https://doi.org/10.1103/PhysRevD.21.2308} {\bibfield  {journal} {\bibinfo
  {journal} {Phys. Rev. D}\ }\textbf {\bibinfo {volume} {21}},\ \bibinfo
  {pages} {2308} (\bibinfo {year} {1980})}\BibitemShut {NoStop}%
\bibitem [{\citenamefont {Cabibbo}\ and\ \citenamefont
  {Marinari}(1982)}]{Cabibbo:1982zn}%
  \BibitemOpen
  \bibfield  {author} {\bibinfo {author} {\bibfnamefont {N.}~\bibnamefont
  {Cabibbo}}\ and\ \bibinfo {author} {\bibfnamefont {E.}~\bibnamefont
  {Marinari}},\ }\bibfield  {title} {\bibinfo {title} {{A New Method for
  Updating SU(N) Matrices in Computer Simulations of Gauge Theories}},\ }\href
  {https://doi.org/10.1016/0370-2693(82)90696-7} {\bibfield  {journal}
  {\bibinfo  {journal} {Phys. Lett. B}\ }\textbf {\bibinfo {volume} {119}},\
  \bibinfo {pages} {387} (\bibinfo {year} {1982})}\BibitemShut {NoStop}%
\end{thebibliography}%

\end{document}